\documentclass[12pt,english]{article}
\usepackage{mathptmx}

\usepackage[T1]{fontenc}
\usepackage[latin9]{inputenc}
\usepackage[letterpaper]{geometry}
\usepackage{color}
\usepackage{babel}
\usepackage{float}
\usepackage{mathtools}
\usepackage{amsmath}
\usepackage{graphicx}
\usepackage{setspace}
\usepackage[authoryear]{natbib}
\usepackage[unicode=true,pdfusetitle,
 bookmarks=true,bookmarksnumbered=false,bookmarksopen=false,
 breaklinks=true,pdfborder={0 0 0},pdfborderstyle={},backref=false,colorlinks=true]
 {hyperref}
\hypersetup{
 citecolor=DarkBlue,linkcolor=DarkBlue,urlcolor=DarkBlue}

\makeatletter
\usepackage[svgnames]{xcolor}
\usepackage{bbm}

\@ifundefined{showcaptionsetup}{}{%
 \PassOptionsToPackage{caption=false}{subfig}}
\usepackage{subfig}
\makeatother

\begin{document}

\title{Heterogeneous Impact of the Minimum Wage:\\
Implications for Changes in Between- and Within-group Inequality\thanks{We are grateful to Garry Barrett, Richard Blundell, Iv\'{a}n Fern\'{a}ndez-Val,
Hidehiko Ichimura, Kengo Kato, Edward Lazear, David Neumark, Whitney
Newey, Ryo Okui, Jesse Rothstein, Aloysius Siow, and conference and
seminar participants in Advances in Econometrics Conference, Asian
and Australasian Society of Labour Economics Inaugural Conference,
Asian Conference on Applied Microeconomics, Econometric Society Asian
Meeting, International Association for Applied Econometrics Annual
Conference, Kansai Labor Economics Workshop, Kyoto Summer Workshop
on Applied Economics, Mini-conference in Microeconometrics, Society
of Labor Economists Annual Meeting, Trans Pacific Labor Seminar, Seoul
National University, Shanghai University of Finance and Economics,
and University of Sydney for comments, questions, and discussions.
Oka gratefully acknowledges financial support from the Australian
Government through the Australian Research Council's Discovery Projects
(project DP190101152). Yamada gratefully acknowledges financial support
from the Kyoto University Foundation, the Murata Science Foundation,
and JSPS KAKENHI grant number: 17H04782.}\textit{\Large{}\medskip{}
}}

\author{Tatsushi Oka\thanks{Monash University. \texttt{tatsushi.oka@monash.edu} }
\and Ken Yamada\thanks{Kyoto University. \texttt{yamada@econ.kyoto-u.ac.jp}}}

\date{July 2019}
\maketitle
\begin{abstract}
\begin{onehalfspace}
Workers who earn at or below the minimum wage in the United States
are mostly either less educated, young, or female. Little is known,
however, concerning the extent to which the minimum wage influences
wage differentials among workers with different observed characteristics
and among workers with the same observed characteristics. This paper
shows that changes in the real value of the minimum wage over recent
decades have affected the relationship of hourly wages with education,
experience, and gender. The results suggest that changes in the real
value of the minimum wage account in part for the patterns of changes
in education, experience, and gender wage differentials and mostly
for the patterns of changes in within-group wage differentials.\bigskip{}
\end{onehalfspace}

\global\long\def\E{\mathbb{E}}
\end{abstract}
\newpage{}

\section{Introduction}

Expectations for the role of the minimum wage in addressing inequality
have increased worldwide with concerns over growing inequality in
recent decades. The minimum wage has been introduced and expanded
in many countries to lift the wages of the lowest paid workers. It
has been pointed out, however, that the minimum wage can cause both
intended and unintended consequences \citep*{Card_Krueger_bk95,Neumark_Wascher_bk08}.
The intended consequences are the beneficial effects on the distributions
of wages and earnings \citep*{DiNardo_Fortin_Lemieux_EM96,Lee_QJE99,Teulings_EJ03,Autor_Manning_Smith_AEJ16,Dube_AEJf}.
The unintended consequences are the adverse effects on employment,
consumer prices, firm value and profitability, and firm entry and
exits \citep*{Aaronson_French_JoLE07,Draca_Machin_VanReenen_AEJ11,Bell_Machin_JoLE18,Aaronson_French_Sorkin_To_IER18}.
Proponents of the policy have typically assumed the view that the
intended effects are substantial and the unintended effects are negligible.
On the other hand, opponents have raised concerns that the unintended
effects are not negligible. Most studies have focused on proving or
disproving the existence of adverse effects of the minimum wage, and
fewer studies have examined the distributional impact of the minimum
wage in recent years \citep*{Card_Krueger_ILRR17}.

\begin{figure}[h]
\caption{Proportion and characteristics of minimum wage workers\label{fig: proportion=000026characteristics}}

\begin{centering}
\subfloat[How many workers earn the minimum wage?\label{fig: proportion}]{
\centering{}\includegraphics[scale=0.55]{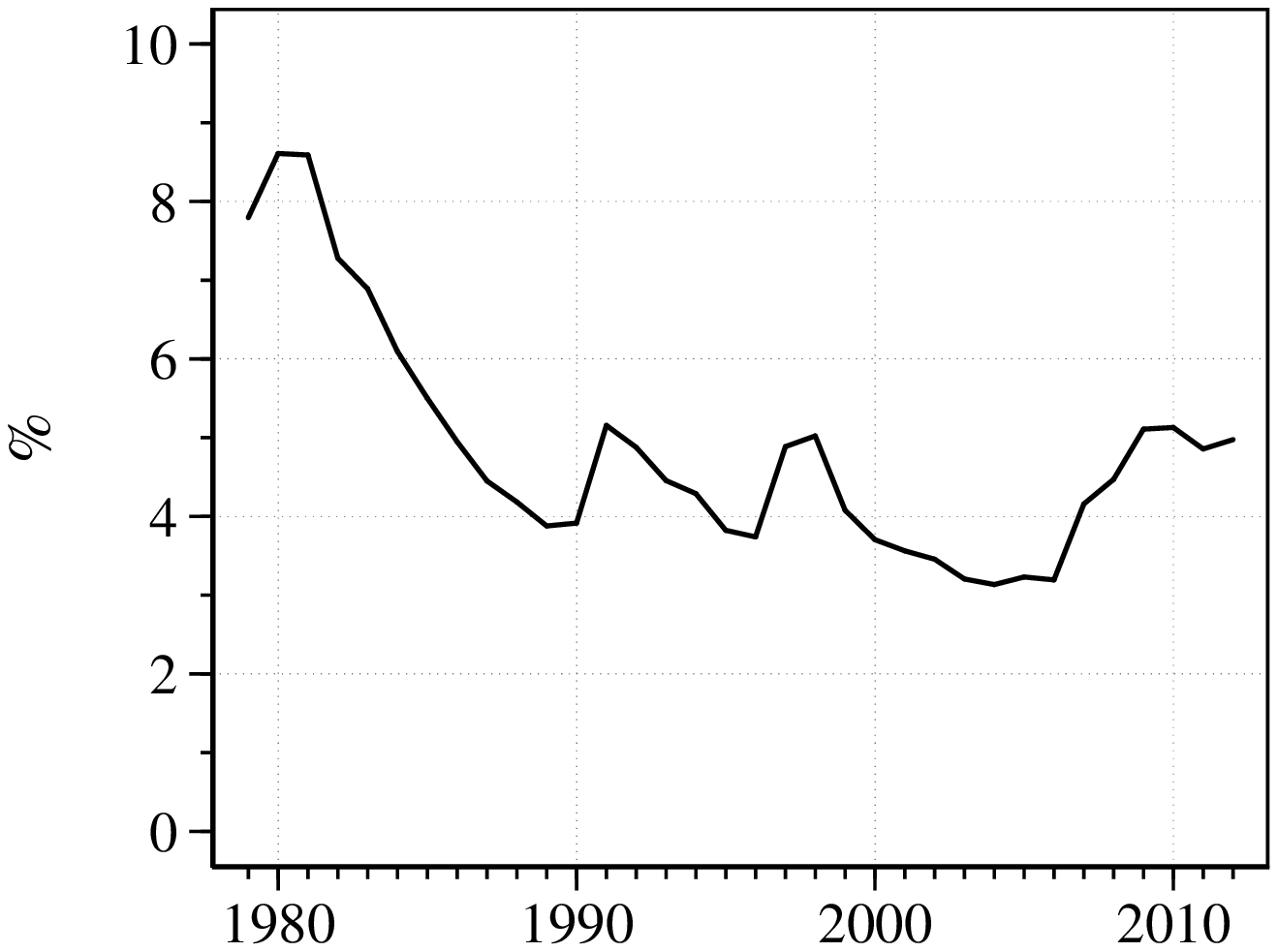}}\subfloat[Who earns the minimum wage?\label{fig: characteristics}]{
\centering{}\includegraphics[scale=0.65]{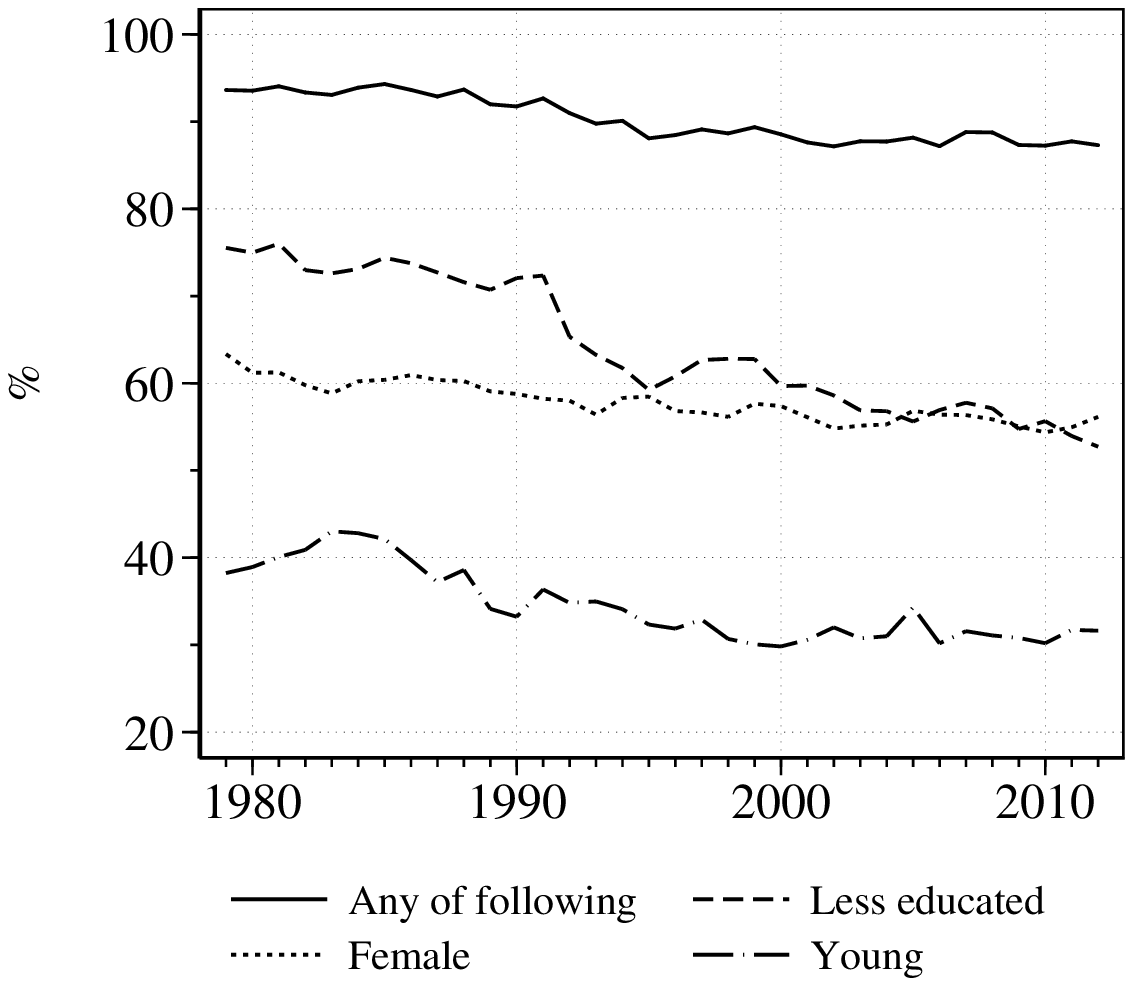}}
\par\end{centering}
\textit{\footnotesize{}Notes}{\footnotesize{}: Figure \ref{fig: proportion}
is reproduced from Figure 2 in \citet{Autor_Manning_Smith_AEJ16}.
In Figure \ref{fig: characteristics}, less-educated workers are those
with a high school degree or less, and young workers are those aged
24 years or less.}{\footnotesize\par}
\end{figure}

The proportion and characteristics of minimum wage workers serve as
starting points for a discussion on the distributional impact of the
minimum wage. According to the Current Population Survey (CPS), the
proportion of workers who earn at or below the minimum wage in the
United States ranges between 3 and 9 percent for the years 1979 to
2012 (Figure \ref{fig: proportion}). Less than 10 percent of workers
have been directly affected by the minimum wage in the U.S. labor
market. The extent to which the minimum wage affects the wage structure
depends on the magnitude of the spillover effects on workers who earn
more than the minimum wage. The minimum wage can exert a substantial
influence on the wage structure if there are strong spillover effects.

Perhaps a less well-known fact is that minimum wage workers are concentrated
in particular demographic groups. Approximately 90 percent of workers
who earned at or below the minimum wage in the United States between
the years 1979 and 2012 were high school graduates or less, younger
than 25 years old, or female (Figure \ref{fig: characteristics}).
The reason was not that the minimum wage policy had been targeted
based on education, experience, or gender, but because the lowest
paid workers were mostly either less educated, young, or female. In
light of this, the minimum wage may affect the relationship of hourly
wages with education, experience, and gender.

Motivated by the fact above, we examine the distributional impact
of the minimum wage in different ways from previous studies. We first
consider a standard wage equation, in which the logarithm of real
hourly wages is determined by education, experience and gender. We
then look at changes in the distribution of wages resulting from the
minimum wage through the lens of the wage equation. We allow the impact
of the minimum wage to be heterogeneous with respect to unobserved,
as well as observed, characteristics of workers, and for this purpose
we adopt a quantile regression approach. Using quantile regression
estimates, we evaluate the contribution of the minimum wage to changes
in between- and within-group inequality.

We show that changes in the real value of the minimum wage over recent
decades have affected the relationship of hourly wages with education,
experience, and gender in the United States. The impact of the minimum
wage is heterogeneous across workers depending on their observed characteristics.
Consequently, changes in the real value of the minimum wage account
in part for the patterns of changes in education, experience, and
gender wage differentials. We further show that changes in the real
value of the minimum wage over recent decades have affected wage differentials
among workers with the same observed characteristics. The impact of
the minimum wage is heterogeneous across quantiles of workers' productivity
not attributable to their observed characteristics. Consequently,
changes in the real value of the minimum wage account mostly for the
patterns of changes in within-group wage differential among workers
with lower levels of experience.

The remainder of the paper is organized as follows. The next section
reviews the related literature. Section \ref{sec: data} describes
the data and institutional background. Section \ref{sec: framework}
presents an econometric framework to evaluate the quantitative contribution
of the minimum wage to changes in between- and within-group inequality.
Section \ref{sec: results} provides the empirical results. The final
section concludes.

\section{Related Literature}

The literature has proven that the minimum wage has an effect on the
distribution of hourly wages in the United States, while the magnitude
and mechanisms of the effect vary across studies \citep{DiNardo_Fortin_Lemieux_EM96,Lee_QJE99,Teulings_EJ03,Autor_Manning_Smith_AEJ16}.
These studies develop and adopt different approaches that take into
account different degrees of heterogeneity and spillovers in the impact
of the minimum wage. \citet{DiNardo_Fortin_Lemieux_EM96} develop
a semiparametric approach to estimating discontinuous changes in the
wage distribution at the minimum wage.\footnote{See also \citet*{Chernozhukov_FernandezVal_Melly_EM13} for related
approaches.} \citet*{Lee_QJE99} develop a semiparametric approach to estimating
heterogeneous effects of the minimum wage across quantiles of the
wage distribution. \citet*{Teulings_EJ03} develops a parametric approach
to estimating the impact of the minimum wage on the wage distribution.
When comparing semiparametric approaches developed by \citet{DiNardo_Fortin_Lemieux_EM96}
and \citet*{Lee_QJE99}, \citeauthor{DiNardo_Fortin_Lemieux_EM96}'s
(1996) approach allows for heterogeneous effects with respect to workers'
observed characteristics, while \citeauthor*{Lee_QJE99}'s (1999)
approach does not. \citeauthor{DiNardo_Fortin_Lemieux_EM96}'s (1996)
approach, however, requires additional assumptions to estimate the
impact of the minimum wage from the cross-sectional distribution of
wages. Consequently, \citeauthor{DiNardo_Fortin_Lemieux_EM96}'s (1996)
approach does not allow for spillover effects, while \citeauthor*{Lee_QJE99}'s
(1999) approach does. The approaches also differ in robustness to
unobserved state and time effects. If there is sufficient variation
in the minimum wage across states over time, \citeauthor*{Lee_QJE99}'s
(1999) approach can separately identify the impact of the minimum
wage from unobserved state and time effects. \citet{Autor_Manning_Smith_AEJ16}
refine and apply \citeauthor*{Lee_QJE99}'s (1999) approach to data
covering a longer period, and develop a test for the presence of spillover
effects under a distributional assumption. However, no study has incorporated
heterogeneous effects across workers with different observed characteristics
in \citeauthor*{Lee_QJE99}'s (1999) approach.

Understanding the sources of changes in between- and within-group
inequality is key to understanding the mechanisms of changes in wage
inequality in the United States \citep*{Lemieux_AER06,Autor_Katz_Kearney_RESTAT08}.
However, little is known concerning the extent to which changes in
between- and within-group wage differentials are attributed to changes
in the real value of the minimum wage. In the literature, changes
in between-group wage differentials have been typically attributed
to changes in technology, workforce composition, and gender discrimination
\citep*[see][for surveys]{Autor_Katz_HLE99,Blau_Kahn_JEL17}. There
is no consensus on the quantitative contribution of the minimum wage
to changes in between-group wage differentials. \citet{DiNardo_Fortin_Lemieux_EM96}
and \citet*{Lee_QJE99} conclude that changes in the educational wage
differential are attributable only to a small extent to changes in
the real value of the minimum wage, while \citet*{Teulings_EJ03}
concludes that changes in the educational wage differential are attributable
to a large extent to changes in the real value of the minimum wage.
\citet{DiNardo_Fortin_Lemieux_EM96} demonstrate that the minimum
wage was an important factor in accounting for changes in wage inequality
in the 1980s. However, the literature identifying the sources of changes
in within-group wage differentials have been less conclusive than
the literature identifying the sources of changes in between-group
wage differentials \citep{Lemieux_AER06,Autor_Katz_Kearney_RESTAT08}.

\section{Data\label{sec: data}}

The data used in our analysis are repeated cross-sectional data from
the Current Population Survey Merged Outgoing Rotation Group. We construct
variables in the same way as in \citet{Autor_Manning_Smith_AEJ16},
and focus on the period between 1979 and 2012 to ensure the comparability
of results. We restrict the sample to workers aged between 18 and
64 including males and females, full-time and part-time workers, but
excluding self-employed workers, in the same way as in \citet{Autor_Manning_Smith_AEJ16}.
We, however, add in the sample individuals for whom we cannot observe
wages. The yearly sample size ranges from 142,000 to 235,000. Following
\citet{DiNardo_Fortin_Lemieux_EM96}, \citet*{Lee_QJE99}, and \citet{Autor_Manning_Smith_AEJ16},
we weight each individual according to the sampling weight multiplied
by hours worked.

\begin{figure}[h]
\caption{The statutory minimum wage, 1979\textendash 2012\label{fig: statutory_minimum}}

\begin{centering}
\subfloat[Low minimum wage states (17 states)\label{fig: low}]{
\centering{}\includegraphics[scale=0.55]{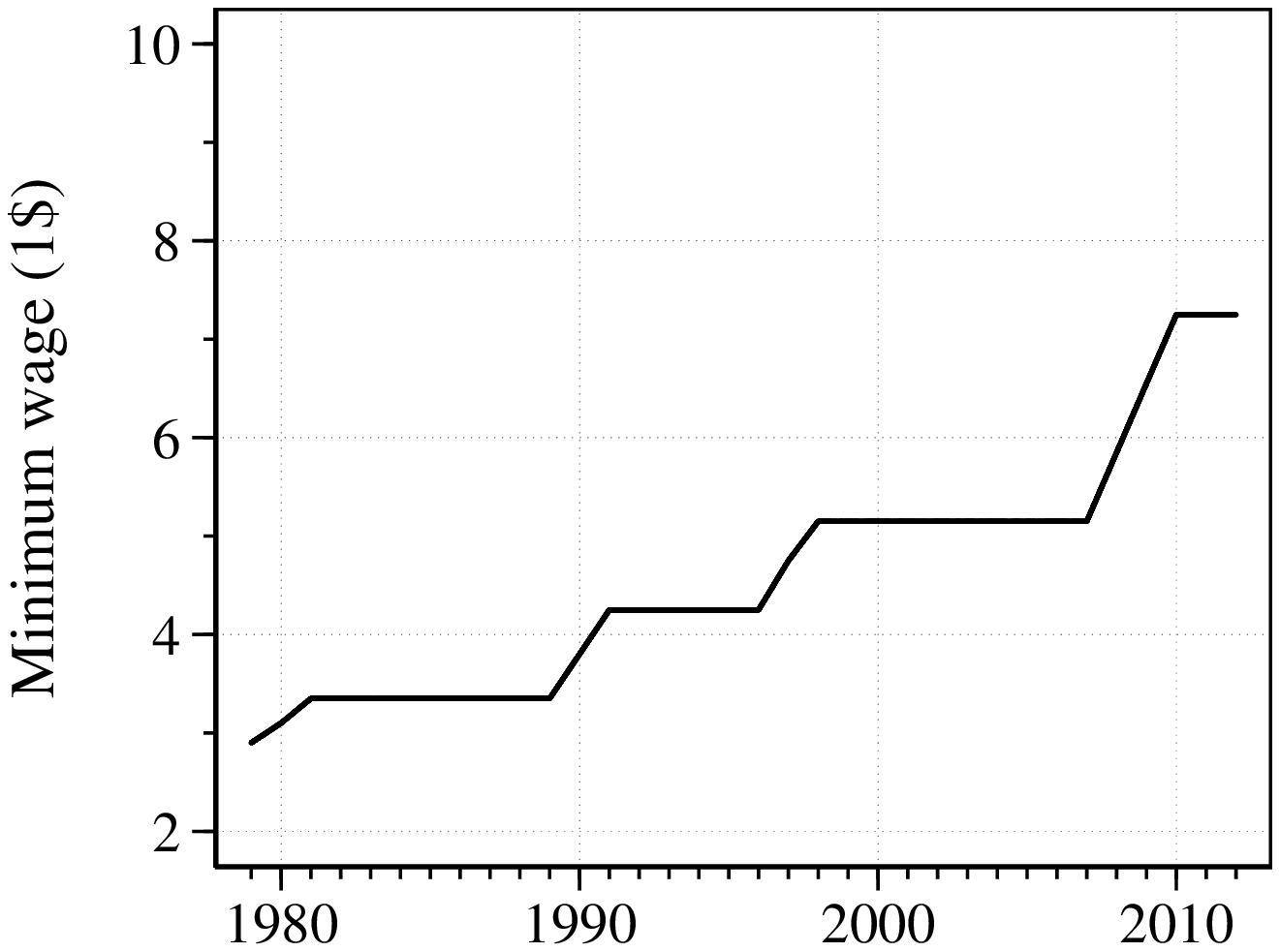}}\subfloat[Medium minimum wage states (17 states)\label{fig: medium}]{
\centering{}\includegraphics[scale=0.55]{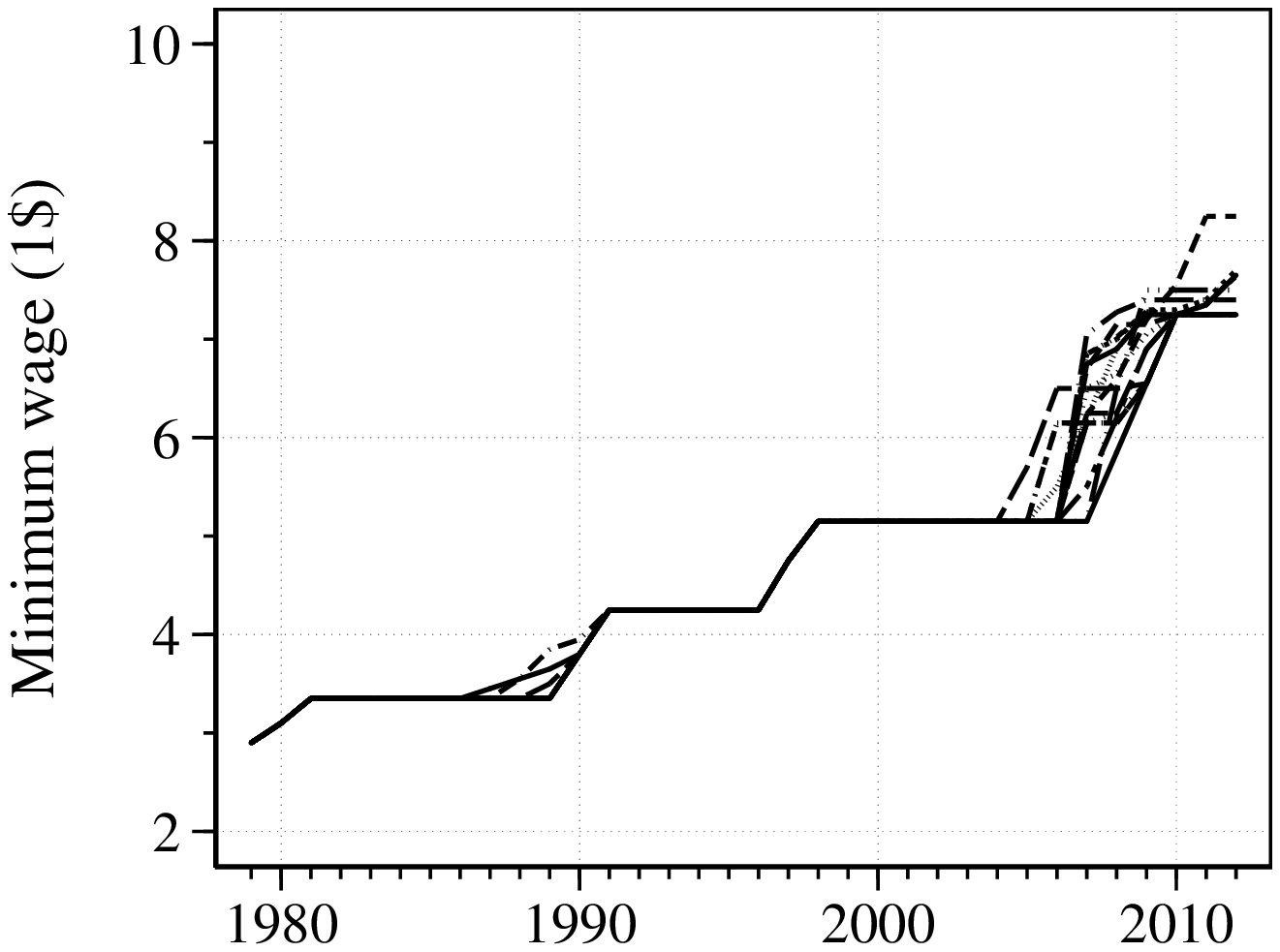}}
\par\end{centering}
\begin{centering}
\subfloat[High minimum wage states (16 states)\label{fig: high}]{
\centering{}\includegraphics[scale=0.55]{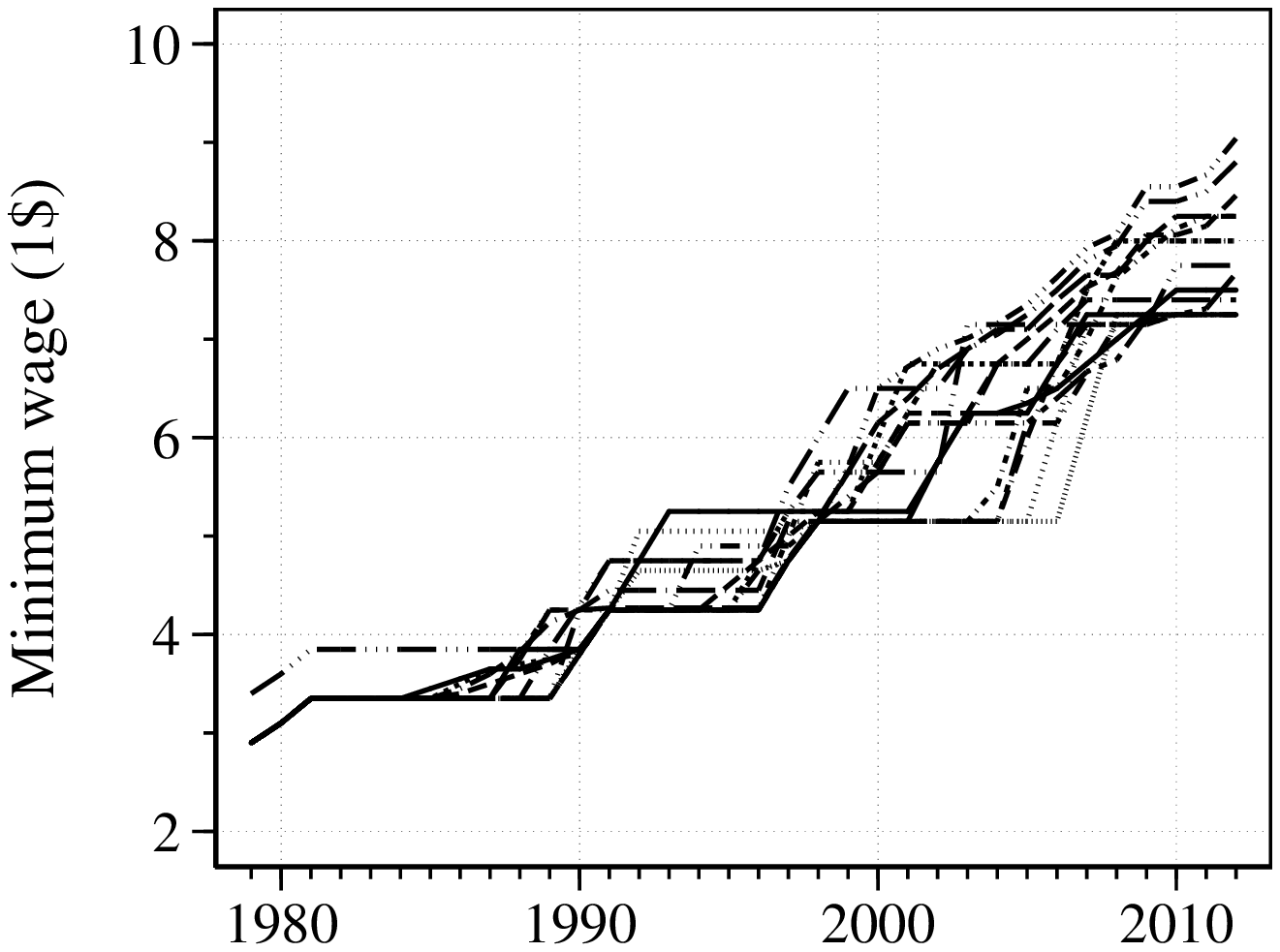}}
\par\end{centering}
\textit{\footnotesize{}Notes}{\footnotesize{}: Panel (a) includes
Alabama, Georgia, Idaho, Indiana, Kansas, Louisiana, Mississippi,
North Dakota, Nebraska, Oklahoma, South Carolina, South Dakota, Tennessee,
Texas, Utah, Virginia, and Wyoming. Panel (b) includes Arkansas, Arizona,
Colorado, Kentucky, Maryland, Michigan, Minnesota, Missouri, Montana,
North Carolina, New Hampshire, New Mexico, Nevada, Ohio, Pennsylvania,
Wisconsin, and West Virginia. Panel (c) includes Alaska, California,
Connecticut, Delaware, Florida, Hawaii, Iowa, Illinois, Massachusetts,
Maine, New Jersey, New York, Oregon, Rhode Island, Vermont, and Washington.}{\footnotesize\par}
\end{figure}

Minimum wage laws differ across states and change over time in the
United States. The federal government sets the federal minimum wage
that applies to all states. State governments can set the state minimum
wage higher than the federal minimum wage. The statutory minimum wage
is the maximum of the federal minimum wage and the state minimum wage.

Figure \ref{fig: statutory_minimum} shows the trend in the statutory
minimum wage. For ease of reference, we divide all 50 states evenly
into three groups according to the level of statutory minimum wage.
During the period, 17 states had no state minimum wage (Figure \ref{fig: low}).
The statutory minimum wage equals the federal minimum wage in these
states. The federal minimum wage increased four times: 1979 to 1981,
1989 to 1991, 1996 to 1998, and 2007 to 2010. The remaining 33 states
set their state minimum wages (Figures \ref{fig: medium} and \ref{fig: high}).
The statutory minimum wage has been higher than the federal minimum
wage for many years in these states. In the 1980s there was not much
variation across states or changes over time in the minimum wage.
On the other hand, in the 1990s and the 2000s there was substantial
variation in the minimum wage across states over time.

\begin{figure}[h]
\caption{The real value of the minimum wage, 1979\textendash 2012\label{fig: real_minimum}}

\begin{centering}
\includegraphics[scale=0.55]{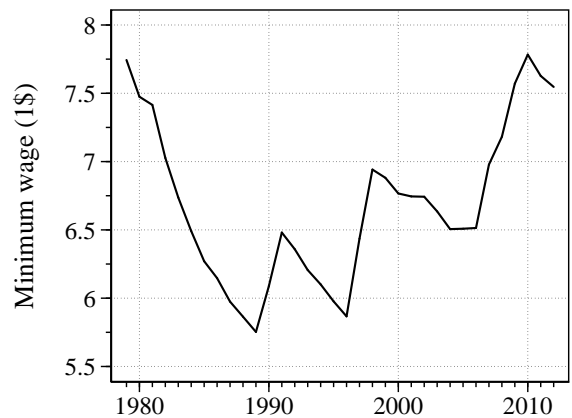}{\footnotesize{}}{\footnotesize\par}
\par\end{centering}
\textit{\footnotesize{}\hspace{10em}Notes}{\footnotesize{}: National
averages are reported.}{\footnotesize\par}
\end{figure}

Figure \ref{fig: real_minimum} shows the national average trend in
the real value of the minimum wage. The statutory minimum wage is
deflated by the personal consumer expenditure price index using 2012
as the base year. During the period, there was a change in the trend
in the year 1989. The real value of the minimum wage fell due to inflation
from 1979 to 1989. Subsequently, the real value of the minimum wage
exhibits an upward trend due to increases in the statutory minimum
wage for the years 1989 to 2012.

\section{Econometric Framework\label{sec: framework}}

In this section, we present our econometric framework. We start by
introducing the (state-level) panel quantile regression model. Then,
we describe the censored quantile regression model. We end this section
by describing our approach to evaluating the quantitative contribution
of the minimum wage to changes in between- and within-group inequality.

\subsection{Model}

The key feature of our model is that it allows the impact of the minimum
wage to be heterogeneous with respect to workers' observed and unobserved
characteristics. This feature is essential for evaluating the contribution
of the minimum wage to changes in between- and within-group inequality.

For the purpose of our analysis, we adopt the quantile regression
approach pioneered by \citet*{Koenker_Bassett_EM78} and developed
by \citet*{Chetverikov_Larsen_Palmer_EM16}. We have a repeated cross
section of individuals $i=1,\ldots,N_{st}$ in states $s=1,\ldots,S$,
and time $t=1,\ldots,T$. For each state and year, the structure of
wages can be expressed using the following quantile regression model:
\begin{equation}
Q_{st}\left(\left.\tau\right|z_{ist}\right)=z_{ist}^{\prime}\alpha_{st}\left(\tau\right)\qquad\text{for \enskip}\tau\in\left(0,1\right),\label{eq: quantile1}
\end{equation}
where $Q_{st}\left(\left.\tau\right|z_{ist}\right)$ is the $\tau$th
conditional quantile of the log of real hourly wages, $w_{ist}$,
given a $J+1$ vector of observed individual characteristics, $z_{ist}$,
for each state $s$ and year $t$. The vector of parameters $\alpha_{st}\left(\tau\right)$
can vary across quantiles $\tau$. The vector $z_{ist}$ includes
a constant term, the linear and quadratic terms in years of education
and of potential experience (age minus education minus six), and
an indicator for being male. There are three reasons we use these
variables. First, they are determined prior to the entry of the labor
market. Second, they are commonly used as regressors in the quantile
regression of wages \citep*{Buchinsky_EM94,Angrist_Chernozhukov_FernandezVal_EM06}.
Finally, and most importantly, they are useful to distinguish minimum
wage workers.\footnote{When we add an indicator of being white in individual characteristics,
we find that the minimum wage has no effect on the racial wage differential.
The proportion of black workers was less than 20 percent among minimum
wage workers throughout the sample period. Even if the linear and
quadratic terms in years of education and years of experience are
interacted with the indicator for being male, the results reported
remain essentially unchanged.} The quantile regression model \eqref{eq: quantile1} is more flexible
than usual in that it allows all intercept and slope coefficients
to vary across states and years.

Given the structure of wages described above, we examine the distributional
impact of the minimum wage by looking at changes in the vector of
coefficients, $\alpha_{st}\left(\tau\right)\equiv\bigl(\alpha_{0st}\left(\tau\right),\alpha_{1st}\left(\tau\right),\ldots,\alpha_{Jst}\left(\tau\right)\bigr)^{\prime}$,
in equation \eqref{eq: quantile1} resulting from changes in the real
value of the minimum wage. We consider the following (state-level)
panel data model:
\begin{equation}
\alpha_{jst}\left(\tau\right)=m_{st}\beta_{j}\left(\tau\right)+x_{st}^{\prime}\gamma_{j}\left(\tau\right)+\epsilon_{jst}\left(\tau\right)\qquad\text{for }j=0,\ldots,J,\label{eq: mean1}
\end{equation}
where $m_{st}$ is the log of the real value of the minimum wage,
and $x_{st}$ is a vector of state-year characteristics. The vector
$x_{st}$ includes state and year dummies and state-specific linear
trends in the same way as in \citet{Autor_Manning_Smith_AEJ16}. A
set of parameters, $\beta\left(\tau\right)=\bigl(\beta_{0}\left(\tau\right),\beta_{1}\left(\tau\right),\ldots,\beta_{J}\left(\tau\right)\bigr)^{\prime}$,
represents the heterogeneous impact of the minimum wage. Note that
the first element of the vector $\boldsymbol{z}_{ist}$ is one. The
second to last elements, $\beta_{1}\left(\tau\right)$ to $\beta_{J}\left(\tau\right)$,
of the vector $\beta\left(\tau\right)$ measure the extent to which
the impact of the minimum wage varies across individuals according
to their observed characteristics. If the impact of the minimum wage
is not heterogeneous with respect to observed characteristics, the
parameter vector is $\beta\left(\tau\right)=\left(\beta_{0}\left(\tau\right),0,\ldots,0\right)^{\prime}$
for a given $\tau$. The quantile $\tau$ can be interpreted as the
position in the distribution of workers' productivity not attributable
to their observed characteristics. If the impact of the minimum wage
is not heterogeneous with respect to unobserved quantiles, the parameter
vector is $\beta\left(\tau\right)=\left(\beta_{0},\beta_{1},\ldots,\beta_{J}\right)^{\prime}$
for all $\tau$.

Following \citet{Chetverikov_Larsen_Palmer_EM16}, equations \eqref{eq: quantile1}
and \eqref{eq: mean1} can be estimated in two steps. In the first
step, we perform separate quantile regressions of $w_{ist}$ by state
$s$ and year $t$ for each quantile $\tau$ using the individual-level
cross-sectional data. We then obtain a set of estimated parameters
$\widehat{\alpha}_{st}\left(\tau\right)=\left(\widehat{\alpha}_{0,st}\left(\tau\right),\widehat{\alpha}_{1,st}\left(\tau\right),\ldots,\widehat{\alpha}_{J,st}\left(\tau\right)\right)^{\prime}$.
In the second step, we perform the linear regression of $\widehat{\alpha}_{jst}\left(\tau\right)$
for each element $j$ and quantile $\tau$ using the state-level panel
data. Relative to several applications discussed in \citet{Chetverikov_Larsen_Palmer_EM16},
we allow for interactions between the treatment variable and individual
characteristics,\footnote{\citet{Koenker_ARE17} recently notes that ``somewhat neglected in
the econometrics literature on treatment response and program evaluation
is the potentially important role of the interactions of covariates
with treatment variables.''} while we assume the exogeneity of the treatment variable. The minimum
wage is commonly assumed to be exogenous in the literature. We, however,
examine the possibility that differences in changes in the real value
of the minimum wage across states may be driven by differences in
changes in unobserved state characteristics.

The approach described above is related to the approach used in \citet*{Lee_QJE99},
who estimates the model of the form:
\begin{equation}
Q_{st}\left(\tau\right)-Q_{st}\left(0.5\right)=\bigl(m_{st}-Q_{st}\left(0.5\right)\bigr)\beta\left(\tau\right)+x_{st}^{\prime}\gamma\left(\tau\right)+\epsilon_{st}\left(\tau\right),\label{eq: lee}
\end{equation}
where $Q_{st}\left(\tau\right)$ is the $\tau$th unconditional quantile
of $w_{ist}$. If the median wage, $Q_{st}\left(0.5\right)$, is absent,
this model corresponds to the case in which all individual characteristics
are excluded from equation \eqref{eq: quantile1}. The main reason
for the use of the median wage is presumably that there was insufficient
variation in the state minimum wage during the period of the author's
analysis, 1979 to 1988.

\subsection{Estimation}

We address the issues of censoring and truncation, building on the
approach described above.

\paragraph{Censoring}

The wage distribution has been left-censored due to the minimum wage
in many states \citep{DiNardo_Fortin_Lemieux_EM96,Lee_QJE99}. This
issue is evident from the data but typically ignored when estimating
the wage equation. The main reason, presumably, is that the magnitude
of the bias due to left-censoring at the minimum wage is negligible
if the interest lies at the mean impact. However, the magnitude of
the bias may not be negligible if the interest lies at the distributional
impact. The left-censoring due to the minimum wage can cause the fitted
wage equation to be flat. In this case, the intercept coefficient
becomes larger, while the slope coefficients become smaller. This
effect is stronger at quantiles closer to the minimum wage. As a likely
consequence, the censoring effect (the impact of the minimum wage
at the minimum wage) may suffer from a downward bias, while the spillover
effect (the impact of the minimum wage above the minimum wage) may
suffer from an upward bias.

In addition, the earnings data from the CPS is right-censored due
to top-coding. This issue has been widely recognized in the literature.
Many studies using the CPS data make some adjustments for top-coding.
\citet*{Hubbard_JHR11} develops a maximum likelihood approach to
addressing this issue under a distributional assumption, and shows
that an increase in top-coded observations causes a serious bias in
the trend in the gender wage differential. The trends in the education
and experience wage differentials are also subject to the influence
of top-coding.\footnote{For the $\tau$th quantile regression, this issue can be solved by
winsorizing, only if the conditional probability of not being censored
given $z_{ist}$ is higher than $\tau$.}

We adopt the censored quantile regression approach developed in \citet*{Powell_JoE86},
\citet*{Chernozhukov_Hong_JASA02}, and \citet*{Chernozhukov_FernandezVal_Kowalski_JoE15}
to address the issue of censoring. This approach is semiparametric
in the sense that it does not require a distributional assumption.
We consider the following censored quantile regression model to deal
with left-censoring due to the minimum wage and right-censoring due
to top-coding.
\begin{equation}
Q_{st}\left(\left.\tau\right|\boldsymbol{z}_{ist}\right)=\begin{cases}
m_{st} & \text{if }w_{ist}\le m_{st},\\
z_{ist}^{\prime}\alpha_{st}\left(\tau\right) & \text{if }m_{st}<w_{ist}<c_{it}\\
c_{it} & \text{if }w_{ist}\ge c_{it},
\end{cases},\label{eq: quantile2}
\end{equation}
where $c_{it}$ denotes the top-coded value.\footnote{The CPS sample is composed of hourly paid workers and monthly paid
workers. Earnings for monthly paid workers are top-coded, while wages
for hourly paid workers are not. For monthly paid workers, earnings
are divided by hours worked to calculate hourly wages. Although the
top-coded value of earnings is constant for a given year, the top-coded
value of wages differs according to hours worked. We, thus, allow
the top-coded value to vary across individuals.} The key concept of this approach is to estimate the quantile regression
model using the subsample of individuals who are unlikely to be left-
or right-censored.\footnote{In practice, it does not matter which values are assigned to the wages
of workers who earn below the minimum wage in the range less than
or equal to the minimum wage. Similarly, it does not matter which
values are assigned to the wages of workers who earn above the top-coded
value in the range greater than or equal to the top-coded value.} Appendix \ref{subsec: procedure} details the estimation procedure.

\paragraph{Missing wages}

There are diverse views on the employment effect of the minimum wage
\citep*{Card_Krueger_bk95,Neumark_Wascher_bk08}. Given the importance
of this issue, a valid question may be whether changes in the wage
distribution are due in part to a potential loss of employment resulting
from a rise in the minimum wage. For the sake of discussion, we suppose
that workers lose their jobs in the order of those with the lowest
to highest productivity. In this case, percentile wages can mechanically
increase even without any actual increase in wages. This implies that
if the sample is restricted to employed individuals, the censoring
effect and the spillover effect might be subject to an upward bias.
The magnitude of the bias depends on the magnitude of the employment
effect. We control for potential bias by imputing the wages of non-employed
individuals.

Our approach builds on the quantile imputation approach developed
in \citet*{Yoon_wp10} and \citet*{Wei_HQR17}. For the purpose of
imputation, we use the censored quantile regression model, instead
of the standard quantile regression, to take into account left- and
right-censoring. In the process of imputation, we assume that non-employed
individuals are less productive than median employed individuals,
as is common in the literature on the gender wage differential \citep*{Johnson_Kitamura_Neal_AERPP00}.\footnote{The results reported remain essentially unchanged if we assume that
non-employed individuals are less productive than 30 or 70 percent
of employed individuals.} We allow for selection on unobservables in that sense. Appendix \ref{subsec: procedure}
details the imputation procedure. Appendix \ref{subsec: impact} provides
the results without imputation.

\paragraph{Procedure}

The estimation procedure is divided into three stages. First, we estimate
the censored quantile regression model \eqref{eq: quantile2} using
the sample of employed individuals and impute the wages of individuals
for whom we cannot observe wages. Second, we estimate the censored
quantile regression model \eqref{eq: quantile2} using the sample
of employed and non-employed individuals, and obtain the estimates
for intercept and slope coefficients $\widehat{\alpha}_{jst}\left(\tau\right)$
in the wage equation for $j=0$, $1$, $\ldots$, $5$, $s=1$, $2$,
$\ldots$, $50$, $t=1979$, $1980$, $\ldots$, $2012$, and $\tau=0.04$,
$0.05$, $\ldots$, $0.97$. Both in the first and second stages,
we perform the separate regressions by state and year for each quantile.
Finally, we estimate the linear regression model \eqref{eq: mean1}
of $\widehat{\alpha}_{jst}\left(\tau\right)$ using the state-level
panel data.

\paragraph{Inference}

\citet{Chetverikov_Larsen_Palmer_EM16} derive the asymptotic properties
of estimators for parameters in equation \eqref{eq: mean1}. The authors
show that estimation errors from the individual-level quantile regression
are asymptotically negligible, if the size of the sample used in the
individual-level quantile regression is sufficiently large relative
to the size of the sample used in the state-level linear regression.
Because the sample size may not be sufficiently large in the least
populous states, we choose to report bootstrapped confidence intervals.
We construct bootstrapped intervals from 500,000 bootstrap estimates
obtained by repeating the individual-level censored quantile regression
500 times and then repeating the state-level linear regression 1,000
times for each quantile regression estimate. We allow for arbitrary
forms of heteroscedasticity and serial correlation.

\paragraph{Specification checks}

As is common when estimating the impact of the minimum wage on the
wage distribution \citep{DiNardo_Fortin_Lemieux_EM96,Lee_QJE99,Teulings_EJ03,Autor_Manning_Smith_AEJ16},
we focus primarily on the contemporaneous effect of the minimum wage.
We estimate the following model in which we add the lag and lead variables,
$m_{s,t-1}$ and $m_{s,t+1}$, to assess the validity of the model
specification.
\begin{equation}
\alpha_{jst}\left(\tau\right)=m_{s,t-1}\beta_{j,-1}\left(\tau\right)+m_{st}\beta_{j,0}\left(\tau\right)+m_{s,t+1}\beta_{j,+1}\left(\tau\right)+x_{st}^{\prime}\gamma_{j}\left(\tau\right)+\epsilon_{jst}\left(\tau\right)\quad\text{for }j=0,\ldots,J.\label{eq: mean2}
\end{equation}
If model \eqref{eq: mean1} is correctly specified, we expect two
restrictions to be satisfied. First, the long-term effect, $\beta_{j,-1}\left(\tau\right)+\beta_{j,0}\left(\tau\right)$,
in model \eqref{eq: mean2}, would be the same as the contemporaneous
effect, $\beta_{j}\left(\tau\right)$, in model \eqref{eq: mean1}.
This restriction will be valid if the policy effect is well captured
by the contemporaneous effect. Second, there would be no leading effect
in model \eqref{eq: mean2}; that is, $\beta_{j,+1}\left(\tau\right)=0$.
This restriction will not hold if changes in the real value of the
minimum wage are driven by changes in unobserved state characteristics.
We, thus, examine whether the long-term effect differs from the contemporaneous
effect, and whether the leading effect differs from zero.

\subsection{Measures of inequality}

The aim of this paper is to evaluate the quantitative contribution
of the minimum wage to changes in between- and within-group inequality.
Here, we provide the definition of the two types of inequality, and
describe the way to measure the contribution of the minimum wage along
the lines of the model described above.

\begin{figure}
\caption{Inequality measures}

\begin{centering}
\subfloat[Between-group inequality\label{fig: between}]{
\centering{}\includegraphics[scale=0.85]{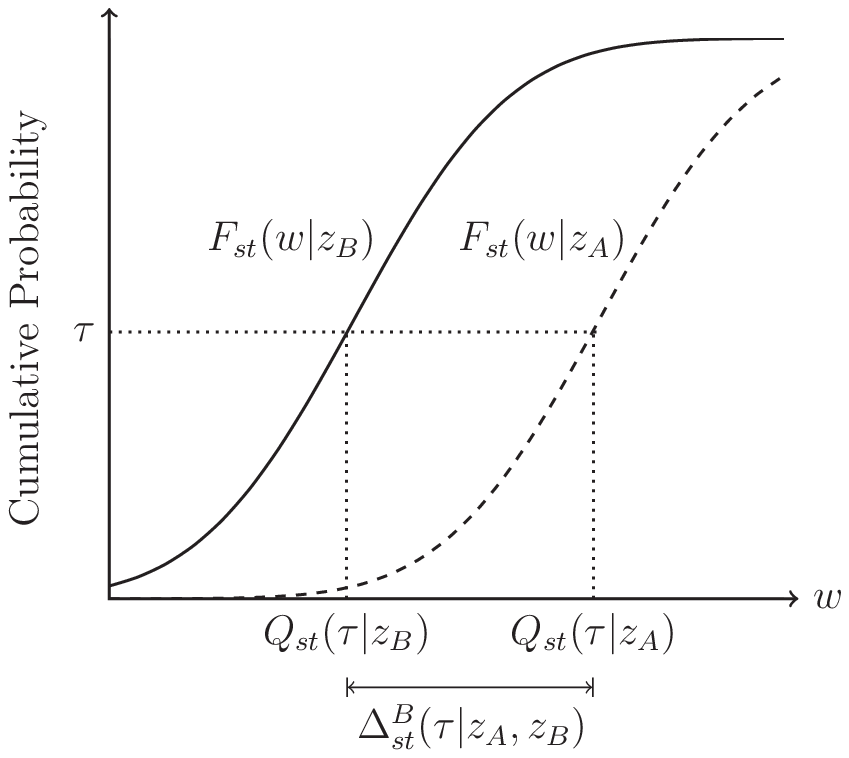}}\subfloat[Within-group inequality\label{fig: within}]{
\centering{}\includegraphics[scale=0.85]{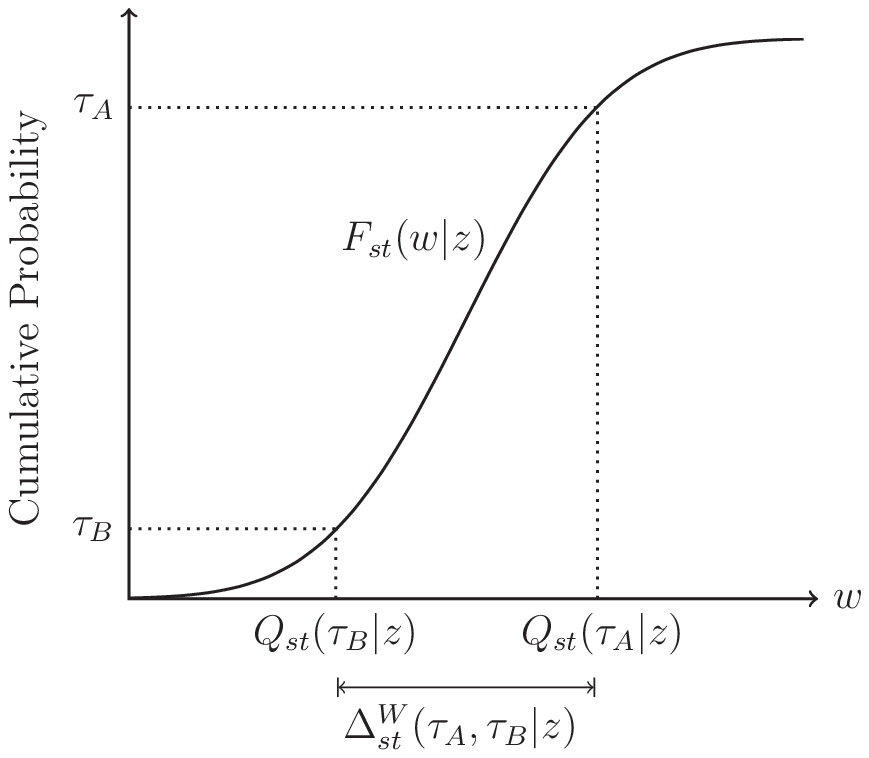}}
\par\end{centering}
\textit{\footnotesize{}Notes}{\footnotesize{}: The conditional quantile
function $Q_{st}\left(\left.\tau\right|z\right)$ is an inverse of
$F_{st}\left(\left.w\right|z\right)$, where $F_{st}\left(\left.\cdot\right|z\right)$
is the conditional distribution function of $w_{ist}$ given $z_{ist}=z$
in state $s$ and year $t$.}{\footnotesize\par}
\end{figure}

Between-group inequality is the wage differential among workers with
different observed characteristics. Consider two groups of workers,
one of which consists of workers with individual characteristics,
$z_{ist}=z_{A}$, and the other consists of workers with individual
characteristics, $z_{ist}=z_{B}$. Between-group inequality can be
defined as:
\begin{equation}
\Delta_{st}^{B}\left(\left.\tau\right|z_{A},z_{B}\right)\coloneqq Q_{st}\left(\left.\tau\right|z_{A}\right)-Q_{st}\left(\left.\tau\right|z_{B}\right)\label{eq: between1}
\end{equation}
for a given quantile $\tau$ (see Figure \ref{fig: between} for graphical
description). Let $\widetilde{\Delta}_{st}^{B}$ denote the counterfactual
between-group wage differential if the real value of the minimum wage
were kept constant at a certain level. The contribution of the minimum
wage can be measured by taking the difference between the actual wage
differential and the counterfactual wage differential:
\begin{equation}
\Delta_{st}^{B}\left(\left.\tau\right|z_{A},z_{B}\right)-\widetilde{\Delta}_{st}^{B}\left(\left.\tau\right|z_{A},z_{B}\right).\label{eq: between2}
\end{equation}

Within-group inequality is the wage differential among workers with
the same observed characteristics. Consider a range between two quantiles,
$\tau_{A}$ and $\tau_{B}$, as a measure of inequality. Within-group
inequality can be defined as:
\begin{equation}
\Delta_{st}^{W}\left(\left.\tau_{A},\tau_{B}\right|z\right)\coloneqq Q_{st}\left(\left.\tau_{A}\right|z\right)-Q_{st}\left(\left.\tau_{B}\right|z\right)\label{eq: within1}
\end{equation}
for a group of workers with individual characteristics, $z_{ist}=z$
(see Figure \ref{fig: within} for graphical description). Let $\widetilde{\Delta}_{st}^{W}$
denote the counterfactual within-group wage differential if the real
value of the minimum wage is kept constant at a certain level. The
contribution of the minimum wage can be measured by taking the difference
between the actual wage differential and the counterfactual wage differential:
\begin{equation}
\Delta_{st}^{W}\left(\left.\tau_{A},\tau_{B}\right|z\right)-\widetilde{\Delta}_{st}^{W}\left(\left.\tau_{A},\tau_{B}\right|z\right).\label{eq: within2}
\end{equation}

\section{Results\label{sec: results}}

Our results are divided into two parts. The first part is a collection
of the results regarding the impact of the minimum wage on the wage
structure. The second part is a collection of the results regarding
the contribution of the minimum wage to changes in between- and within-group
inequality.

\subsection{Impact on the wage structure}

\begin{figure}[h]
\caption{Impact of the minimum wage on the wage structure\label{fig: estimates}}

\begin{centering}
\subfloat[Intercept\label{fig: intercept}]{
\centering{}\includegraphics[scale=0.6]{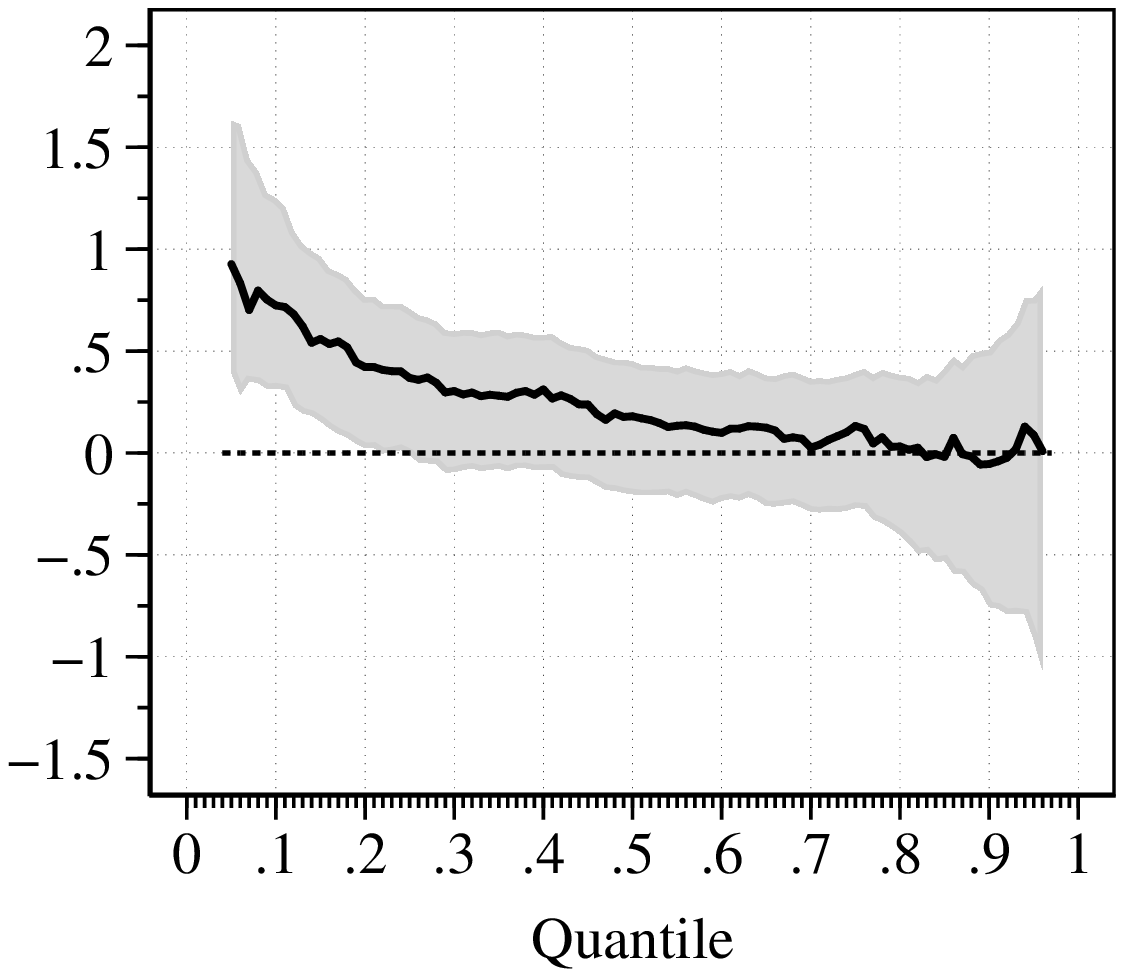}}\subfloat[Education\label{fig: education}]{
\centering{}\includegraphics[scale=0.6]{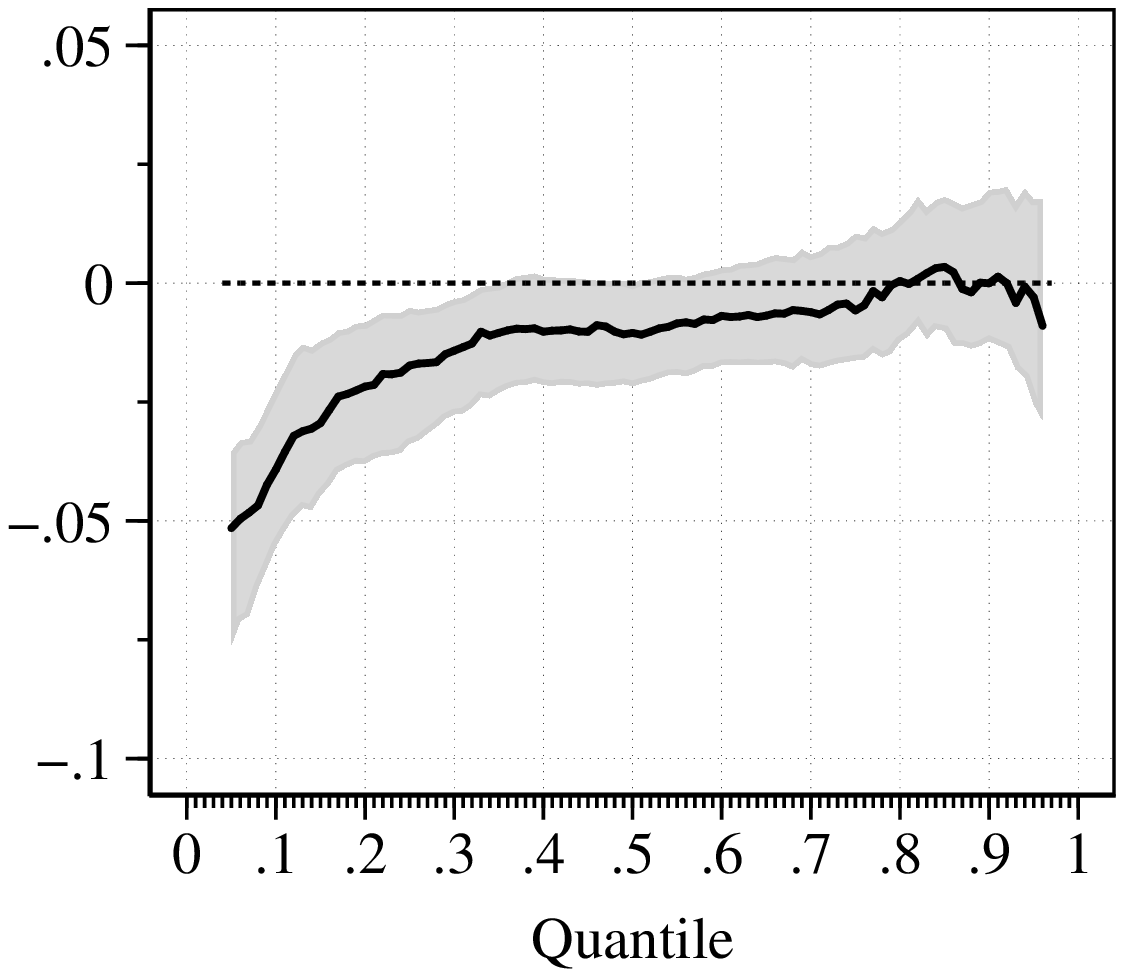}}
\par\end{centering}
\begin{centering}
\subfloat[Experience\label{fig: experience}]{
\centering{}\includegraphics[scale=0.6]{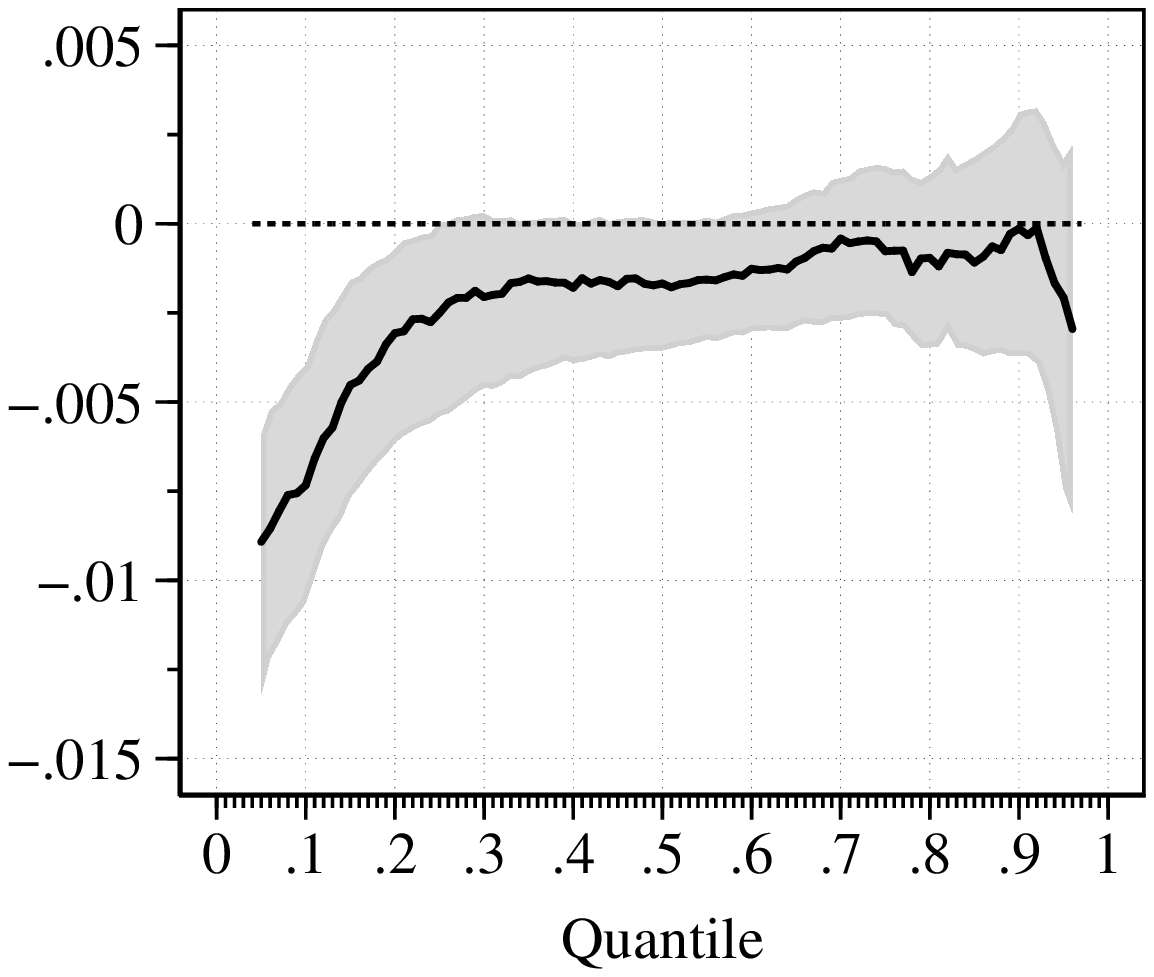}}\subfloat[Gender (male)\label{fig: gender}]{
\centering{}\includegraphics[scale=0.6]{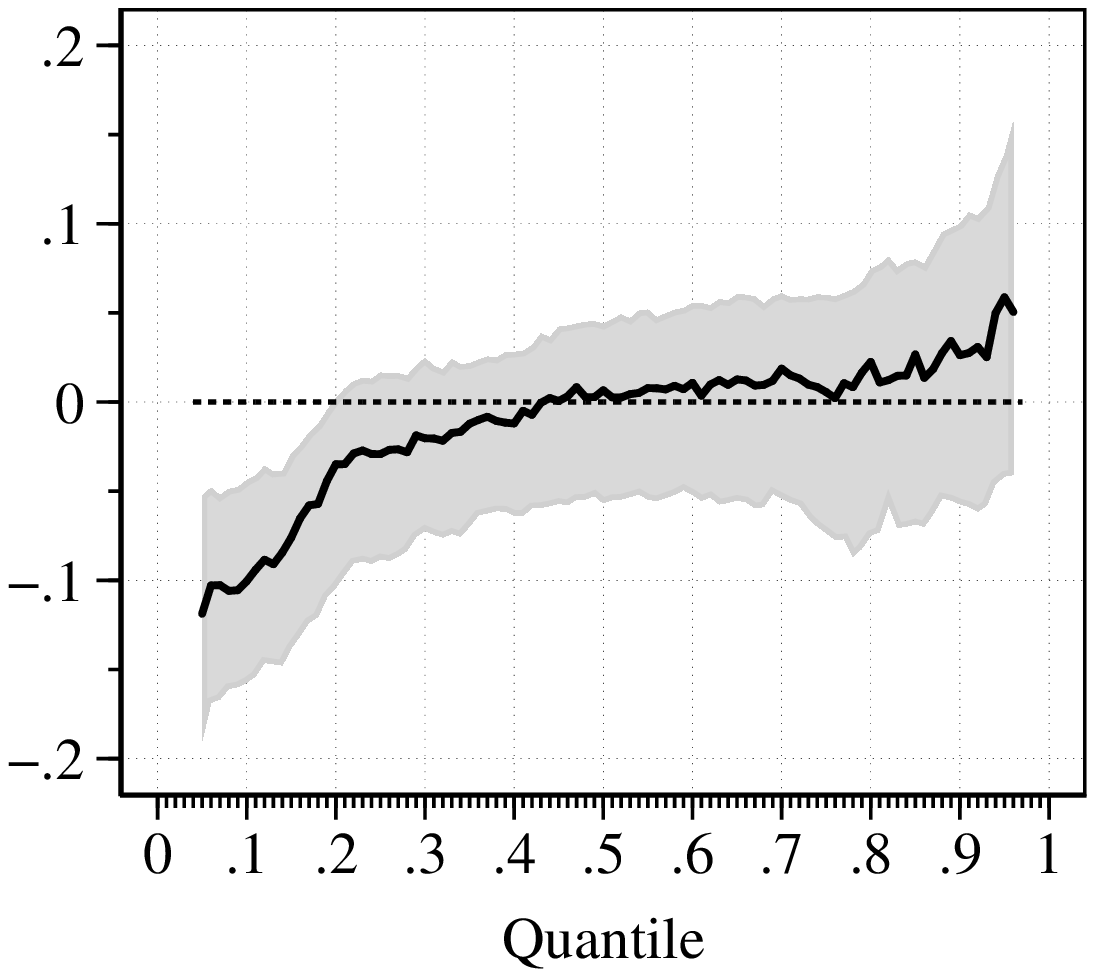}}
\par\end{centering}
\textit{\footnotesize{}Notes}{\footnotesize{}: Estimates of partial
effects in equation \eqref{eq: mean1} are reported. The shaded area
represents the 95 percent confidence interval. See Figure \ref{fig: estimates_band}
for the uniform confidence band.}{\footnotesize\par}
\end{figure}

We first present the results of estimating equation \eqref{eq: mean1}.
Figure \ref{fig: estimates} shows the impact of the minimum wage
on the intercept and slope coefficients in the wage equation across
quantiles. The four panels show the estimates for $\beta_{0}\left(\tau\right)$,
$\beta_{1}\left(\tau\right)+2\beta_{2}\left(\tau\right)\overline{educ}$,
$\beta_{3}\left(\tau\right)+2\beta_{4}\left(\tau\right)\overline{exper}$,
and $\beta_{5}\left(\tau\right)$, respectively, where the bar represents
the sample mean over all states and years. We summarize the impact
of the minimum wage on the coefficients of linear and quadratic terms
in education and experience as the impact on their marginal effects.

Both the intercept and slope coefficients in the wage equation are
affected by the real value of the minimum wage. The intercept coefficient
increases with a rise in the minimum wage (Figure \ref{fig: intercept}),
while the slope coefficients of education, experience, and gender
decrease with a rise in the minimum wage (Figures \ref{fig: education},
\ref{fig: experience}, and \ref{fig: gender}). The former result
implies that a rise in the minimum wage results in a rise in the wages
of the least-skilled workers in terms of observed characteristics.
The latter result implies that a rise in the minimum wage weakens
the relationship of hourly wages with education, experience, and gender.
These results are consistent with the fact that less-educated, less-experienced,
and female workers are more directly affected by a rise in the minimum
wage than more-educated, more-experienced, and male workers. Furthermore,
the magnitude of changes in the intercept and slope coefficients varies
across quantiles. In all cases, the impact of the minimum wage is
greatest at the lowest quantile and gradually declines in absolute
value to zero by the 0.3 quantile. Spillover effects are present but
limited mostly to the first quintile.

\begin{figure}[h]
\caption{Long-term effect of the minimum wage on the wage structure\label{fig: estimates_lag}}

\begin{centering}
\subfloat[Intercept]{
\centering{}\includegraphics[scale=0.6]{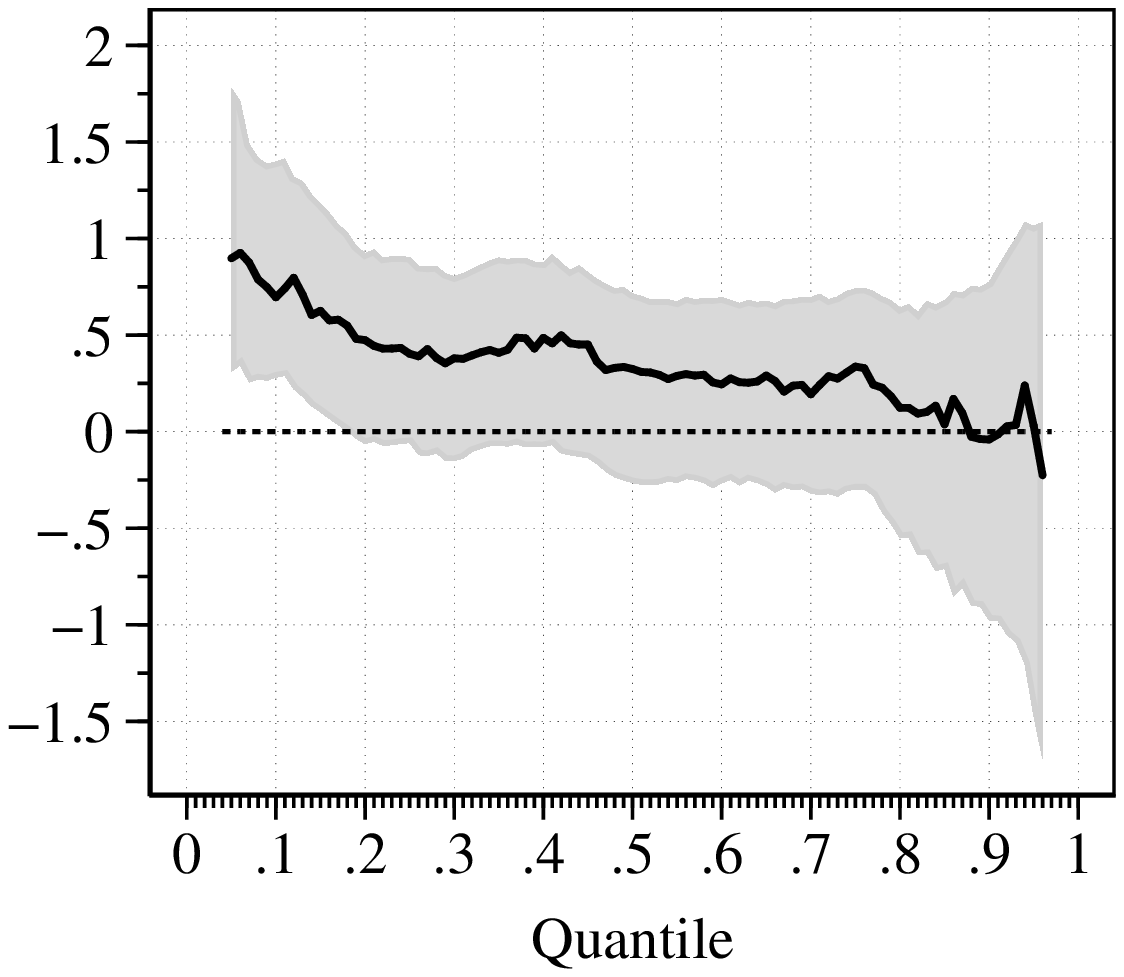}}\subfloat[Education]{
\centering{}\includegraphics[scale=0.6]{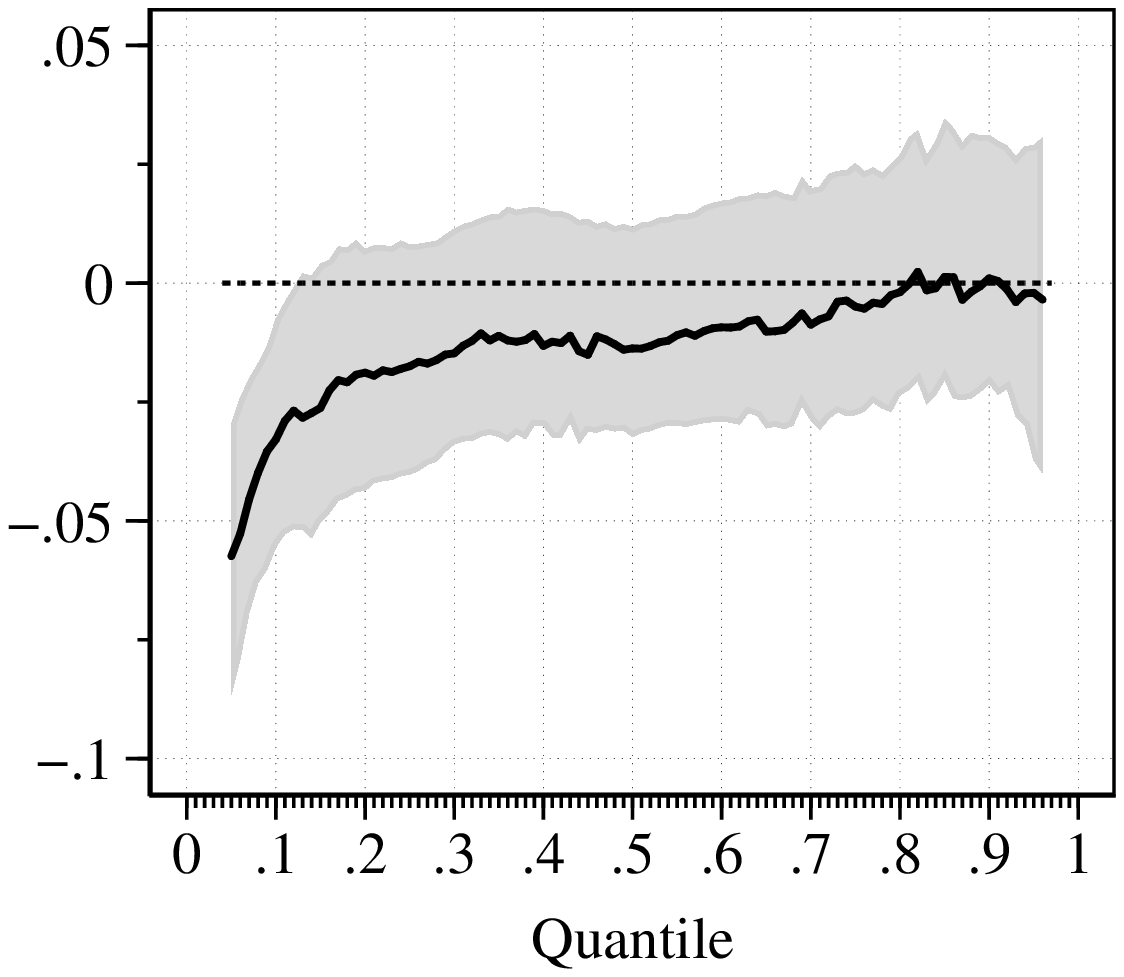}}
\par\end{centering}
\begin{centering}
\subfloat[Experience]{
\centering{}\includegraphics[scale=0.6]{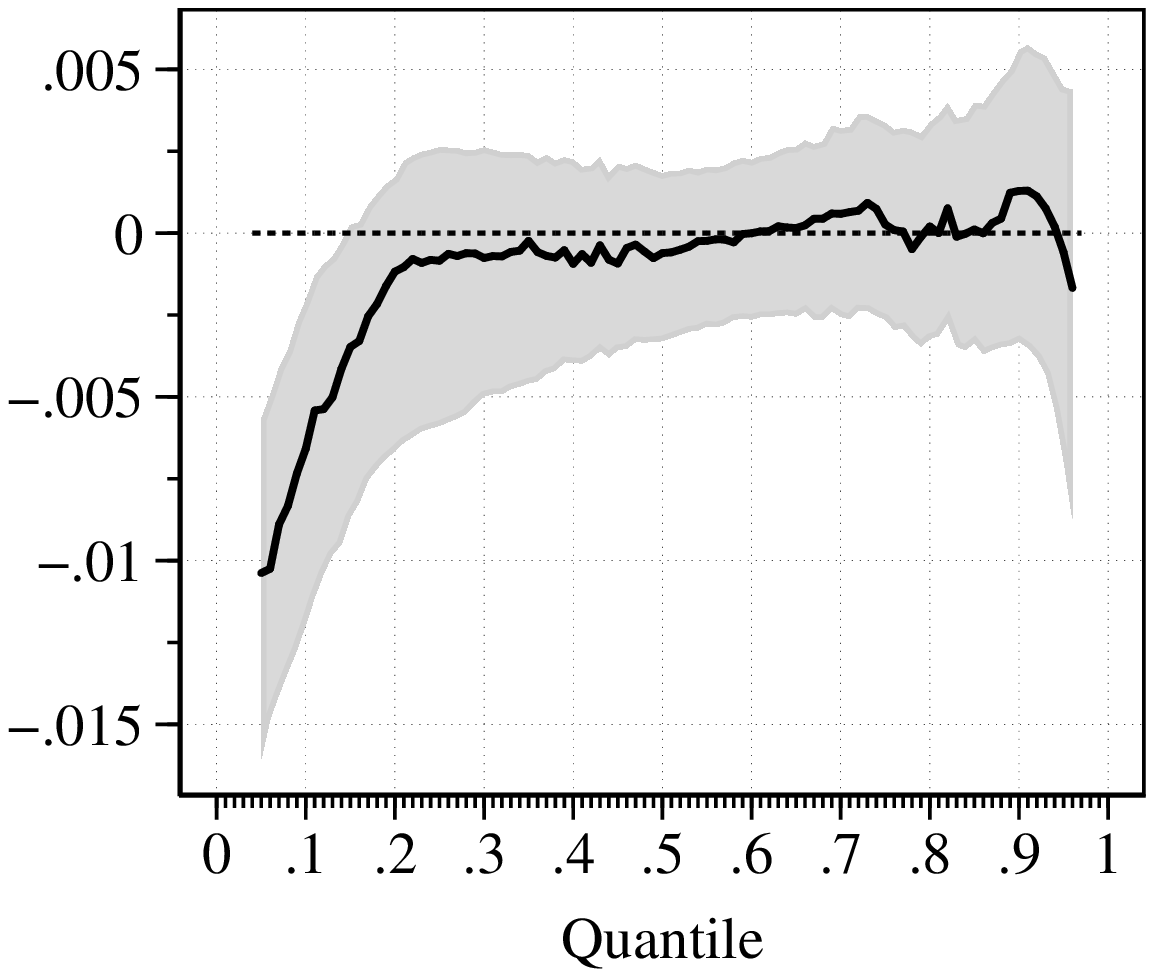}}\subfloat[Gender (male)]{
\centering{}\includegraphics[scale=0.6]{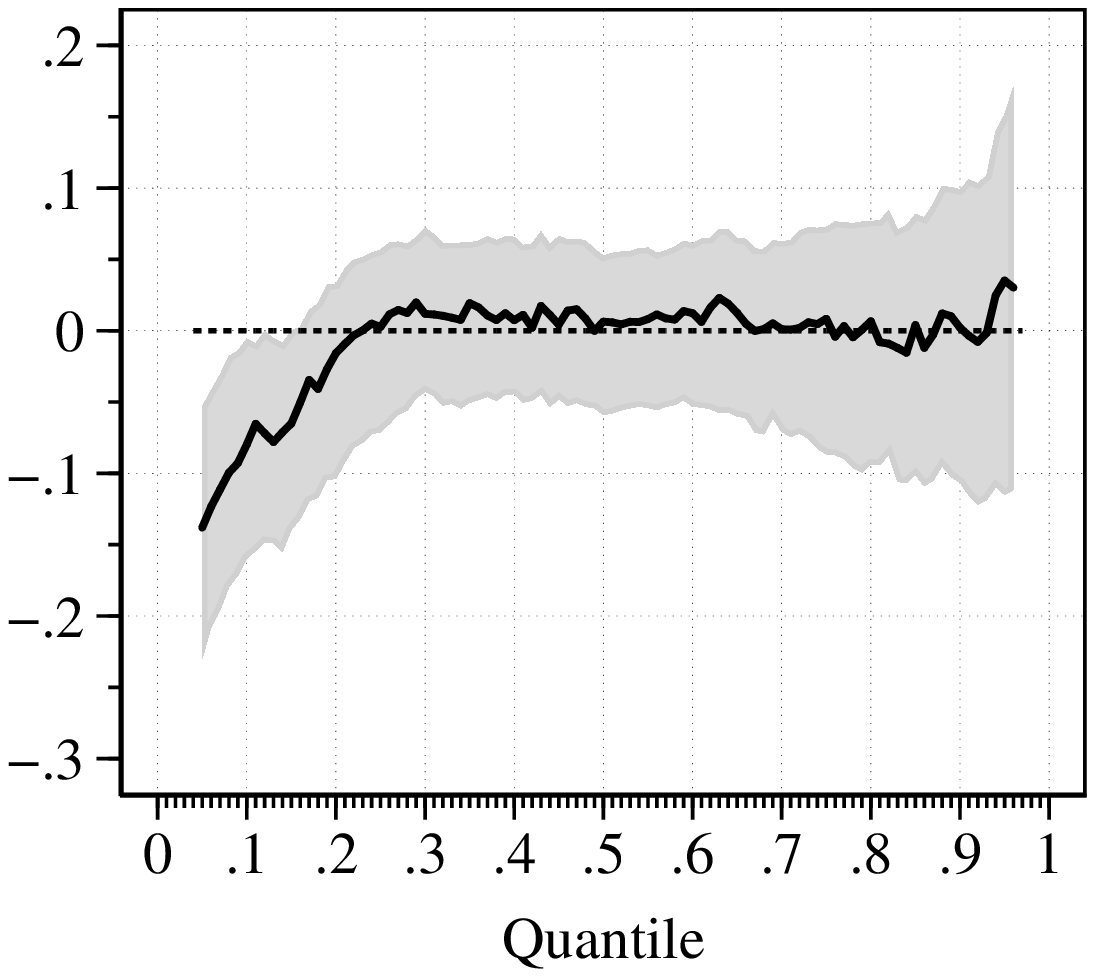}}
\par\end{centering}
\textit{\footnotesize{}Notes}{\footnotesize{}: Estimates of the long-term
effects in equation \eqref{eq: mean2} are reported. The shaded area
represents the 95 percent confidence interval. See Figure \ref{fig: estimates_lag_band}
for the uniform confidence band.}{\footnotesize\par}
\end{figure}

\begin{figure}[h]
\caption{Placebo effect on the wage structure\label{fig: estimates_lead}}

\begin{centering}
\subfloat[Intercept]{
\centering{}\includegraphics[scale=0.6]{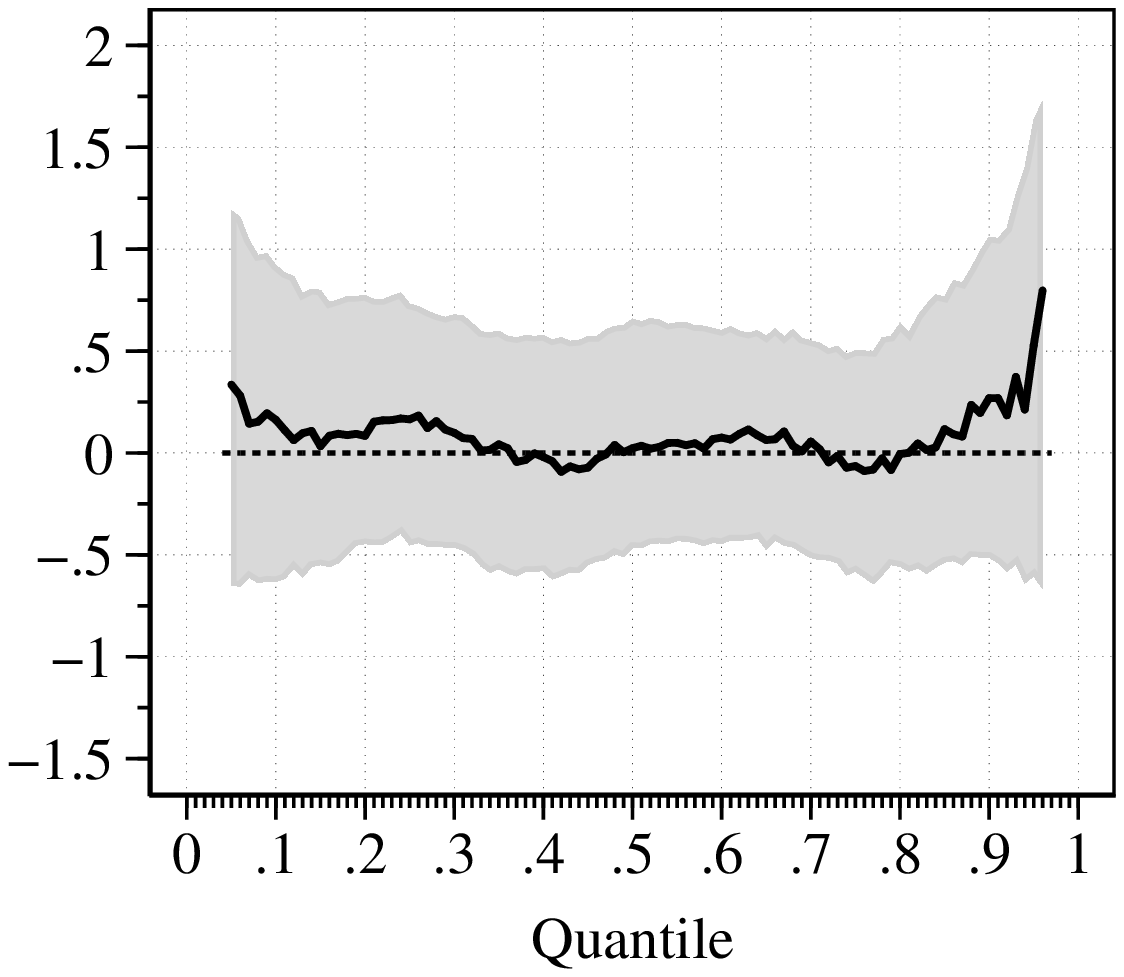}}\subfloat[Education]{
\centering{}\includegraphics[scale=0.6]{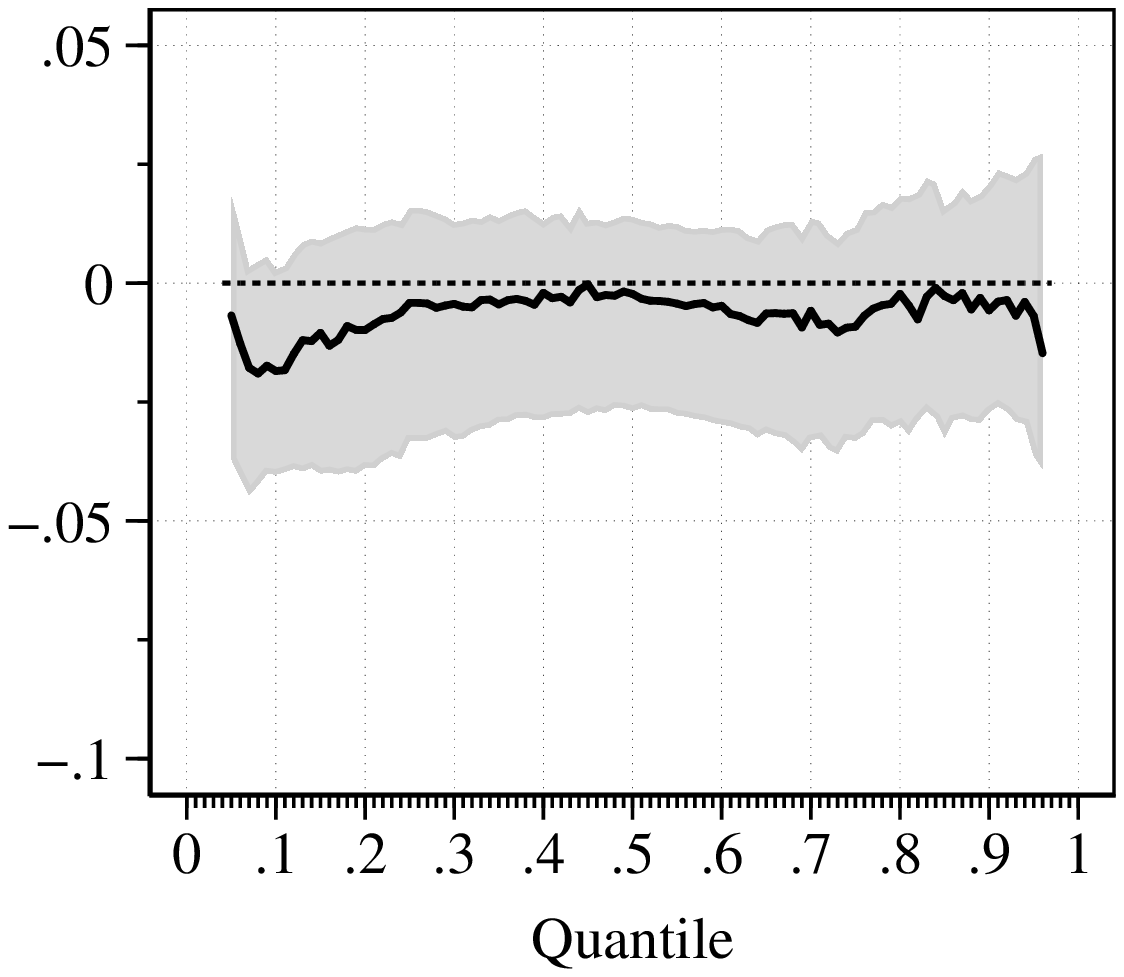}}
\par\end{centering}
\begin{centering}
\subfloat[Experience]{
\centering{}\includegraphics[scale=0.6]{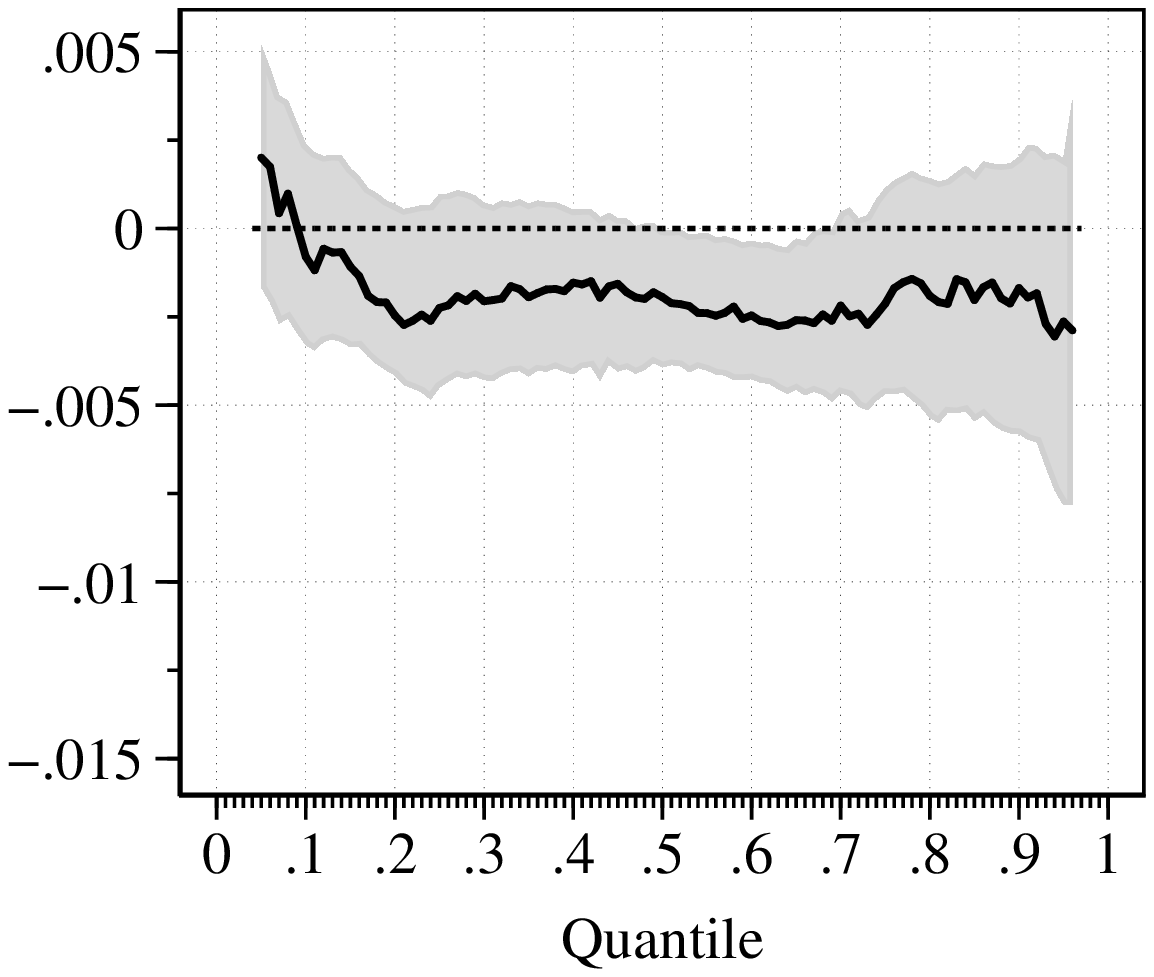}}\subfloat[Gender (male)]{
\centering{}\includegraphics[scale=0.6]{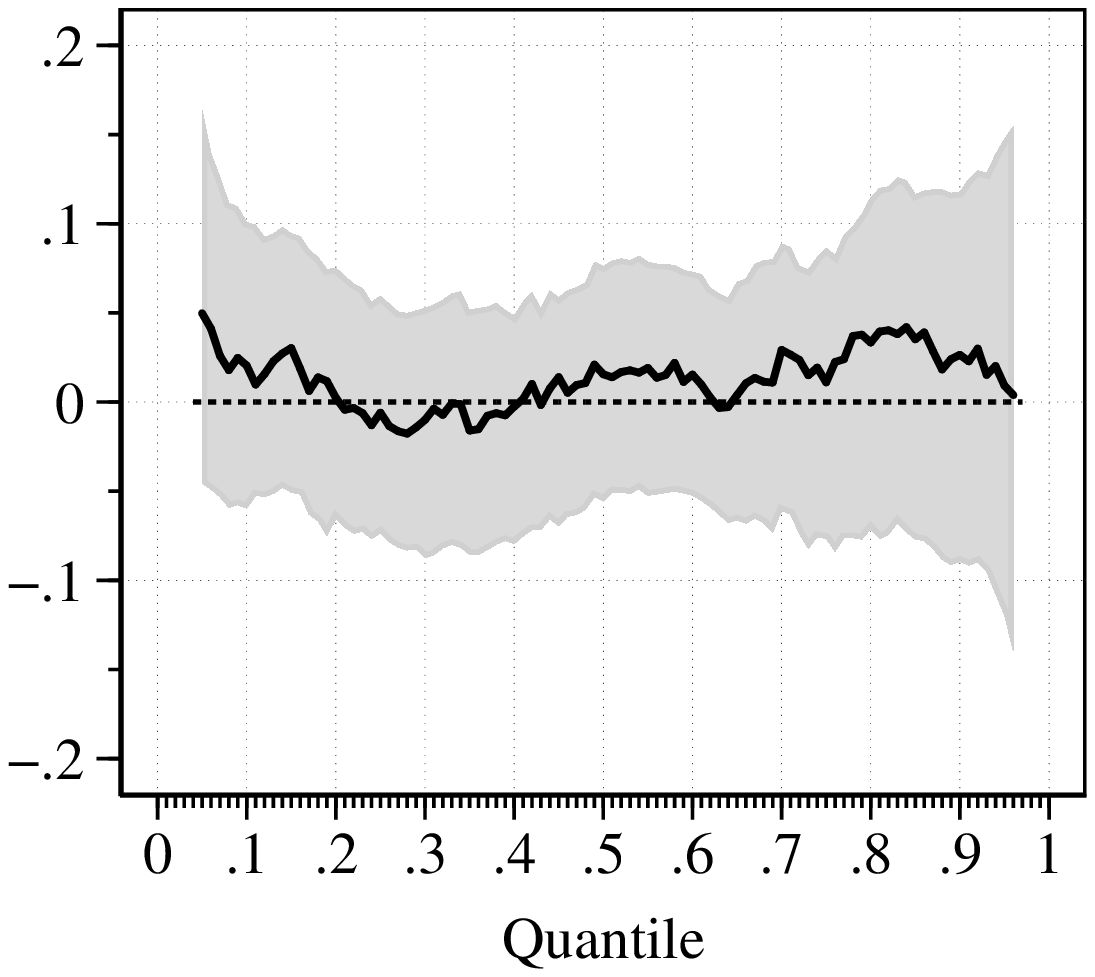}}
\par\end{centering}
\textit{\footnotesize{}Notes}{\footnotesize{}: Estimates of the leading
effects in equation \eqref{eq: mean2} are reported. The shaded area
represents the 95 percent confidence interval. See Figure \ref{fig: estimates_lead_band}
for the uniform confidence band.}{\footnotesize\par}
\end{figure}

\paragraph{Lag and lead}

Before discussing the contribution of the minimum wage to changes
in between- and within-group inequality, we present the results when
estimating the augmented equation \eqref{eq: mean2}. The four panels
in Figure \ref{fig: estimates_lag} show the estimates of the long-term
effects. All estimates remain essentially unchanged, although they
become less precise. Indeed, the long-term effects fall inside the
95 percent confidence intervals of the contemporary effects. The four
panels in Figure \ref{fig: estimates_lead} illustrate the estimates
of the leading (placebo) effects. All estimates are close to zero
for virtually all quantiles, and none of them are statistically significant.
These results support our specification.

\subsection{Contribution to changes in between- and within-group inequality\label{subsec: contribution}}

Finally, we discuss the quantitative contribution of the minimum wage
to changes in between- and within-group inequality. As in Figure \ref{fig: real_minimum},
the real value of the minimum wage declined by 30 log points due to
inflation for the years 1979 to 1989 and subsequently increased by
28 log points due to increases in the statutory minimum wage for the
years 1989 to 2012. Here, we provide the results for workers with
10 years of experience or less, who are subject to the influence of
the minimum wage, for the latter period. Appendix \ref{subsec: changes_1989}
shows the results for the former period.

\subsubsection{Between-group inequality}

\paragraph{Educational wage differential}

\begin{figure}[h]
\caption{Changes in the educational wage differential (16 versus 12 years of
education) due to the minimum wage, 1989\textendash 2012\label{fig: diff_educ}}

\begin{centering}
\subfloat[5 years of experience, males]{
\centering{}\includegraphics[scale=0.6]{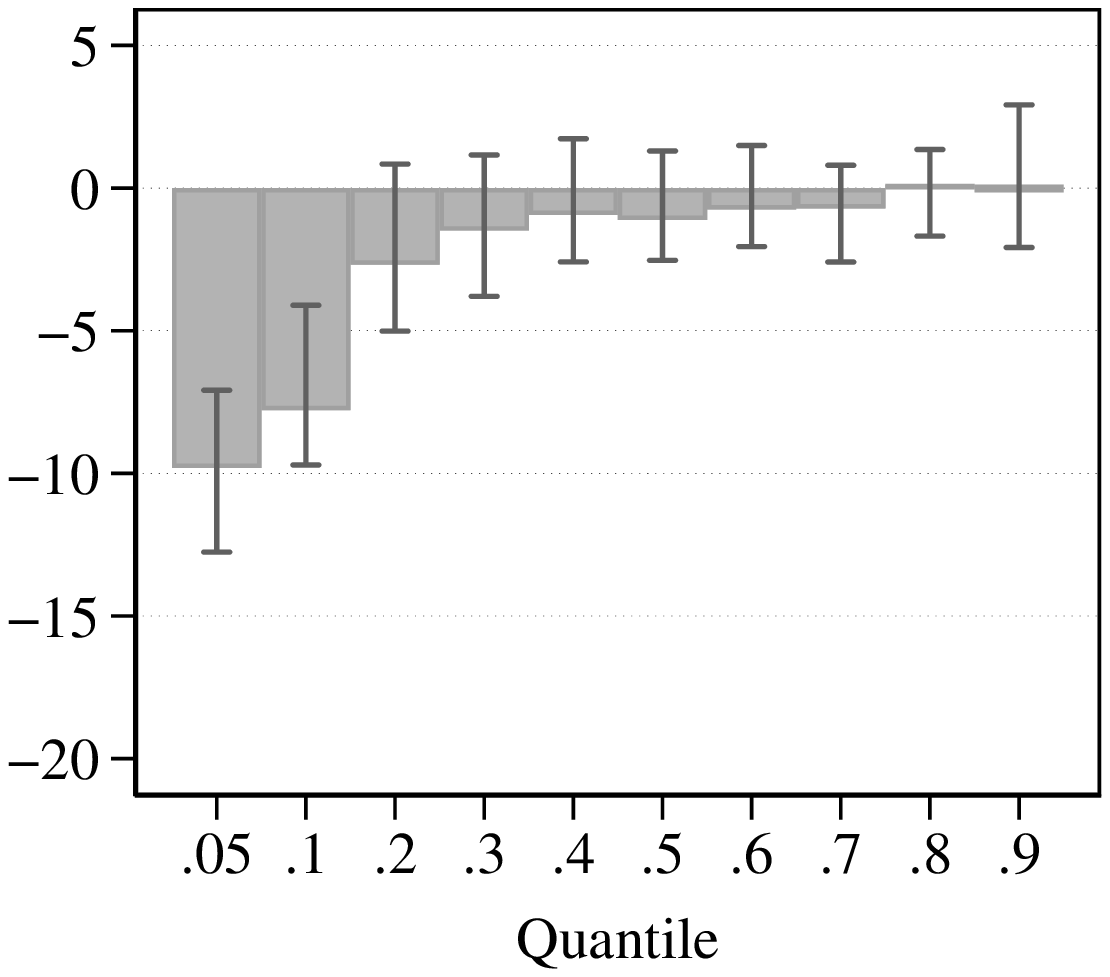}}\subfloat[10 years of experience, males]{
\centering{}\includegraphics[scale=0.6]{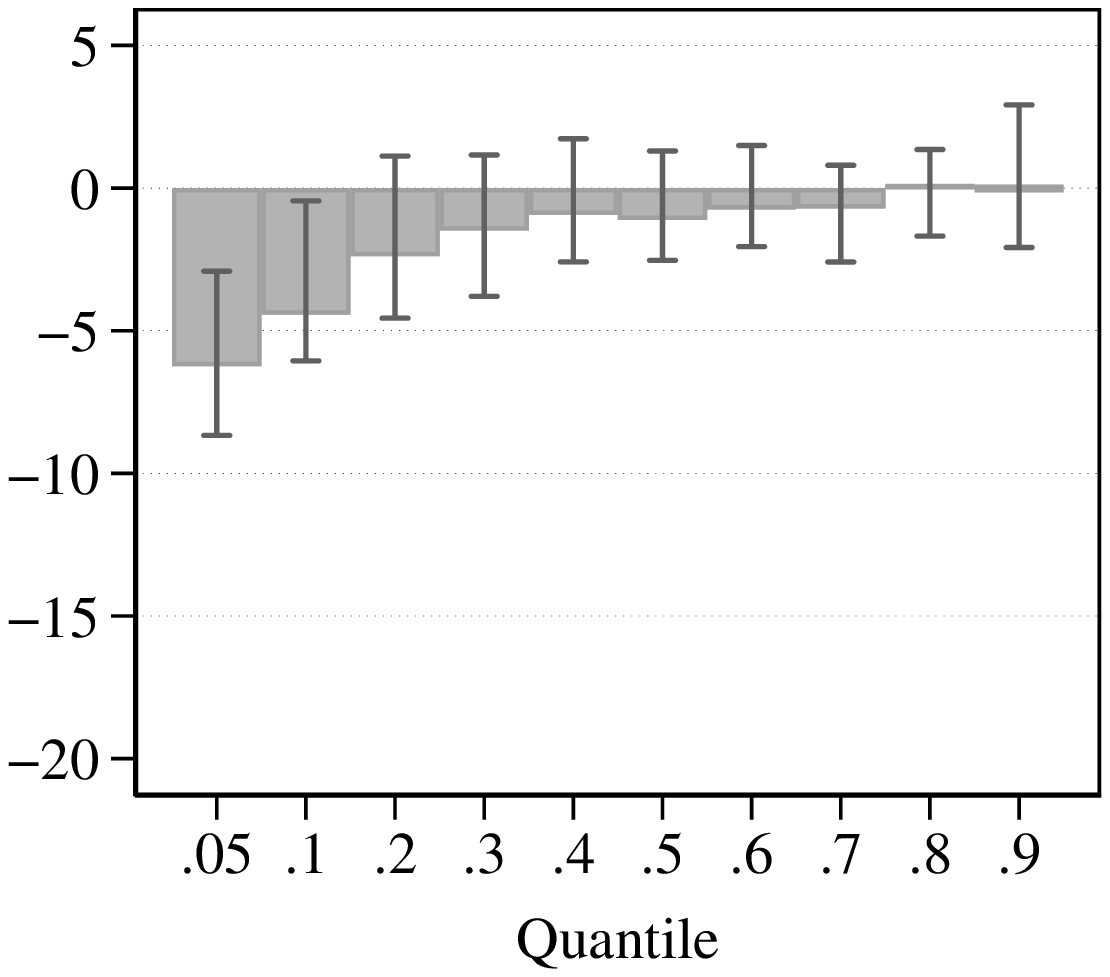}}
\par\end{centering}
\begin{centering}
\subfloat[5 years of experience, females]{
\centering{}\includegraphics[scale=0.6]{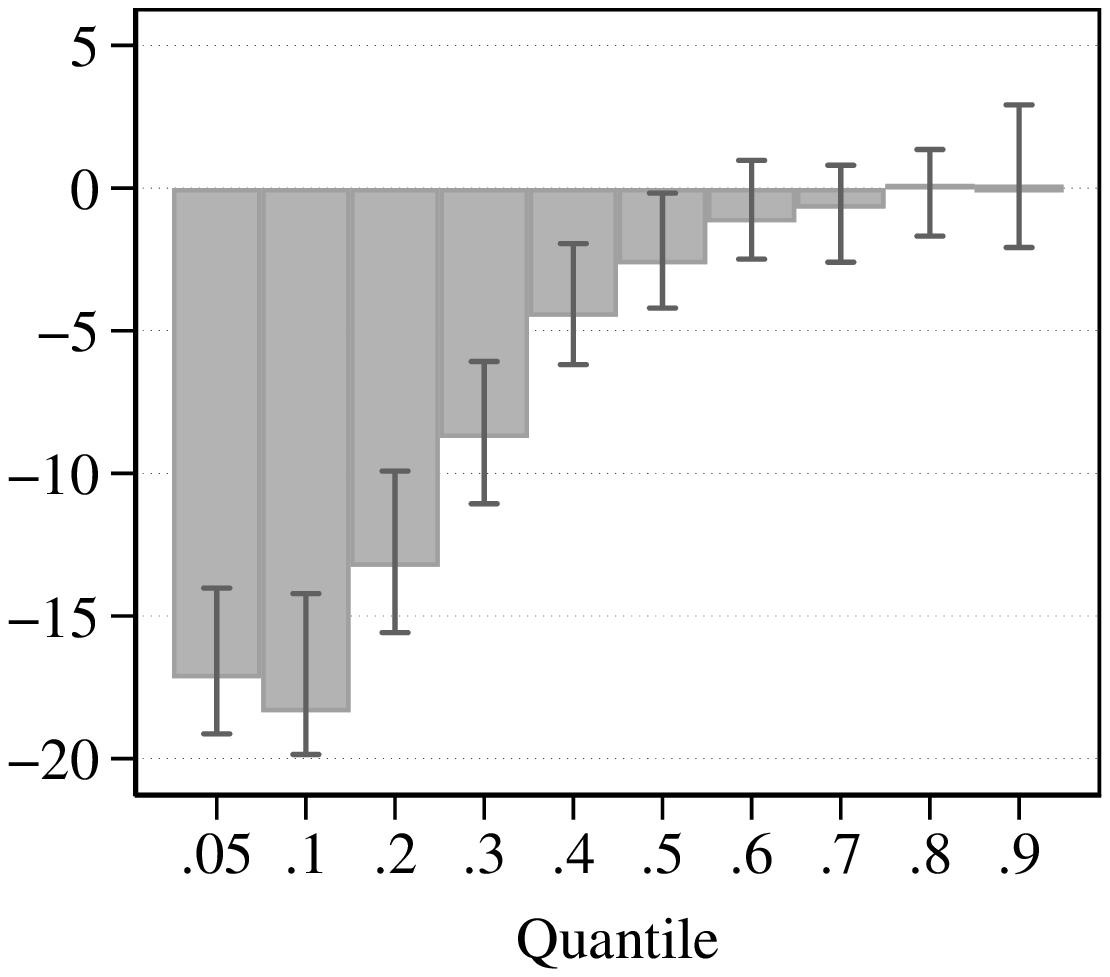}}\subfloat[10 years of experience, females]{
\centering{}\includegraphics[scale=0.6]{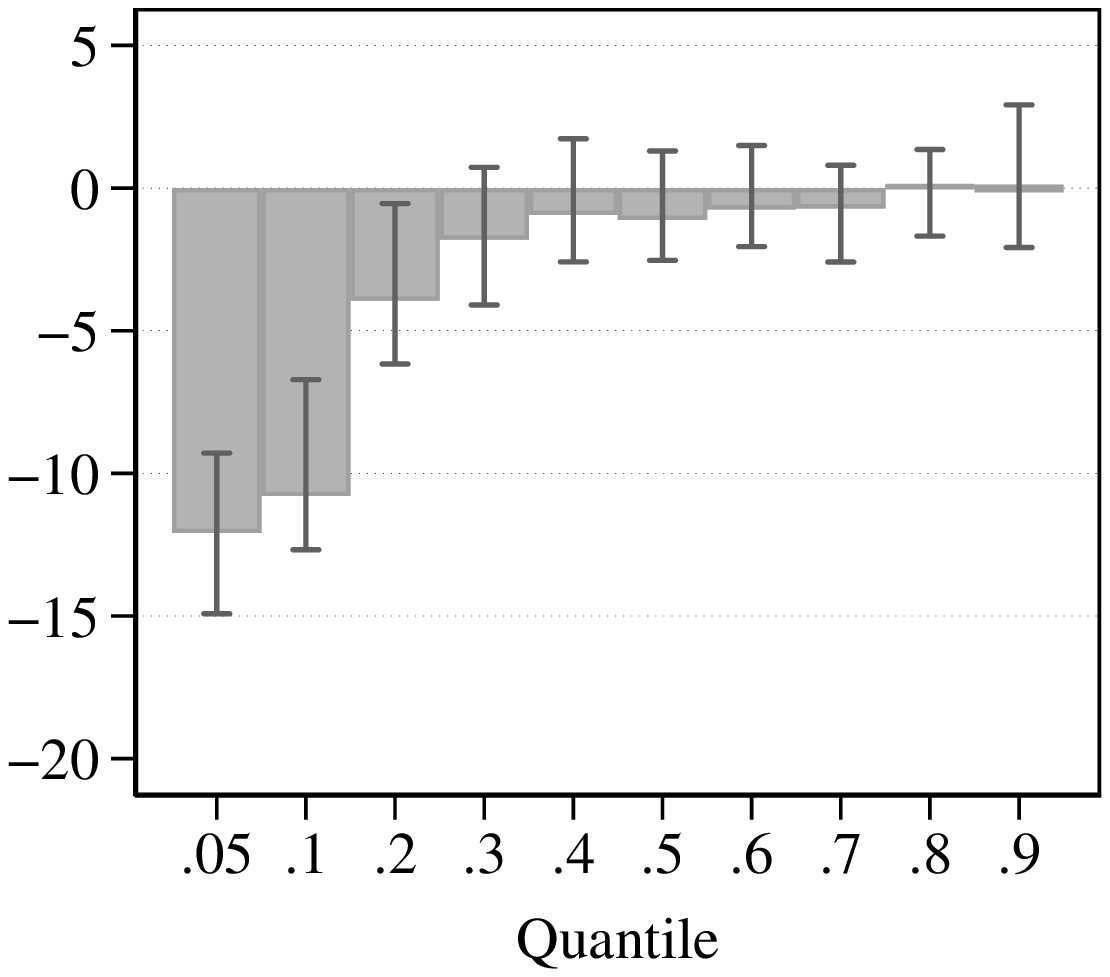}}
\par\end{centering}
\textit{\footnotesize{}Notes}{\footnotesize{}: Estimates of log-point
changes in the educational wage differential due to the minimum wage
are obtained from equation \eqref{eq: between2}. The error bar represents
the 95 percent confidence interval.}{\footnotesize\par}
\end{figure}

We measure the educational wage differential by comparing workers
with 16 years of education (equivalent to college graduates) and those
with 12 years of education (equivalent to high school graduates),
holding experience and gender constant. The four panels in Figure
\ref{fig: diff_educ} show the national means of changes in the educational
wage differential due to increases in the real value of the minimum
wage for the years 1989 to 2012 by experience and gender for each
decile $\tau=0.05$, $0.1$, $0.2$, ..., $0.9$.

The minimum wage contributes to a reduction in the educational wage
differential in the lower quantiles. The contribution of the minimum
wage to a reduction in the educational wage differential is greater
for more-experienced, female workers than less-experienced, male workers.
For each group of workers, the contribution of the minimum wage is
greatest at the 0.05th quantile and gradually declines in absolute
value to zero by the 0.2th to 0.5th quantiles. For female workers
with five years of experience, however, it is slightly greater at
the 0.1th quantile than the 0.05th quantile. The reason is that, at
the 0.05th quantile in this group, both more- and less-educated workers
are affected by a rise in the real value of the minimum wage.

\begin{figure}[h]
\caption{Actual and counterfactual changes in the educational wage differential
(16 versus 12 years of education), 1989\textendash 2012\label{fig: diff_educ_actual}}

\begin{centering}
\subfloat[5 years of experience, males]{
\centering{}\includegraphics[scale=0.6]{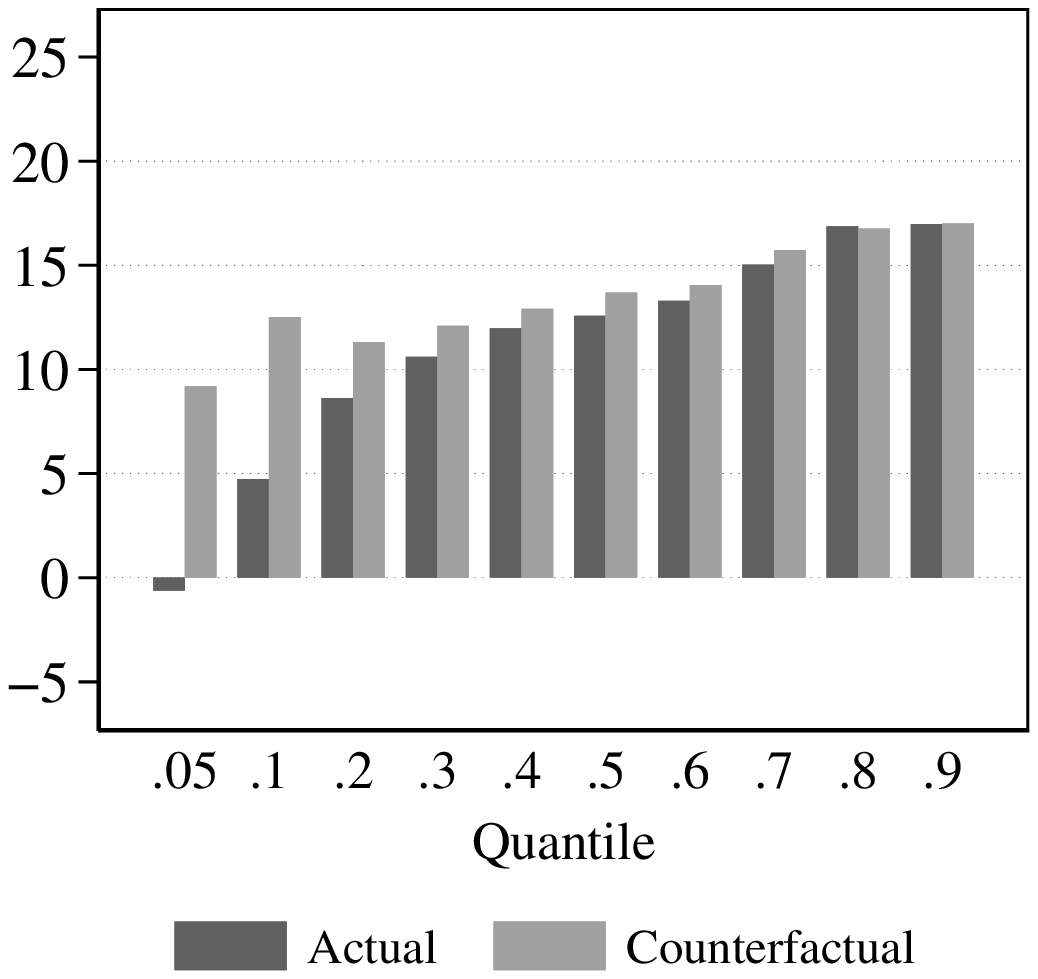}}\subfloat[10 years of experience, males]{
\centering{}\includegraphics[scale=0.6]{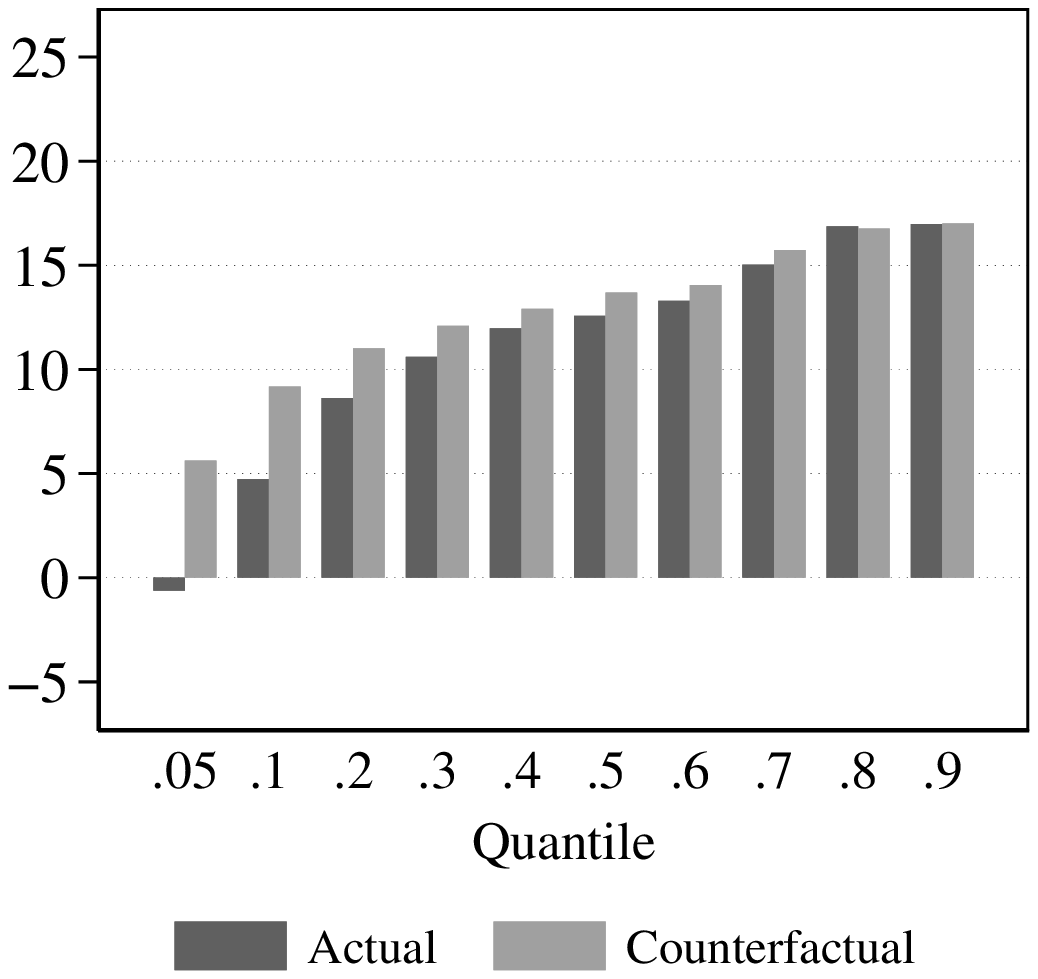}}
\par\end{centering}
\begin{centering}
\subfloat[5 years of experience, females]{
\centering{}\includegraphics[scale=0.6]{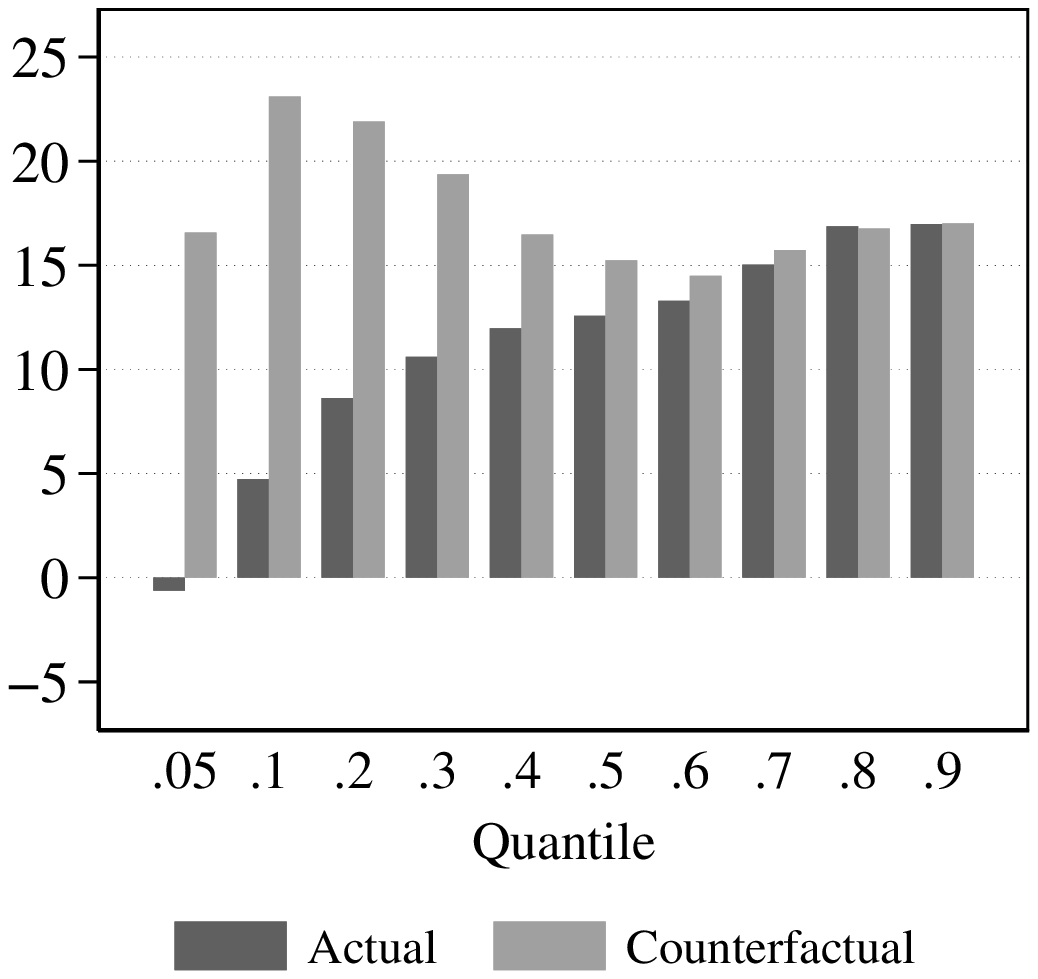}}\subfloat[10 years of experience, females]{
\centering{}\includegraphics[scale=0.6]{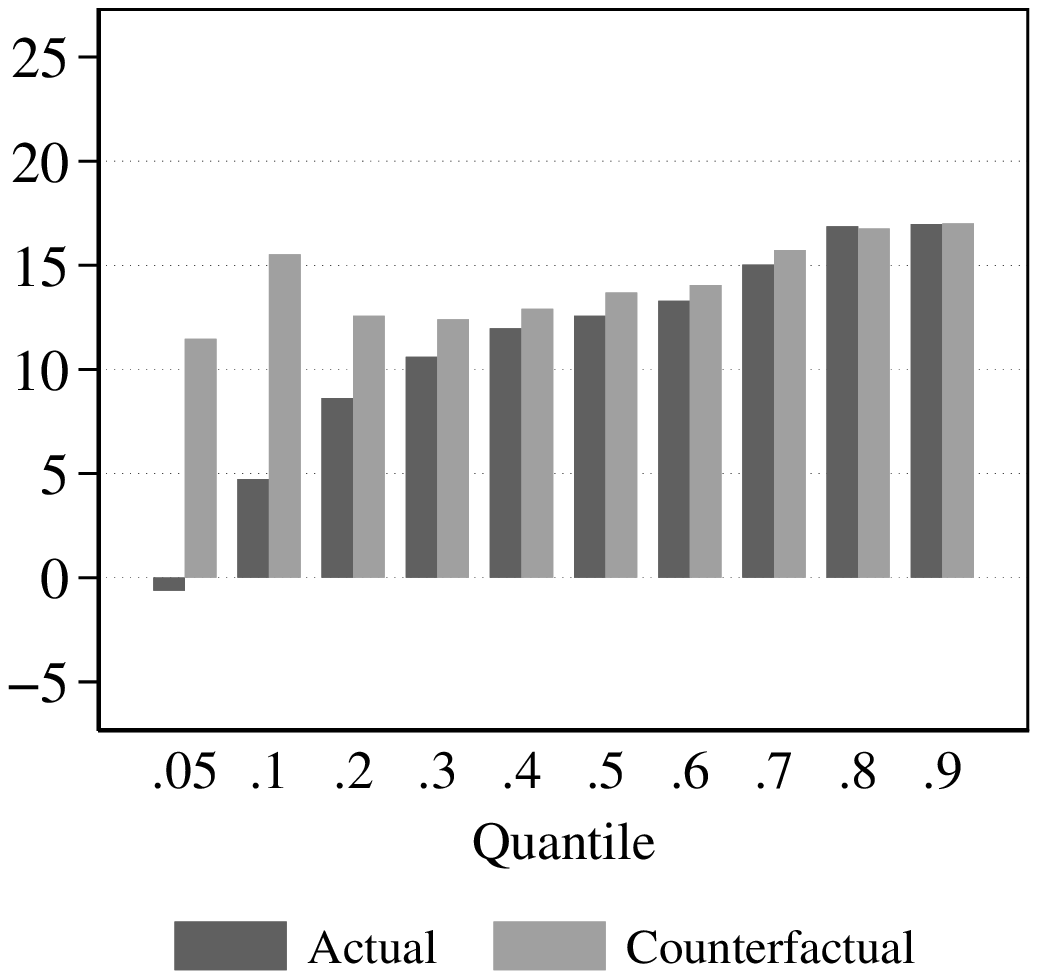}}
\par\end{centering}
\textit{\footnotesize{}Notes}{\footnotesize{}: Actual and counterfactual
log-point changes in the educational wage differential are obtained
from equations \eqref{eq: between1} and \eqref{eq: between2}.}{\footnotesize\par}
\end{figure}

The educational wage differential increased during the period (Figure
\ref{fig: diff_educ_actual}). The trend in the educational wage differential
is known to be important in accounting for the rise in wage inequality
in the United States \citep{Autor_Katz_Kearney_RESTAT08}. The increase
in the educational wage differential is typically attributed in the
literature to skill-biased technological change and compositional
changes in the workforce \citep{Bound_Johnson_AER92,Katz_Murphy_QJE92,Autor_Katz_Kearney_RESTAT08}.
The magnitude of the increase in the educational wage differential
is greater in the higher quantiles than the lower quantiles during
the period, as also shown by \citet*{Buchinsky_EM94} and \citet{Angrist_Chernozhukov_FernandezVal_EM06}.
The educational wage differential did not increase at the 0.05 quantile
and increased only moderately at the 0.1 quantile, while it increased
more in the higher quantiles. If there were no increase in the real
value of the minimum wage, however, the educational wage differential
would increase at the 0.05 quantile and more than double at the 0.1
quantile for all groups. Consequently, in the counterfactual case
in which the real value of the minimum wage is kept constant, the
increase in the educational wage differential is more uniform across
quantiles. Our results indicate that the minimum wage is another factor
in accounting for the patterns of changes in the educational wage
differential.

\paragraph{Experience wage differential}

\begin{figure}[h]
\caption{Changes in the experience wage differential (25 versus 5 years of
experience) due to the minimum wage, 1989\textendash 2012\label{fig: diff_exper1}}

\begin{centering}
\subfloat[12 years of education, males]{
\centering{}\includegraphics[scale=0.6]{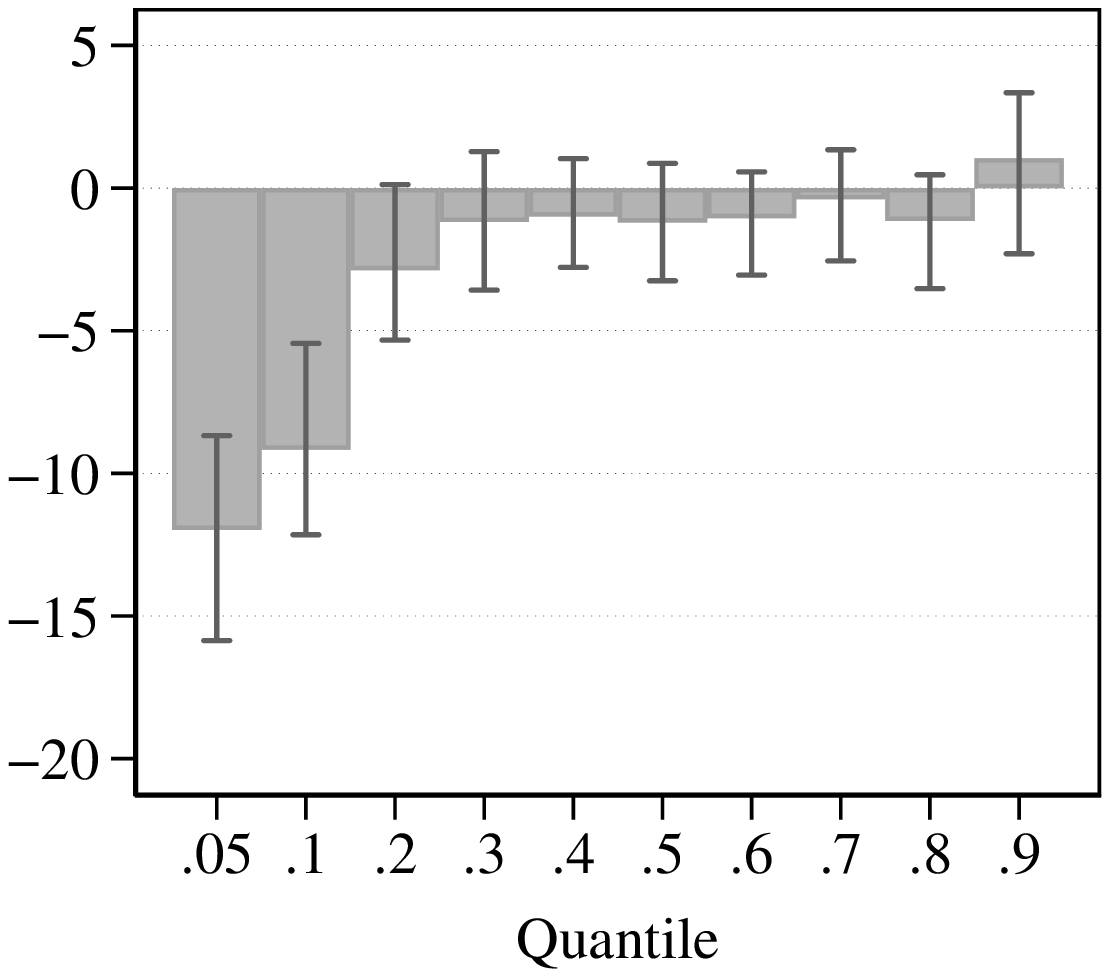}}\subfloat[16 years of education, males]{
\centering{}\includegraphics[scale=0.6]{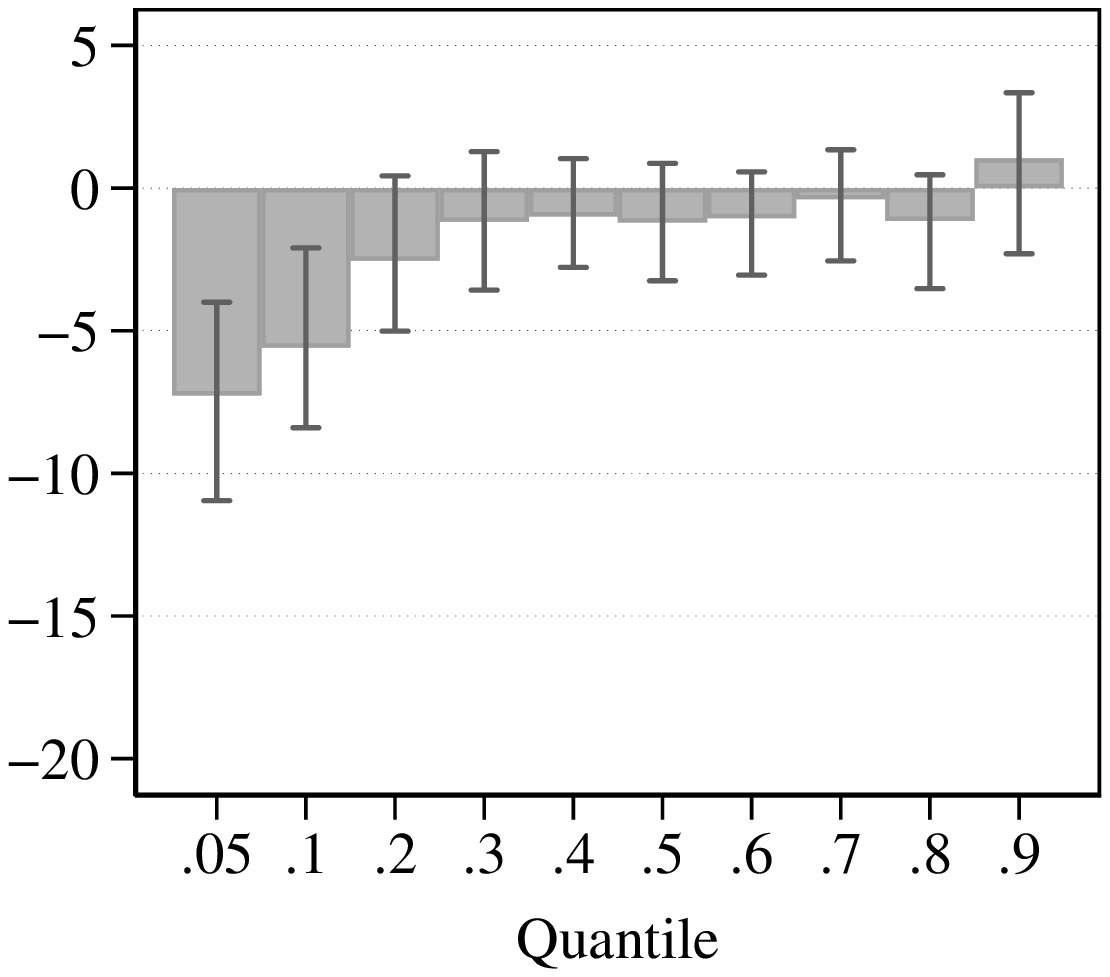}}
\par\end{centering}
\begin{centering}
\subfloat[12 years of education, females]{
\centering{}\includegraphics[scale=0.6]{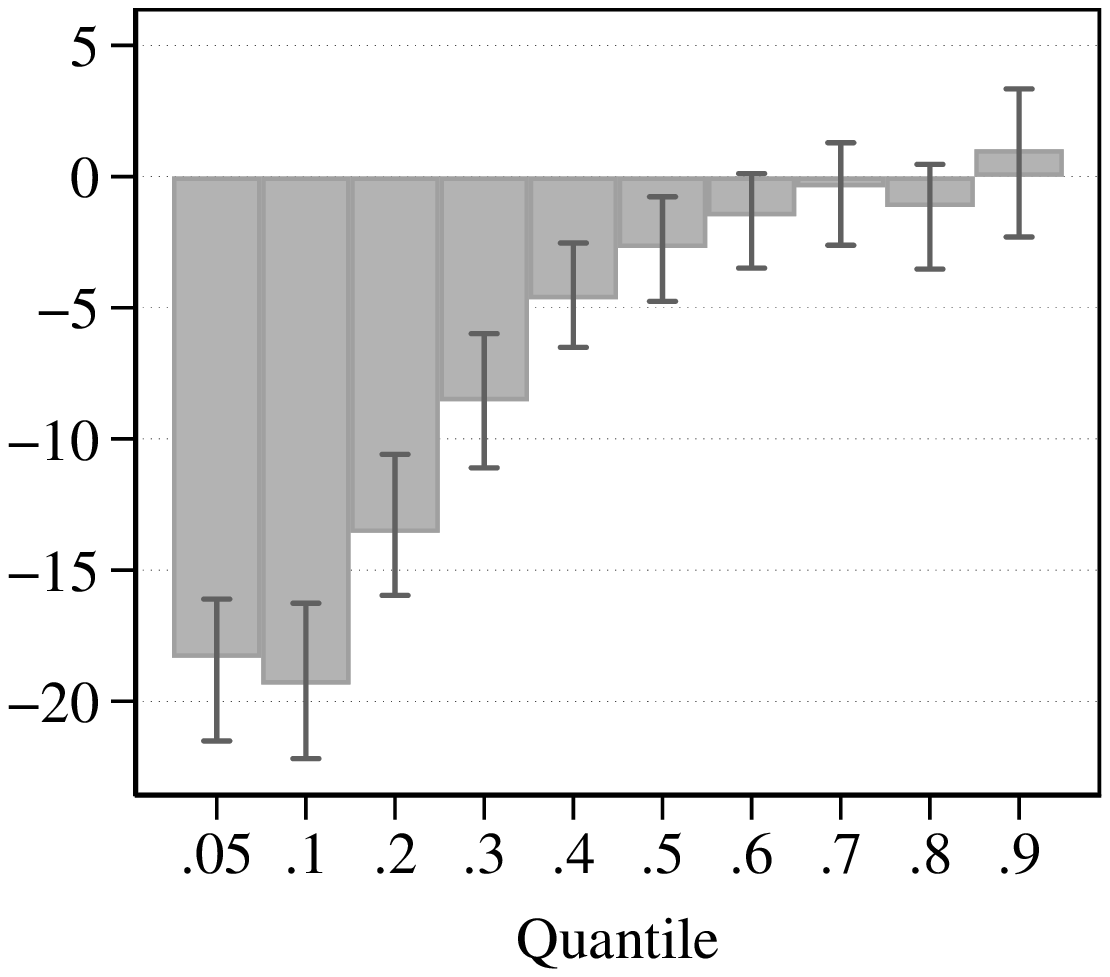}}\subfloat[16 years of education, females]{
\centering{}\includegraphics[scale=0.6]{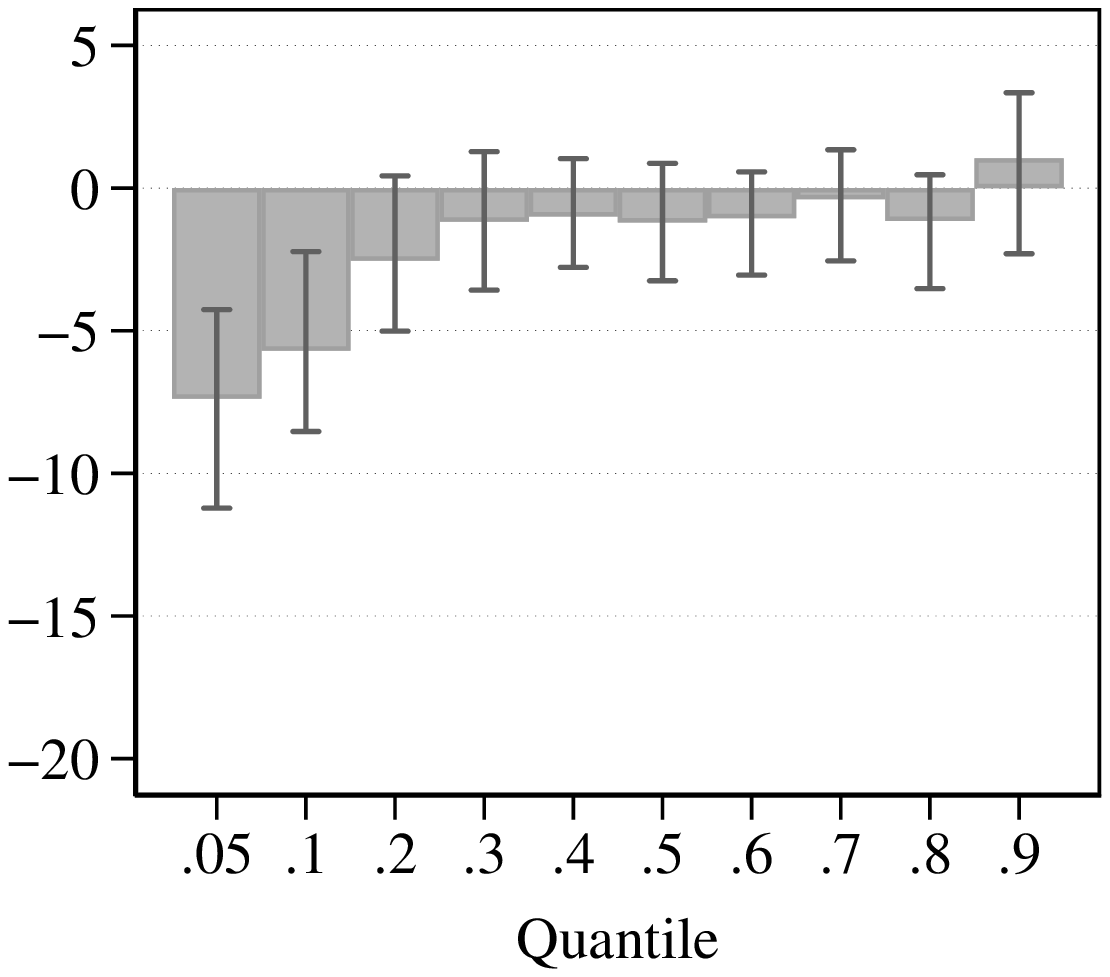}}
\par\end{centering}
\textit{\footnotesize{}Notes}{\footnotesize{}: Estimates of log-point
changes in the experience wage differential due to the minimum wage
are obtained from equation \eqref{eq: between2}. The error bar represents
the 95 percent confidence interval.}{\footnotesize\par}
\end{figure}

We measure the experience wage differential by comparing workers with
25 years of experience and those with five years of experience, holding
education and gender constant. The four panels in Figure \ref{fig: diff_exper1}
show the national means of changes in the experience wage differential
due to increases in the real value of the minimum wage for the years
1989 to 2012 by education and gender.

The minimum wage contributes to a reduction in the experience wage
differential in the lower quantiles. The contribution of the minimum
wage to a reduction in the experience wage differential is greater
for less-educated, female workers than more-educated, male workers.
For each group of workers, the contribution of the minimum wage is
greatest at the 0.05th quantile and gradually declines in absolute
value to zero by the 0.2th to 0.5th quantiles. For female workers
with 12 years of education, however, it is slightly  greater at the
0.1th quantile than the 0.05th quantile. The reason is that, at the
0.05th quantile in this group, both more- and less-experienced workers
are affected by a rise in the real value of the minimum wage.

\begin{figure}[h]
\caption{Actual and counterfactual changes in the experience wage differential
(25 versus 5 years of experience), 1989\textendash 2012\label{fig: diff_exper_actual}}

\begin{centering}
\subfloat[12 years of education, males]{
\centering{}\includegraphics[scale=0.6]{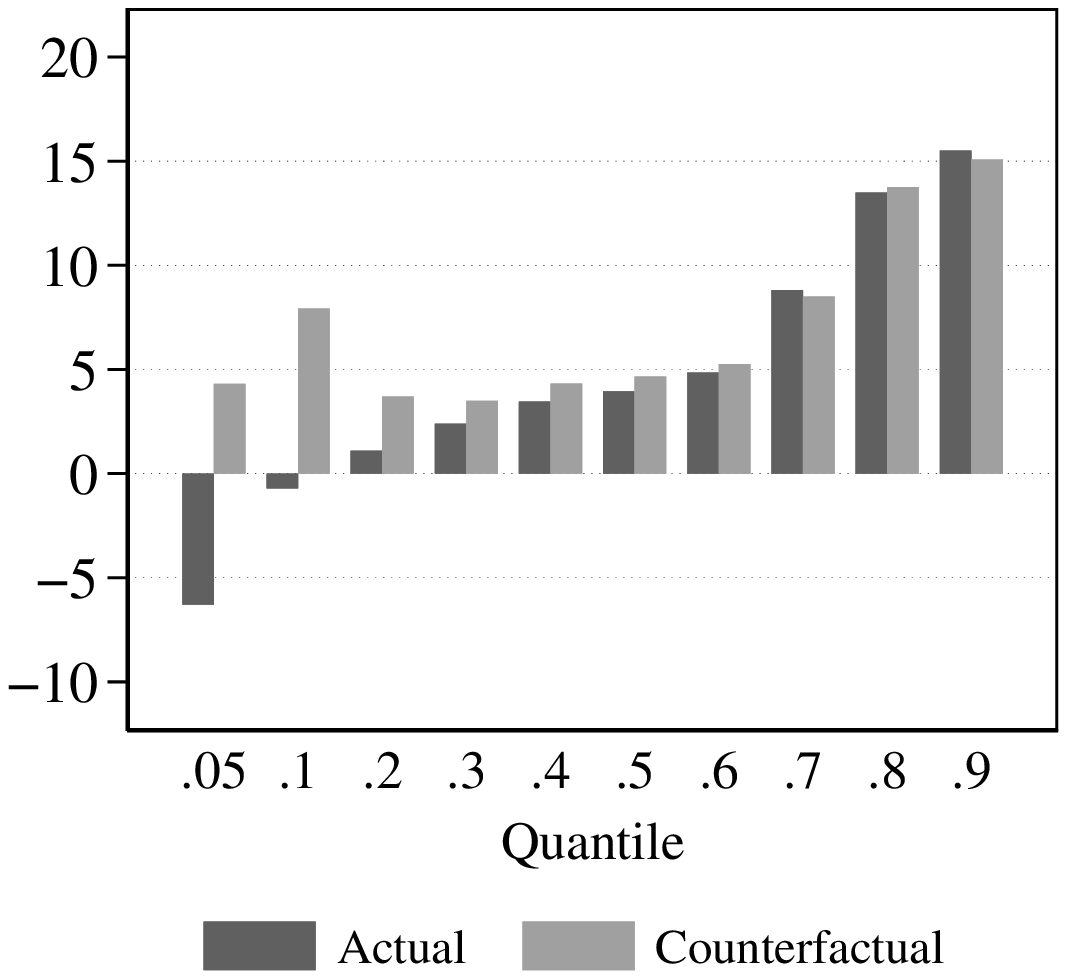}}\subfloat[16 years of education, males]{
\centering{}\includegraphics[scale=0.6]{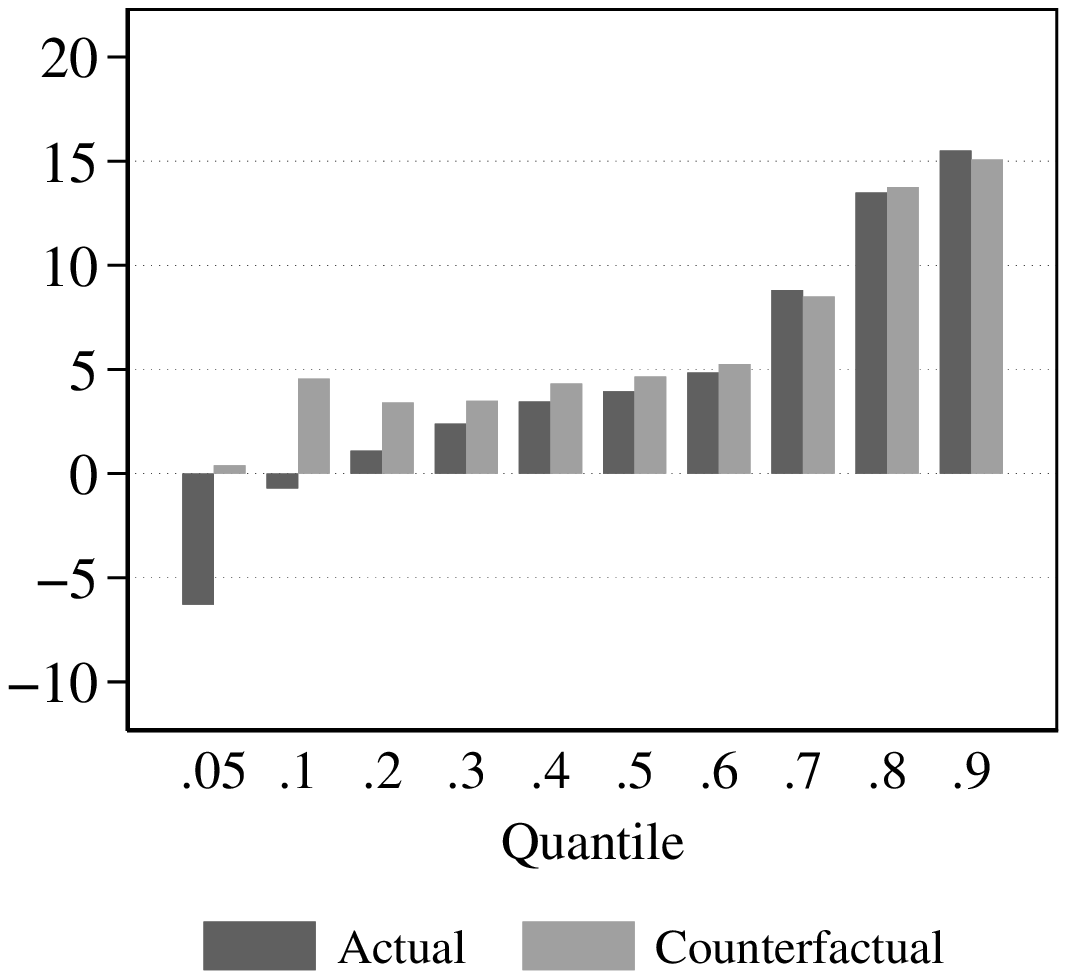}}
\par\end{centering}
\begin{centering}
\subfloat[12 years of education, females]{
\centering{}\includegraphics[scale=0.6]{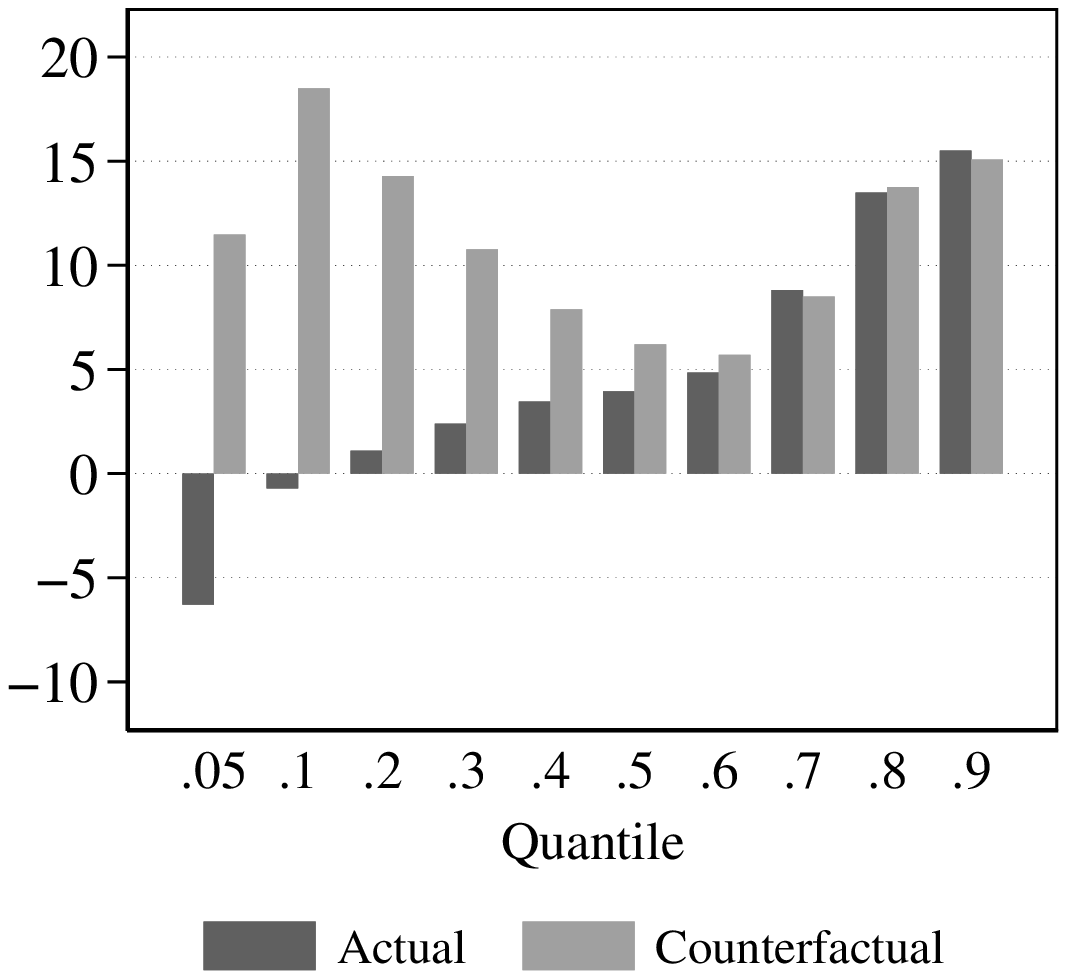}}\subfloat[16 years of education, females]{
\centering{}\includegraphics[scale=0.6]{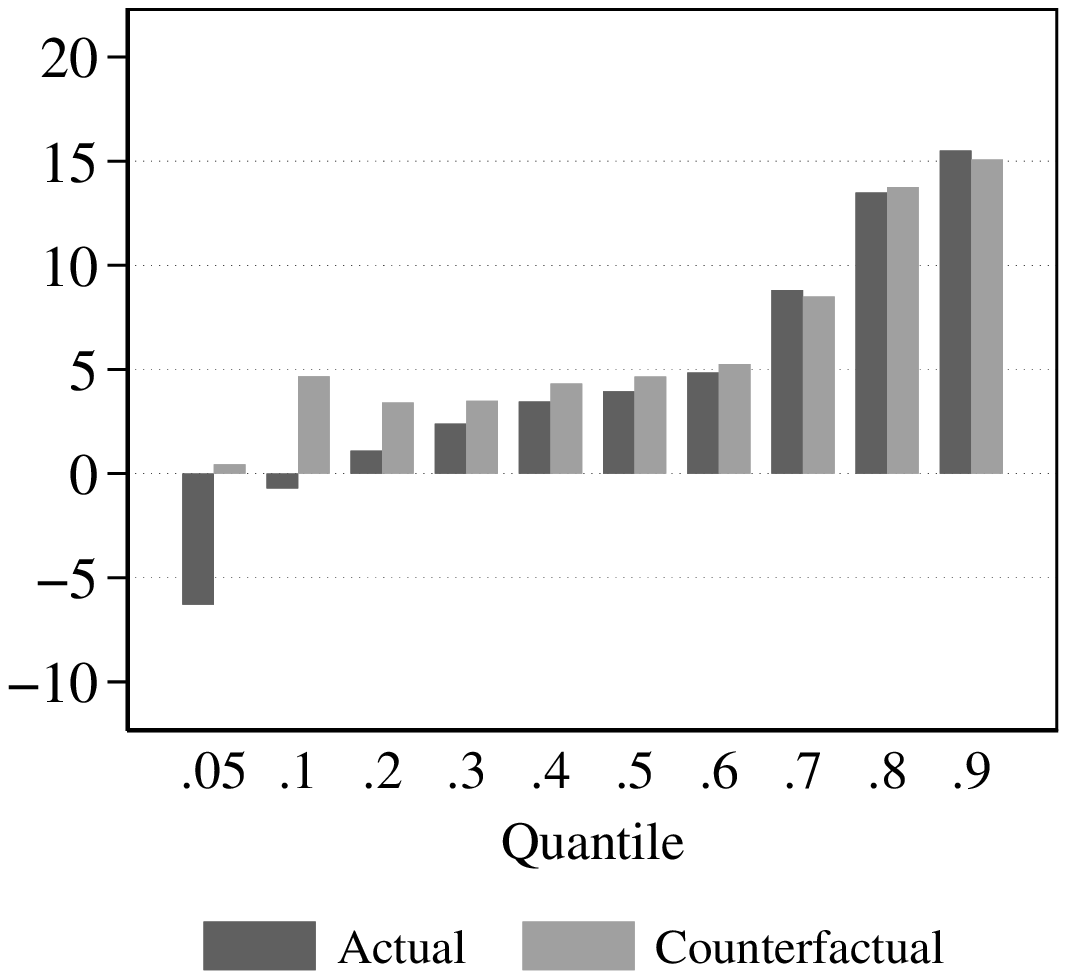}}
\par\end{centering}
\textit{\footnotesize{}Notes}{\footnotesize{}: Actual and counterfactual
log-point changes in the experience wage differential are obtained
from equations \eqref{eq: between1} and \eqref{eq: between2}.}{\footnotesize\par}
\end{figure}

The experience wage differential increased during the period with
the exception of the lowest quantile (Figure \ref{fig: diff_exper_actual}).
Changes in the experience wage differential are typically attributed
in the literature to compositional changes in the workforce \citep*{Welch_JPE79,Jeong_Kim_Manovskii_AER15}.
The magnitude of the increase in the experience wage differential
is greater in the higher quantiles than the lower quantiles during
the period. The experience wage differential declined at the 0.05th
quantile and increased only moderately at the median, while it increased
more at the 0.7th and higher quantiles. If there were no increase
in the real value of the minimum wage, however, the experience wage
differential would increase in the lower as well as higher quantiles.
Consequently, in the counterfactual case in which the real value of
the minimum wage is kept constant, the increase in the educational
wage differential at the 0.1th quantile is at least as high as the
increase in the median for all groups. Our results indicate that the
minimum wage is another factor in accounting for the patterns of changes
in the experience wage differential.

\paragraph{Gender wage differential}

\begin{figure}[h]
\caption{Changes in the gender wage differential (males versus females) due
to the minimum wage, 1989\textendash 2012\label{fig: diff_gender}}

\begin{centering}
\subfloat[12 years of education, 5 years of experience]{
\centering{}\includegraphics[scale=0.6]{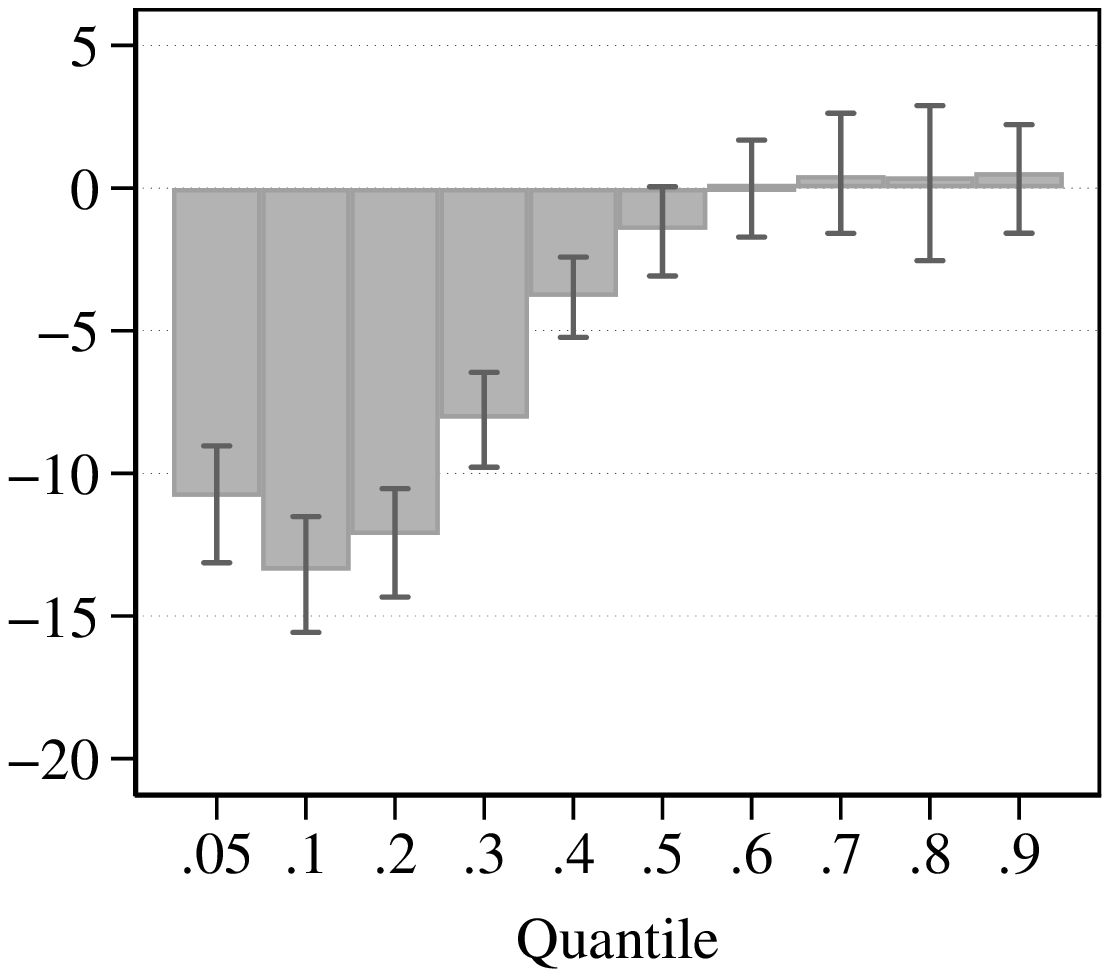}}\subfloat[12 years of education, 10 years of experience]{
\centering{}\includegraphics[scale=0.6]{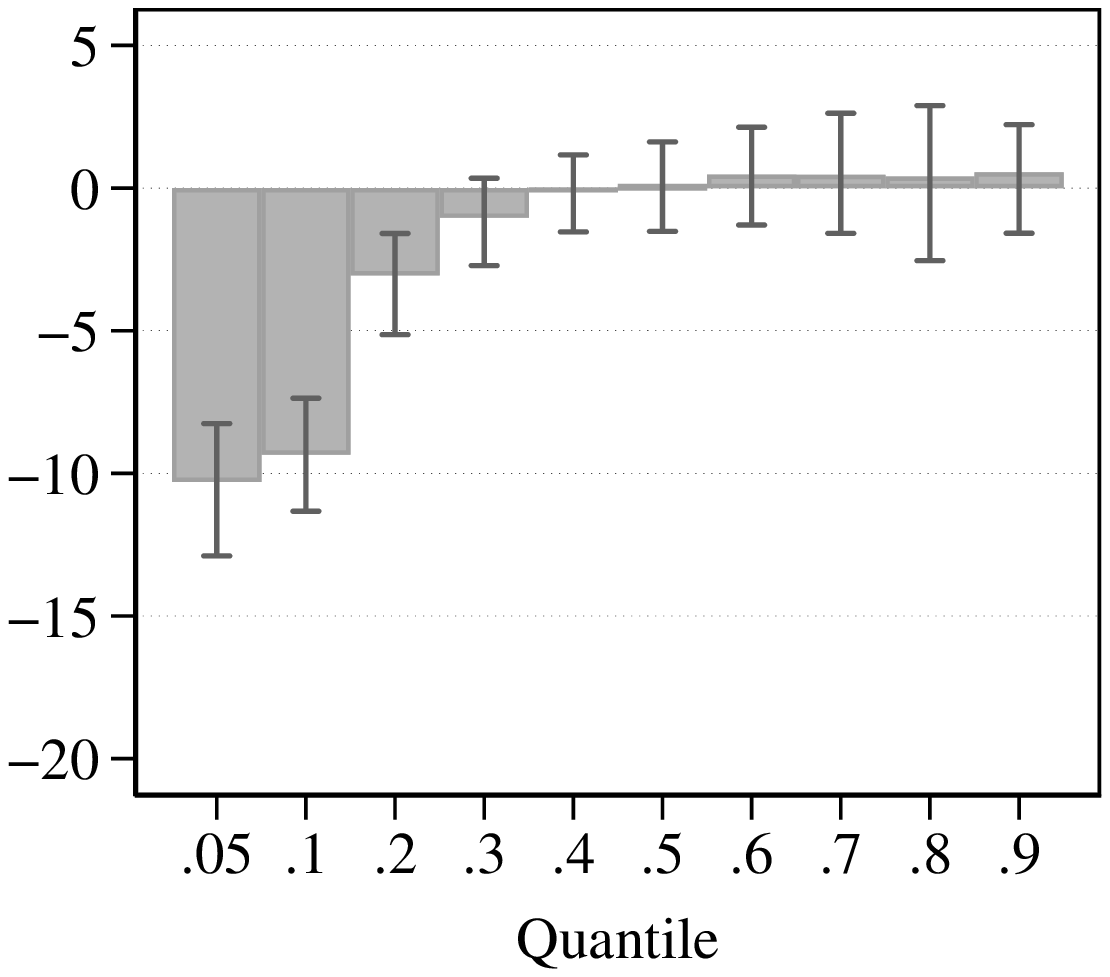}}
\par\end{centering}
\begin{centering}
\subfloat[16 years of education, 5 years of experience]{
\centering{}\includegraphics[scale=0.6]{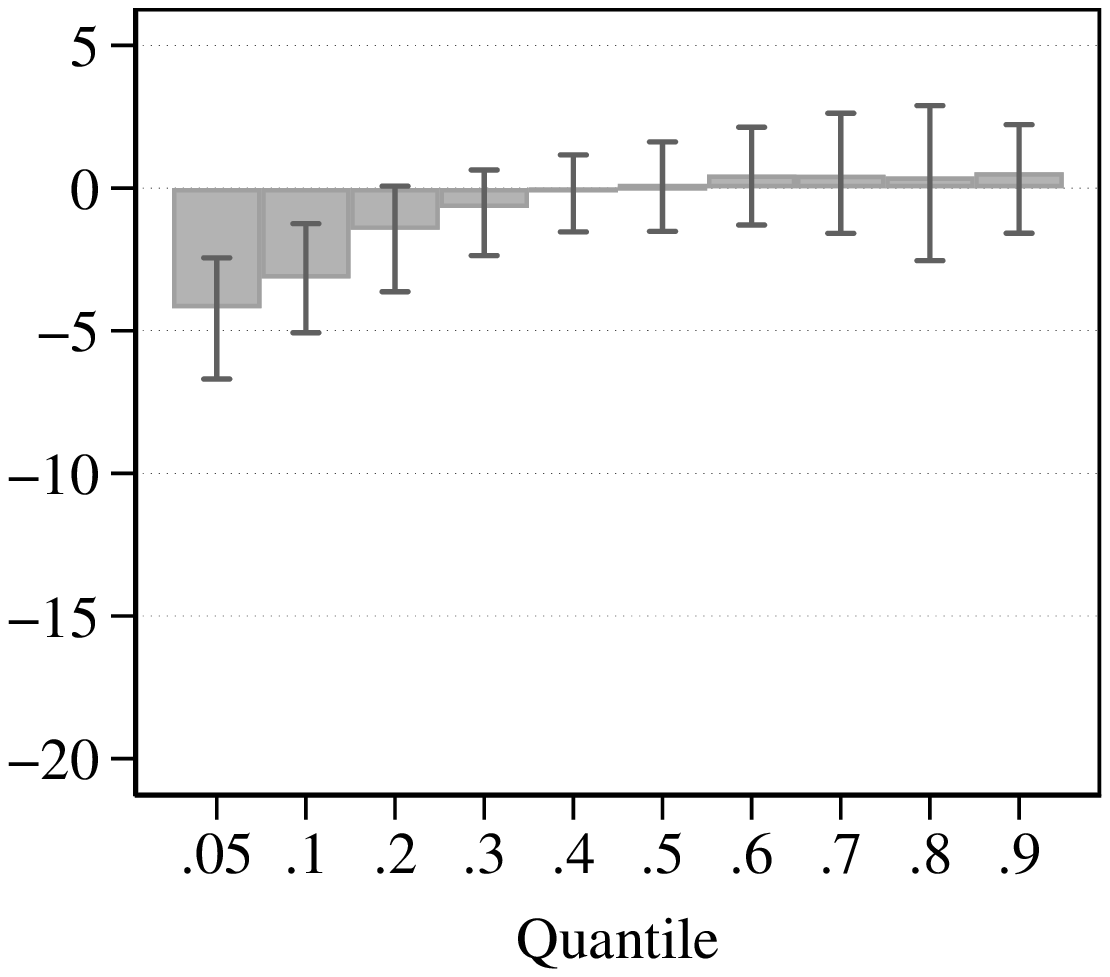}}\subfloat[16 years of education, 10 years of experience]{
\centering{}\includegraphics[scale=0.6]{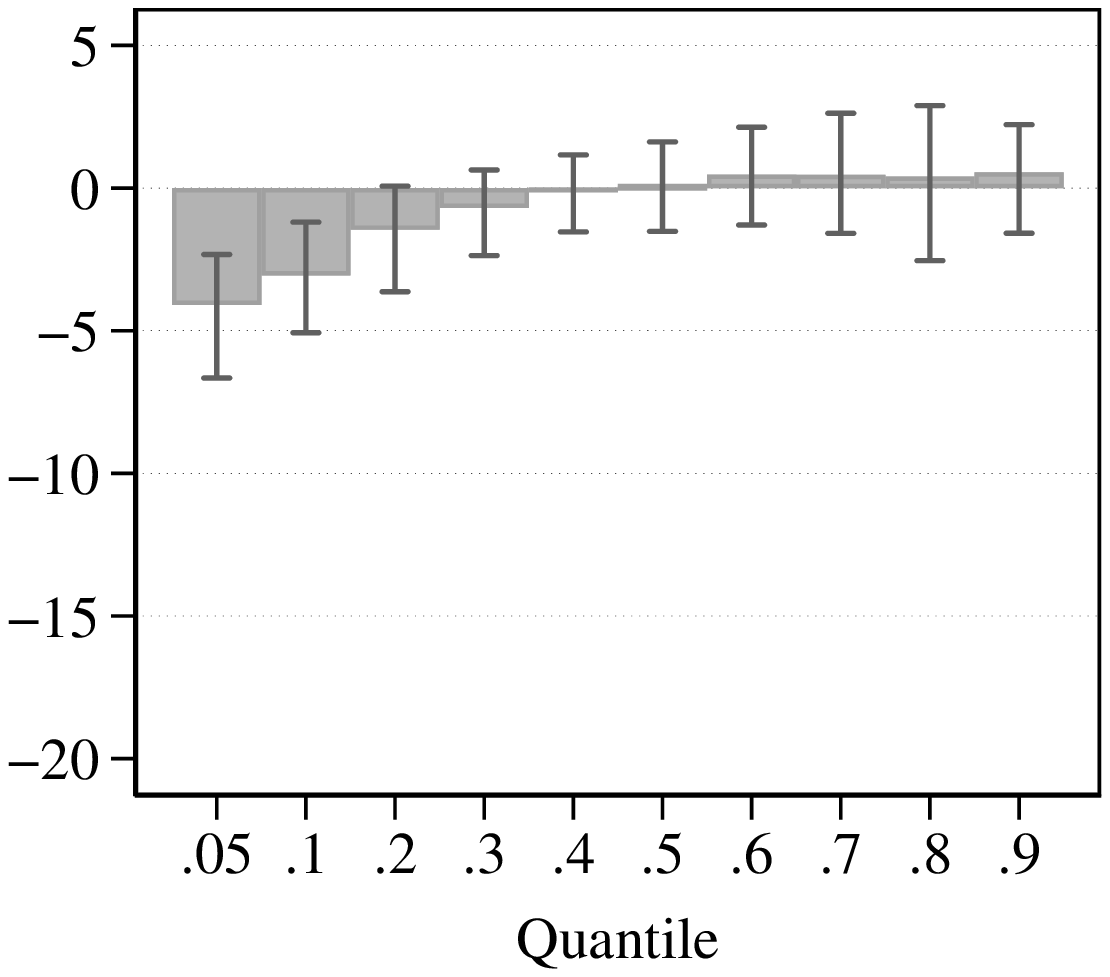}}
\par\end{centering}
\textit{\footnotesize{}Notes}{\footnotesize{}: Estimates of log-point
changes in the gender wage differential due to the minimum wage are
obtained from equation \eqref{eq: between2}. The error bar represents
the 95 percent confidence interval.}{\footnotesize\par}
\end{figure}

We measure the gender wage differential by comparing male workers
and female workers, holding education and experience constant. The
four panels in Figure \ref{fig: diff_gender} show the national means
of changes in the gender wage differential due to increases in the
real value of the minimum wage for the years 1989 to 2012 by education
and experience.

The minimum wage contributes to a reduction in the gender wage differential
in the lower quantiles. The contribution of the minimum wage to a
reduction in the gender wage differential is greater for less-educated,
less-experienced workers than more-educated, more-experienced workers.
For each group of workers, the contribution of the minimum wage is
greatest at the 0.05th quantile and gradually declines in absolute
value to zero by the 0.2th to 0.5th quantiles. For workers with 12
years of education and 5 years of experience, however, it is slightly
greater at the 0.1th quantile than the 0.05th quantile. The reason
is that, at the 0.05th quantile in this group, both male and female
workers are affected by a rise in the real value of the minimum wage.
For workers with 16 years of education, however, the contribution
of the minimum wage is only modest across quantiles.

\begin{figure}[h]
\caption{Actual and counterfactual changes in the gender wage differential
(males versus females), 1989\textendash 2012\label{fig: diff_gender_actual}}

\begin{centering}
\subfloat[12 years of education, 5 years of experience]{
\centering{}\includegraphics[scale=0.6]{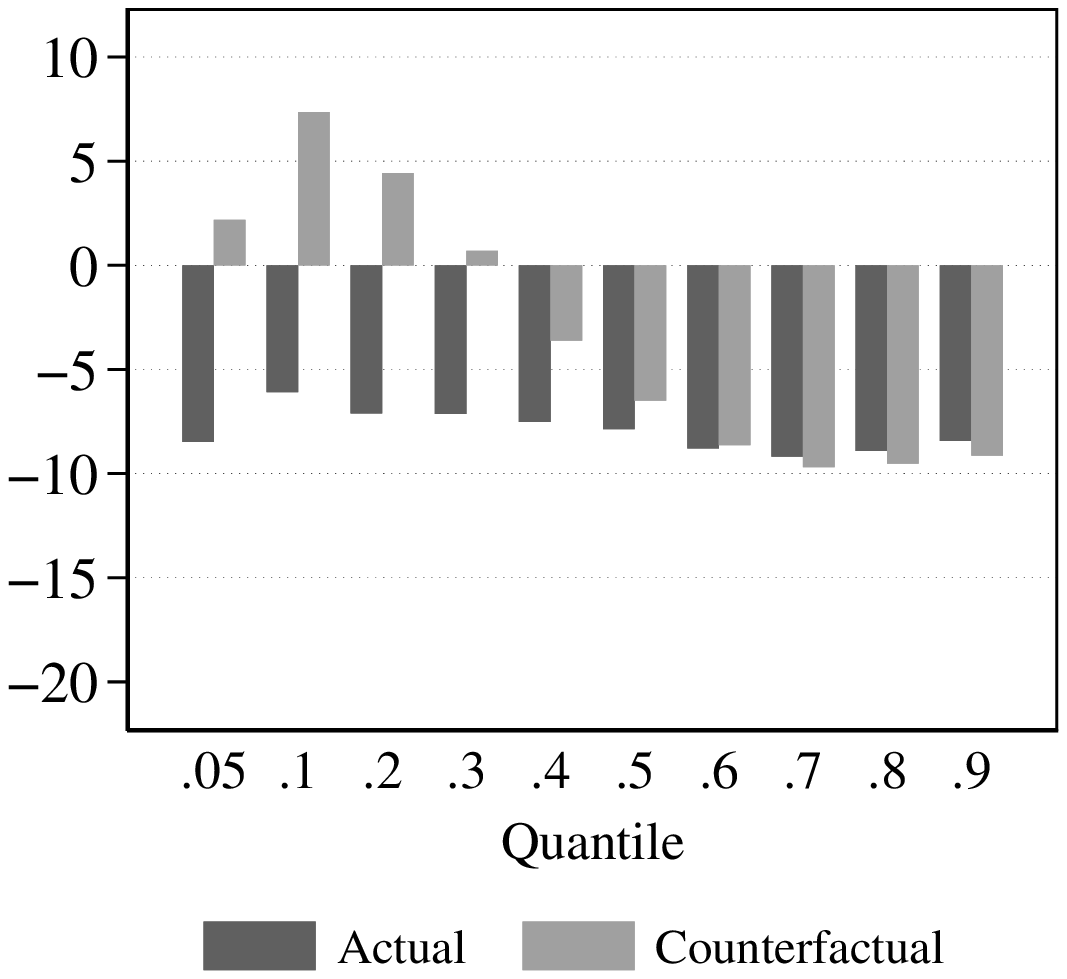}}\subfloat[12 years of education, 10 years of experience]{
\centering{}\includegraphics[scale=0.6]{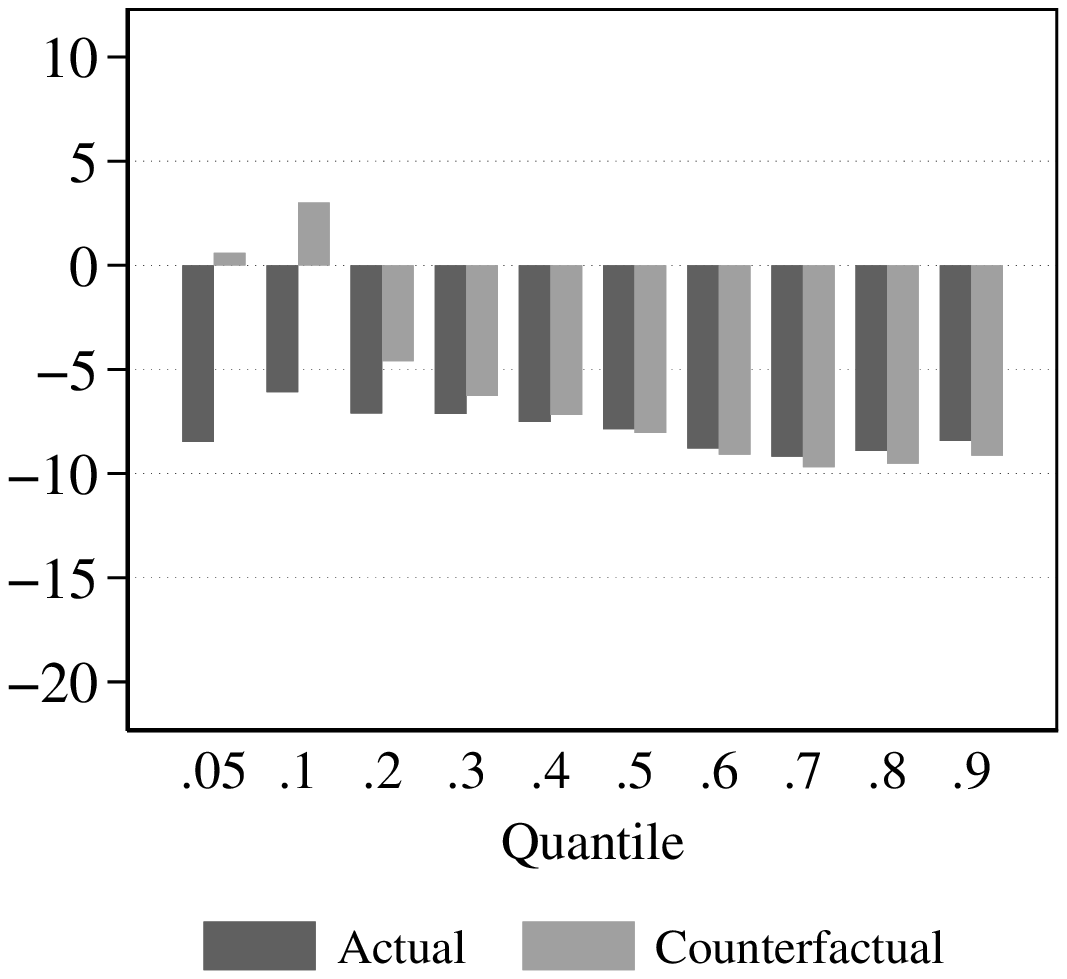}}
\par\end{centering}
\begin{centering}
\subfloat[16 years of education, 5 years of experience]{
\centering{}\includegraphics[scale=0.6]{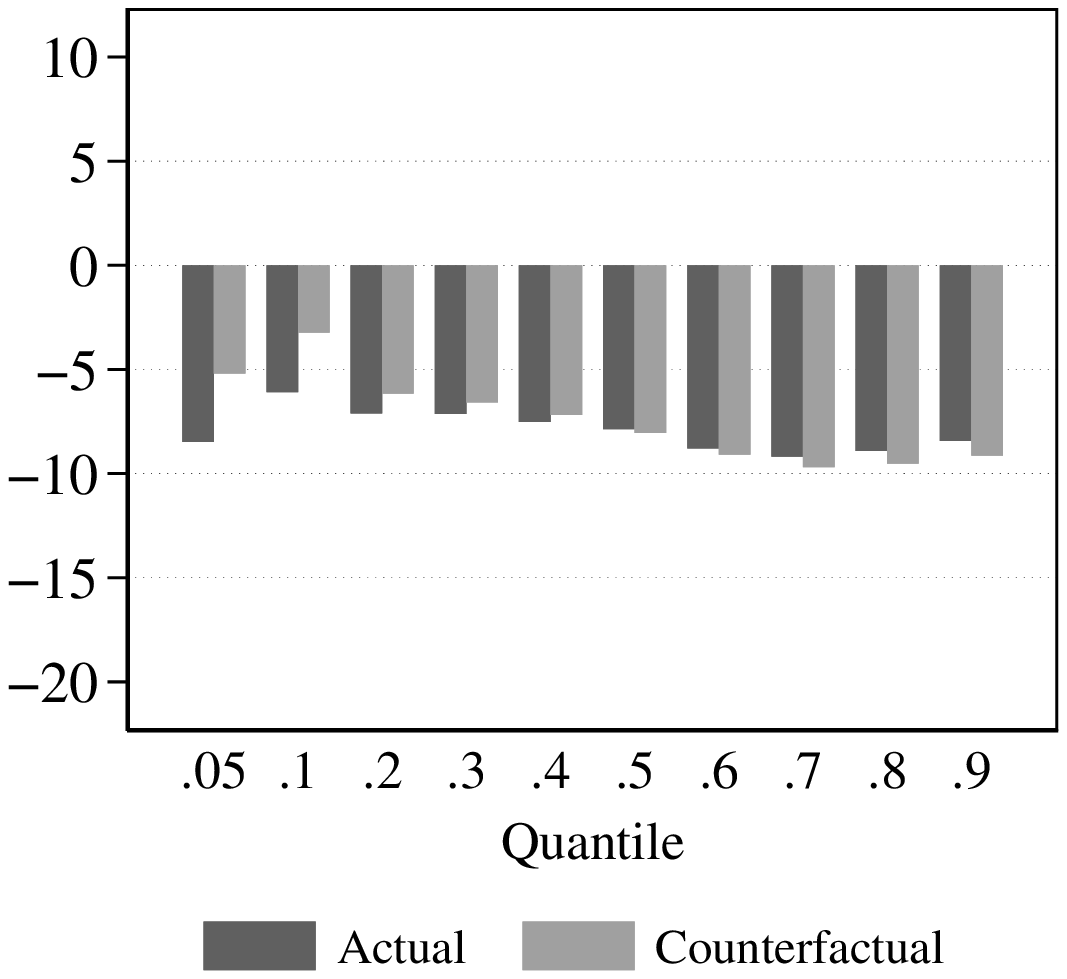}}\subfloat[16 years of education, 10 years of experience]{
\centering{}\includegraphics[scale=0.6]{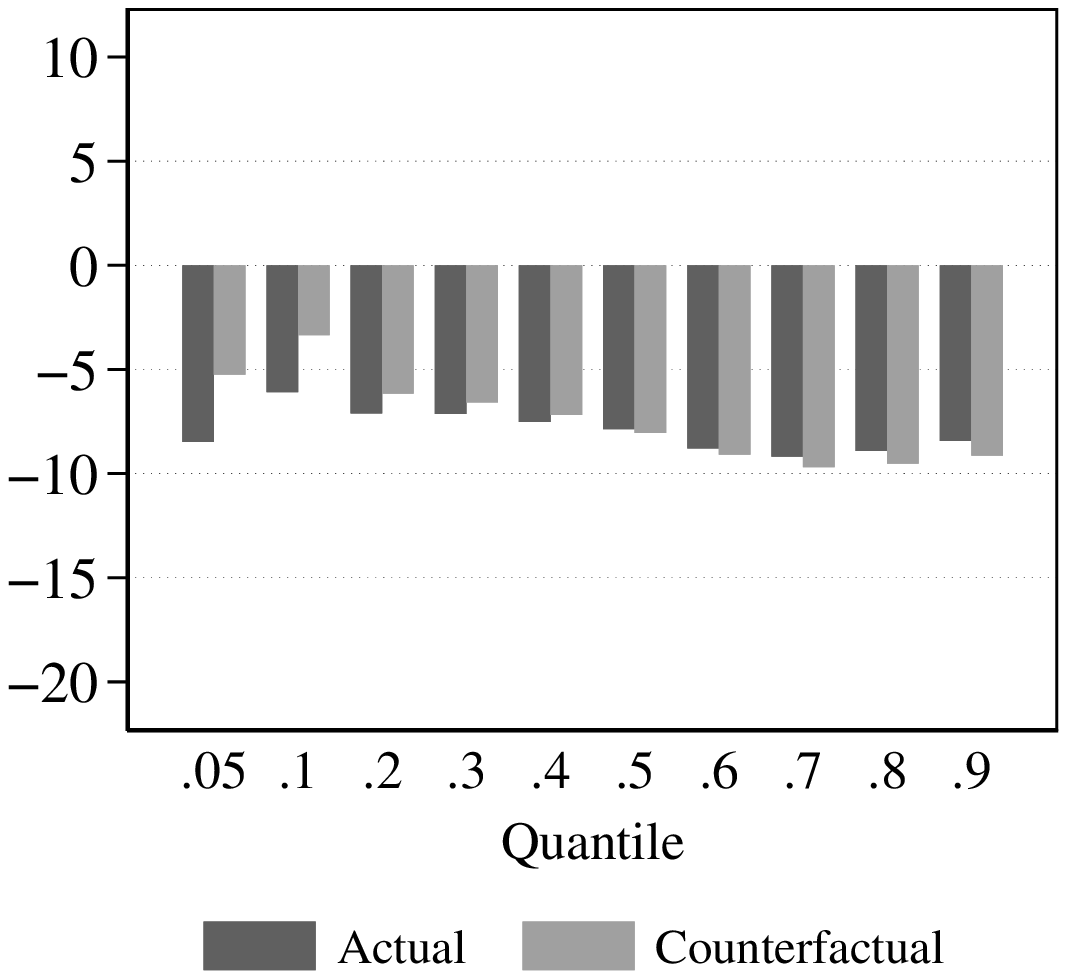}}
\par\end{centering}
\textit{\footnotesize{}Notes}{\footnotesize{}: Actual and counterfactual
log-point changes in the gender wage differential are obtained from
equations \eqref{eq: between1} and \eqref{eq: between2}.}{\footnotesize\par}
\end{figure}

The gender wage differential declined during the period (Figure \ref{fig: diff_gender_actual}).
Changes in the gender wage differential are typically attributed in
the literature to changes in workforce composition and gender discrimination
\citep{Blau_Kahn_JEL17}. Differently from the education and experience
wage differentials, the magnitude of the change in the gender wage
differential is almost uniform across quantiles. If there were no
increase in the real value of the minimum wage, however, the gender
wage differential would decline less in the lower quantiles. For workers
with 12 years of education, the gender wage differential would not
decline but could increase in the lower quantiles. Consequently, in
the counterfactual case in which the real value of the minimum wage
is kept constant, the decline in the gender wage differential is less
in the lower quantiles than the higher quantiles for all groups. Our
results indicate that the minimum wage is another factor in accounting
for the patterns of changes in the gender wage differential.

\subsubsection{Within-group inequality}

\begin{figure}[H]
\caption{Changes in the 90/10, 50/10, and 50/20 within-group differentials
due to the minimum wage, 1989\textendash 2012\label{fig: diff_within}}

\begin{centering}
\subfloat[90/10, males]{
\centering{}\includegraphics[scale=0.6]{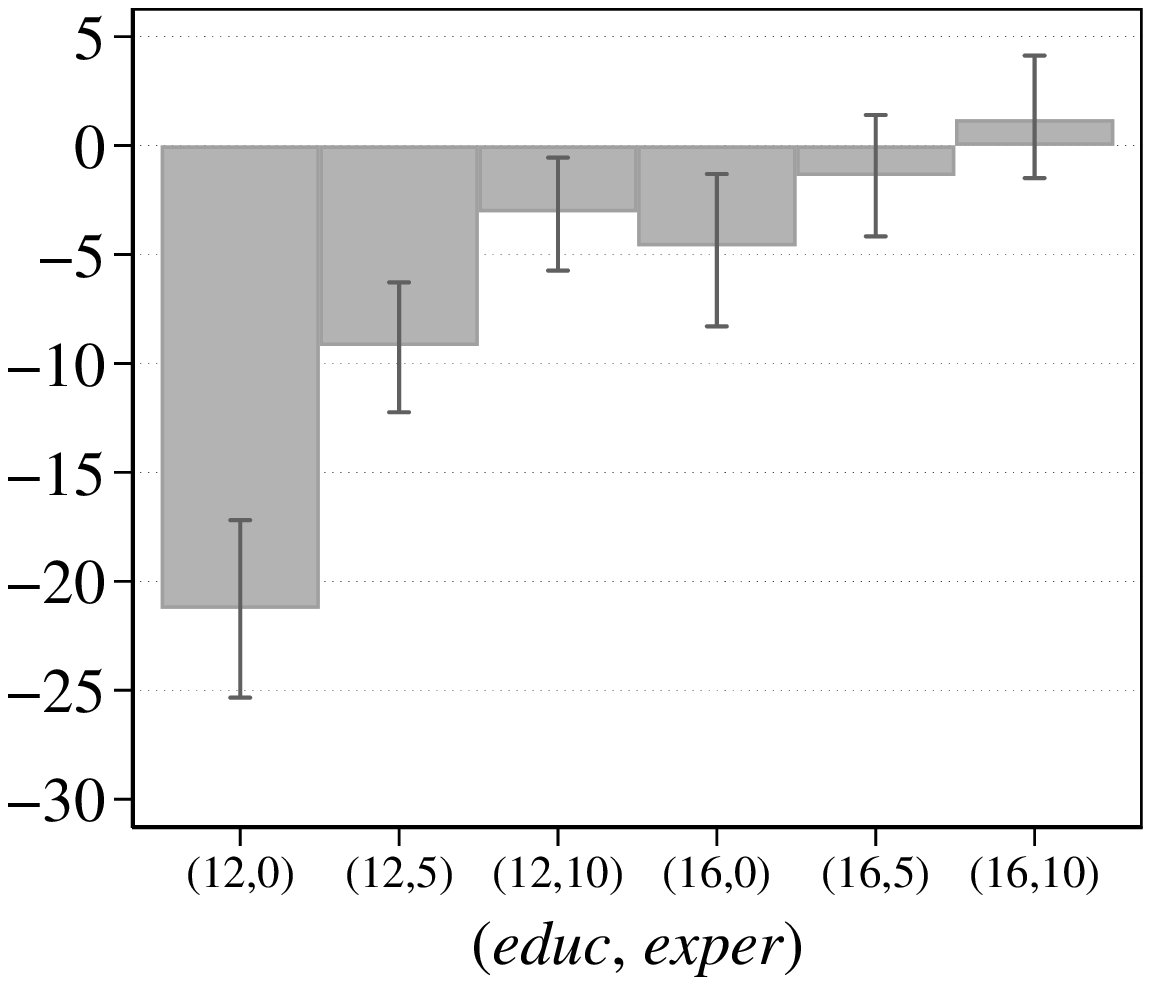}}\subfloat[90/10, females]{
\centering{}\includegraphics[scale=0.6]{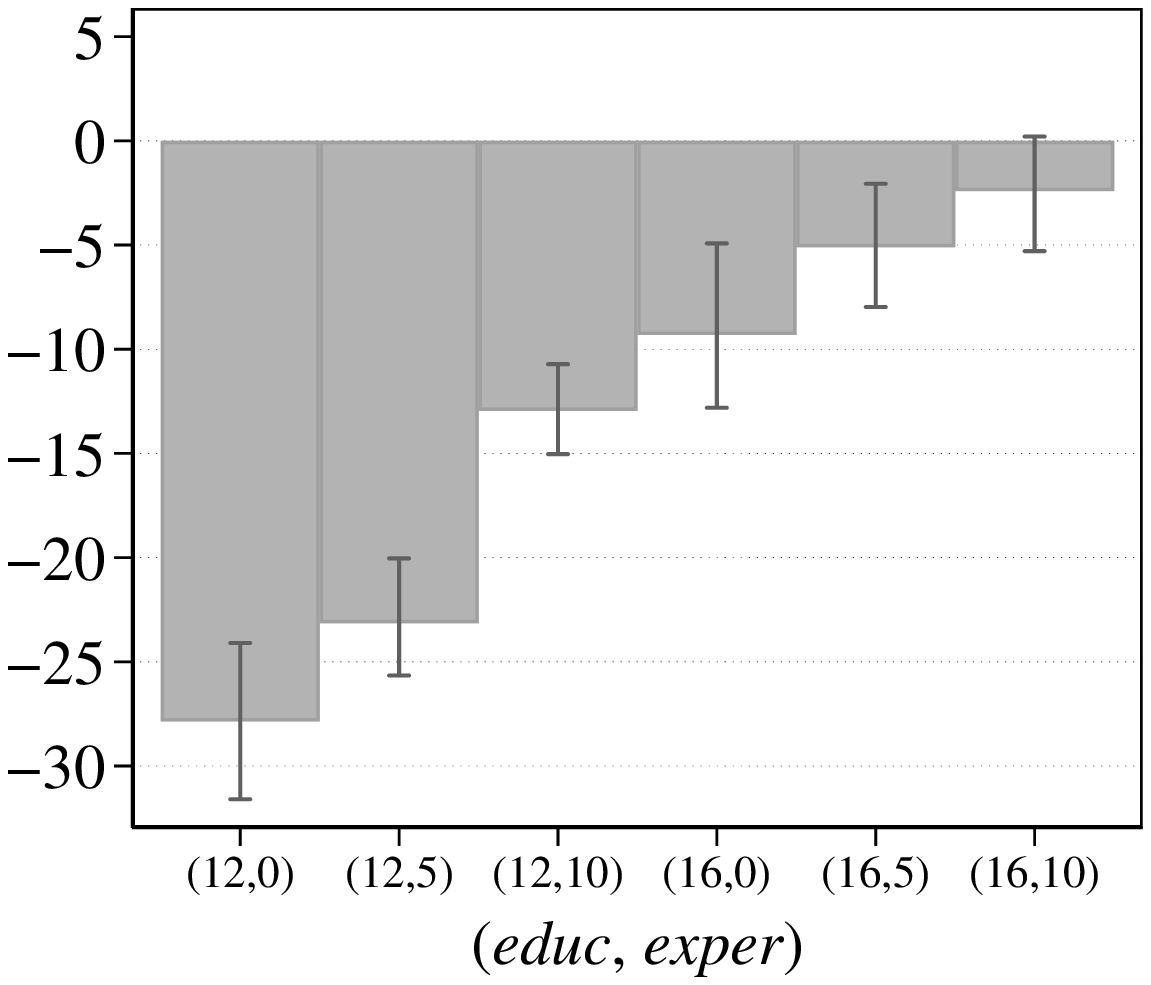}}
\par\end{centering}
\begin{centering}
\subfloat[50/10, males]{
\centering{}\includegraphics[scale=0.6]{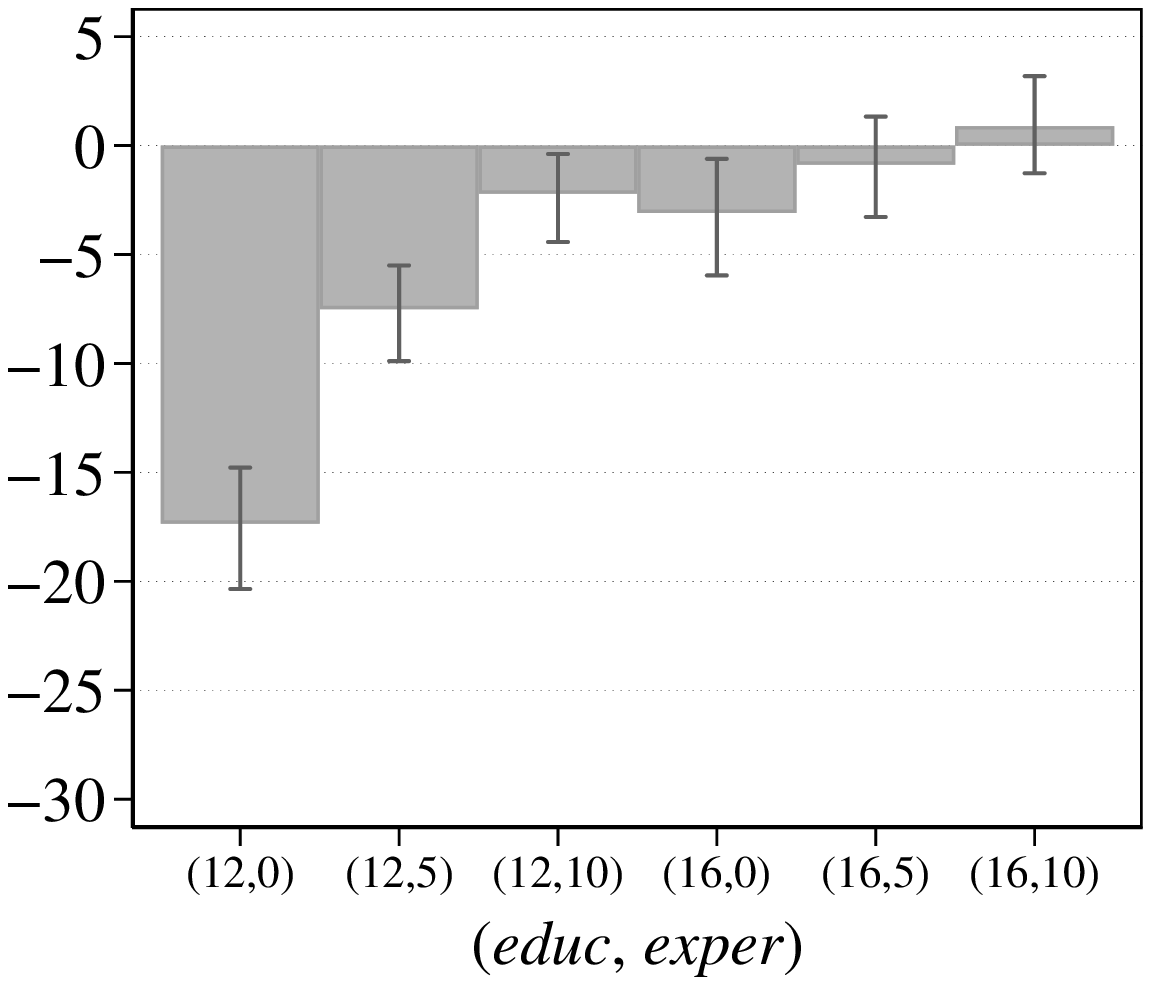}}\subfloat[50/10, females]{
\centering{}\includegraphics[scale=0.6]{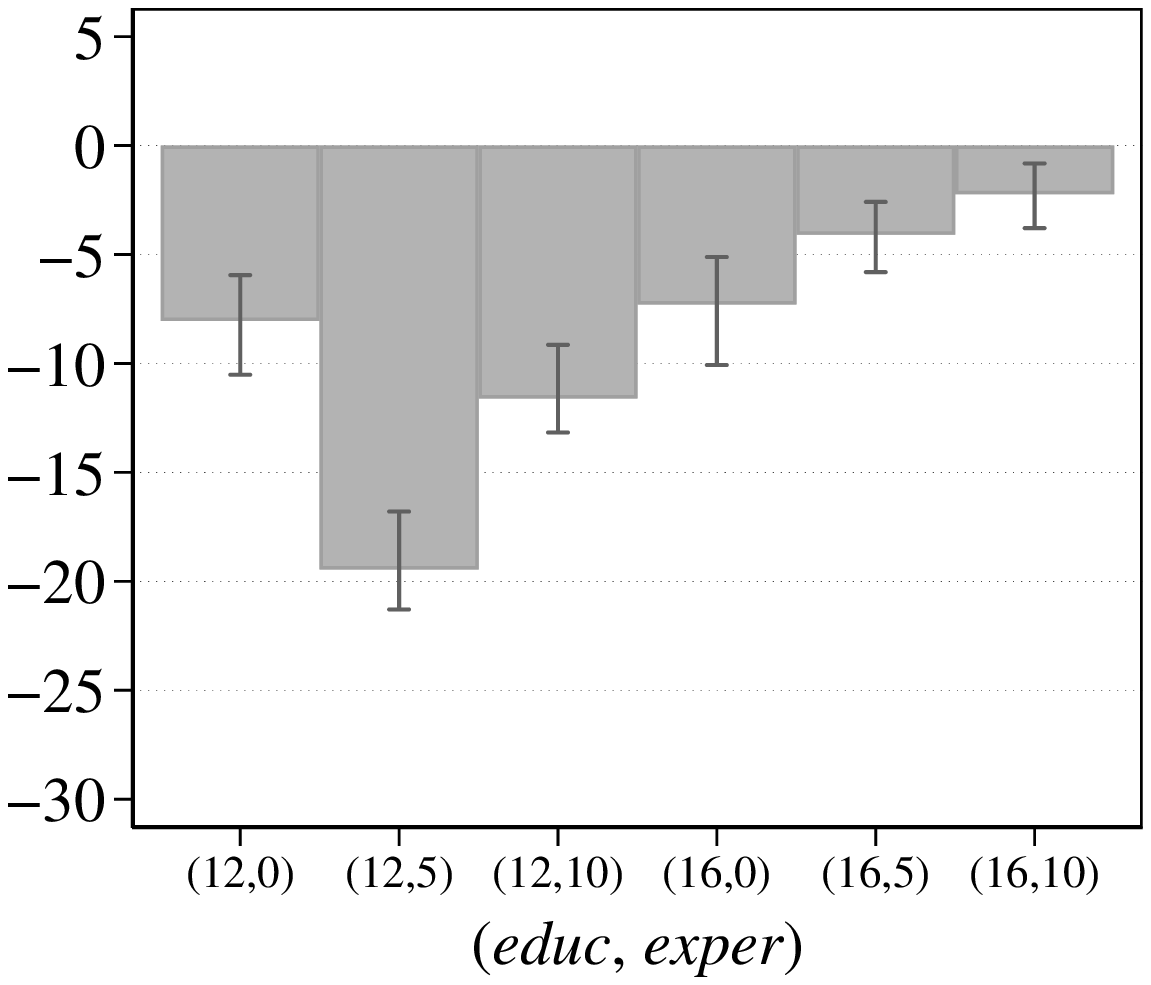}}
\par\end{centering}
\begin{centering}
\subfloat[50/20, males]{
\centering{}\includegraphics[scale=0.6]{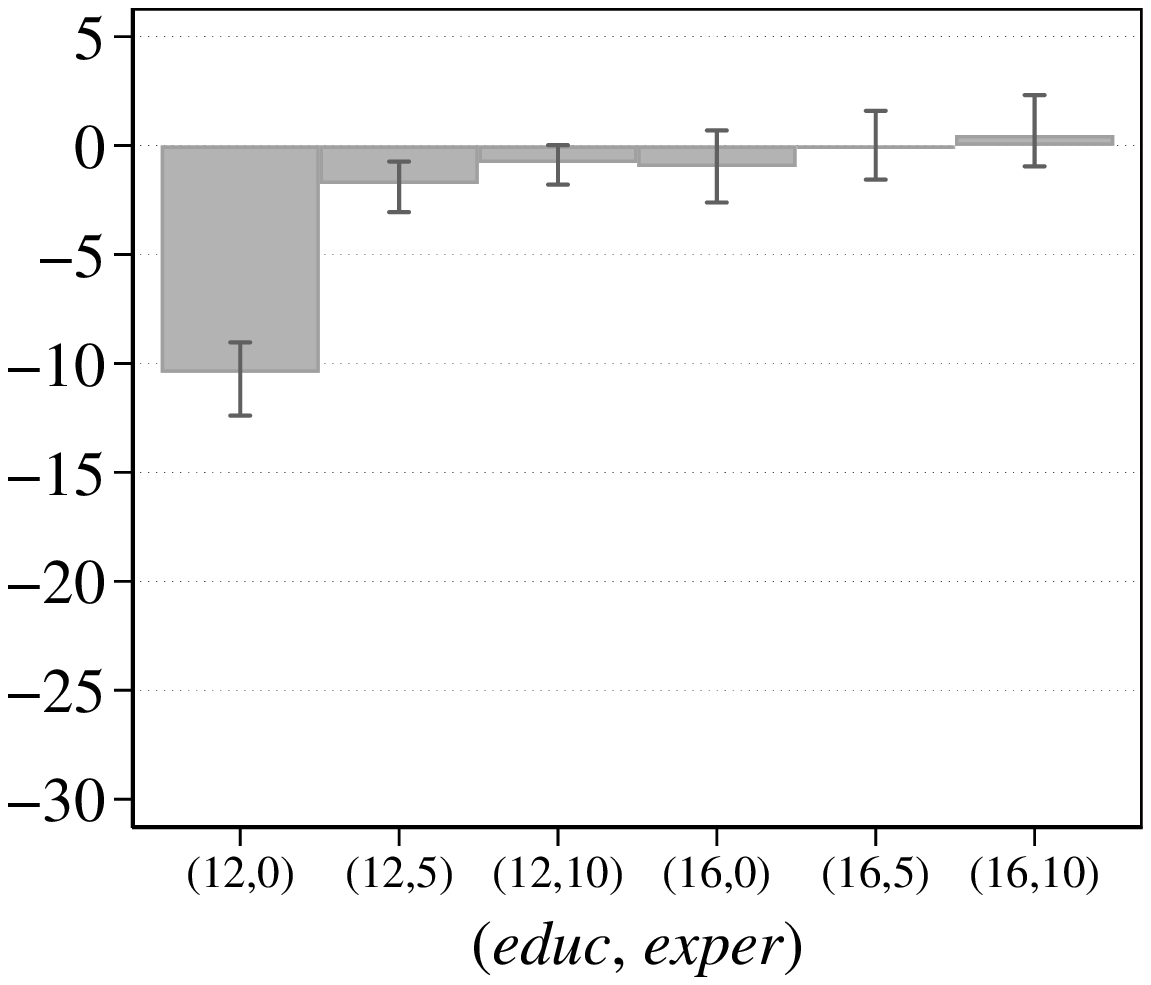}}\subfloat[50/20, females]{
\centering{}\includegraphics[scale=0.6]{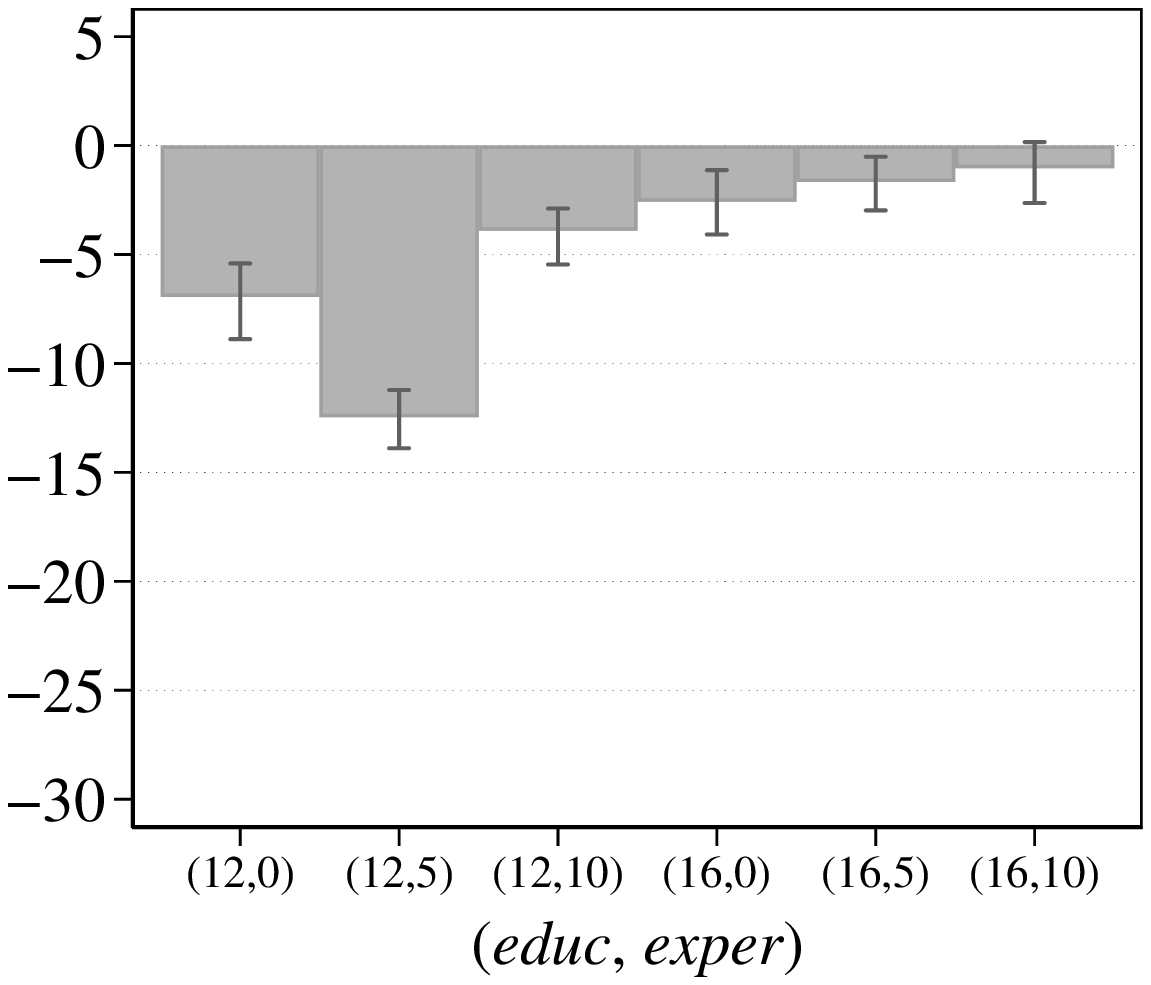}}
\par\end{centering}
\textit{\footnotesize{}Notes}{\footnotesize{}: Estimates of log-point
changes in the within-group wage differentials due to the minimum
wage are obtained from equation \eqref{eq: within2}. The error bar
represents the 95 percent confidence interval.}{\footnotesize\par}
\end{figure}

The four panels in Figure \ref{fig: diff_within} show the national
means of changes in the 90/10 and 50/10 within-group wage differentials
due to increases in the real value of the minimum wage for the years
1989 to 2012 by education, experience, and gender.

\begin{figure}[H]
\caption{Actual and counterfactual changes in the 90/10, 50/10, and 50/20 within-group
differentials, 1989\textendash 2012\label{fig: diff_within_actual}}

\begin{centering}
\subfloat[90/10, males]{
\centering{}\includegraphics[scale=0.6]{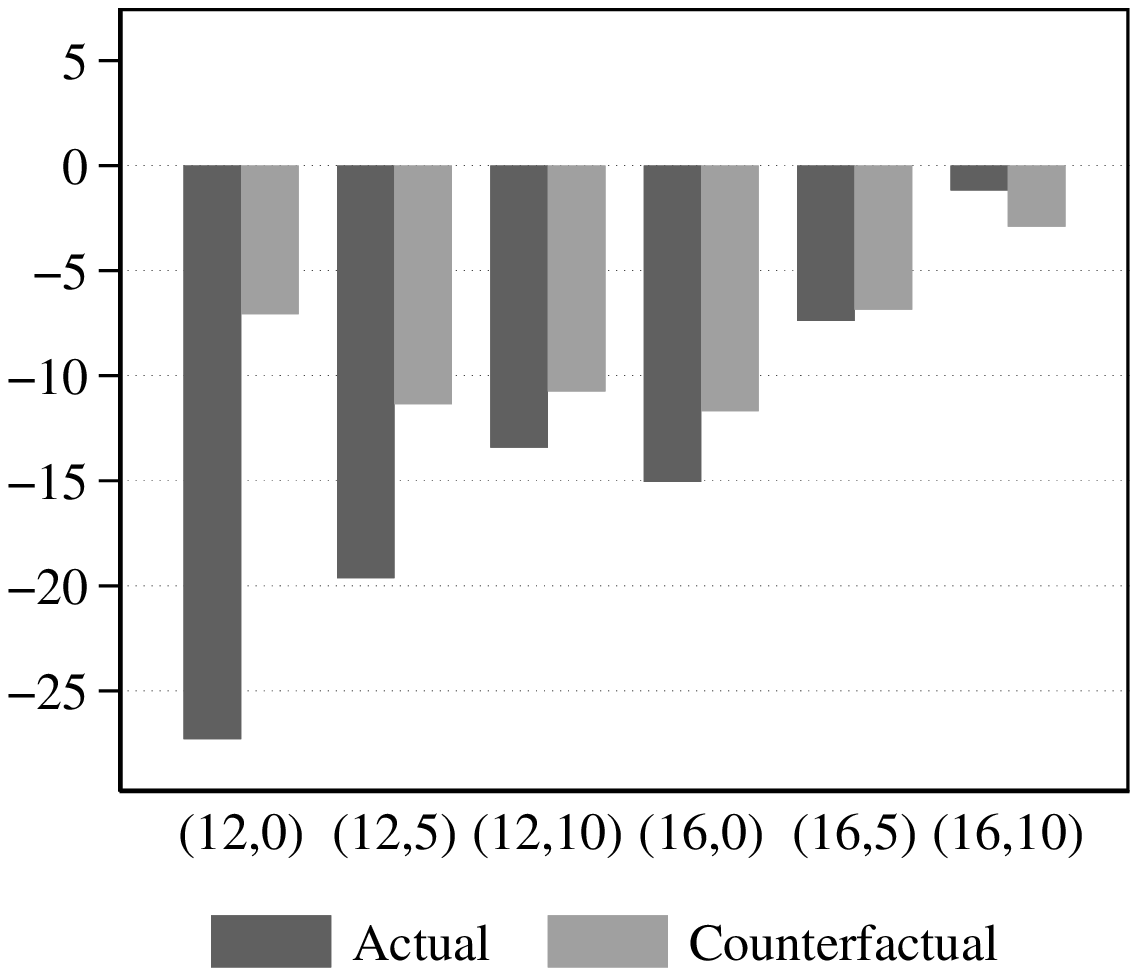}}\subfloat[90/10, females]{
\centering{}\includegraphics[scale=0.6]{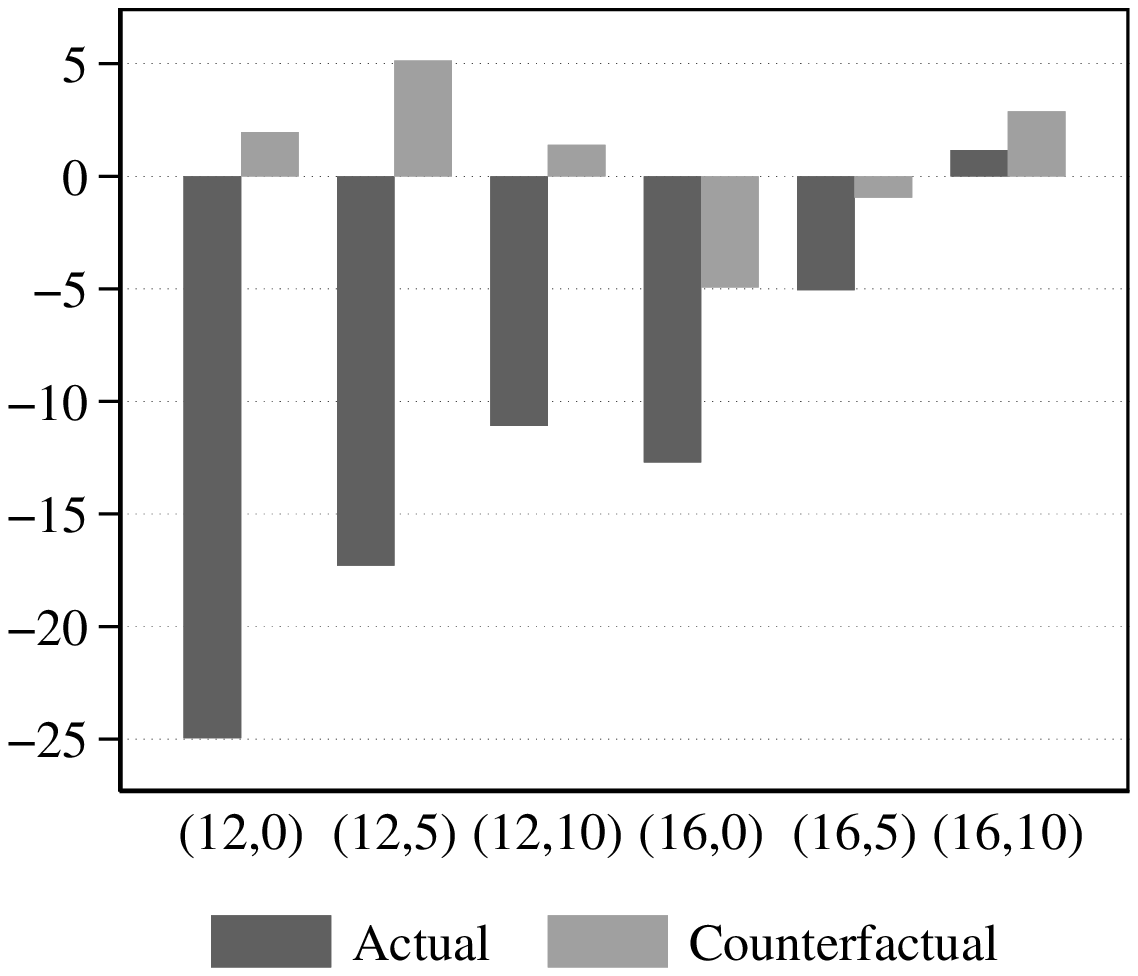}}
\par\end{centering}
\begin{centering}
\subfloat[50/10, males]{
\centering{}\includegraphics[scale=0.6]{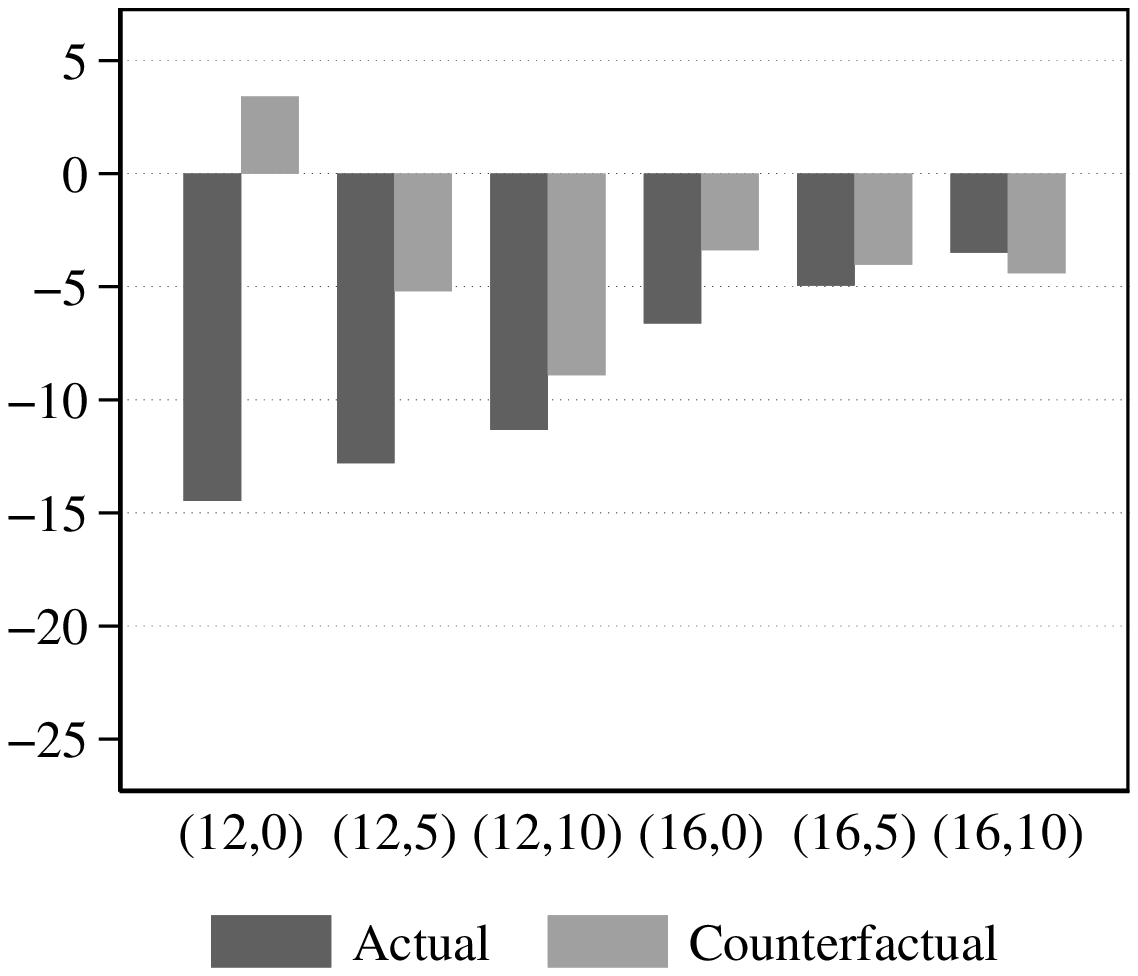}}\subfloat[50/10, females]{
\centering{}\includegraphics[scale=0.6]{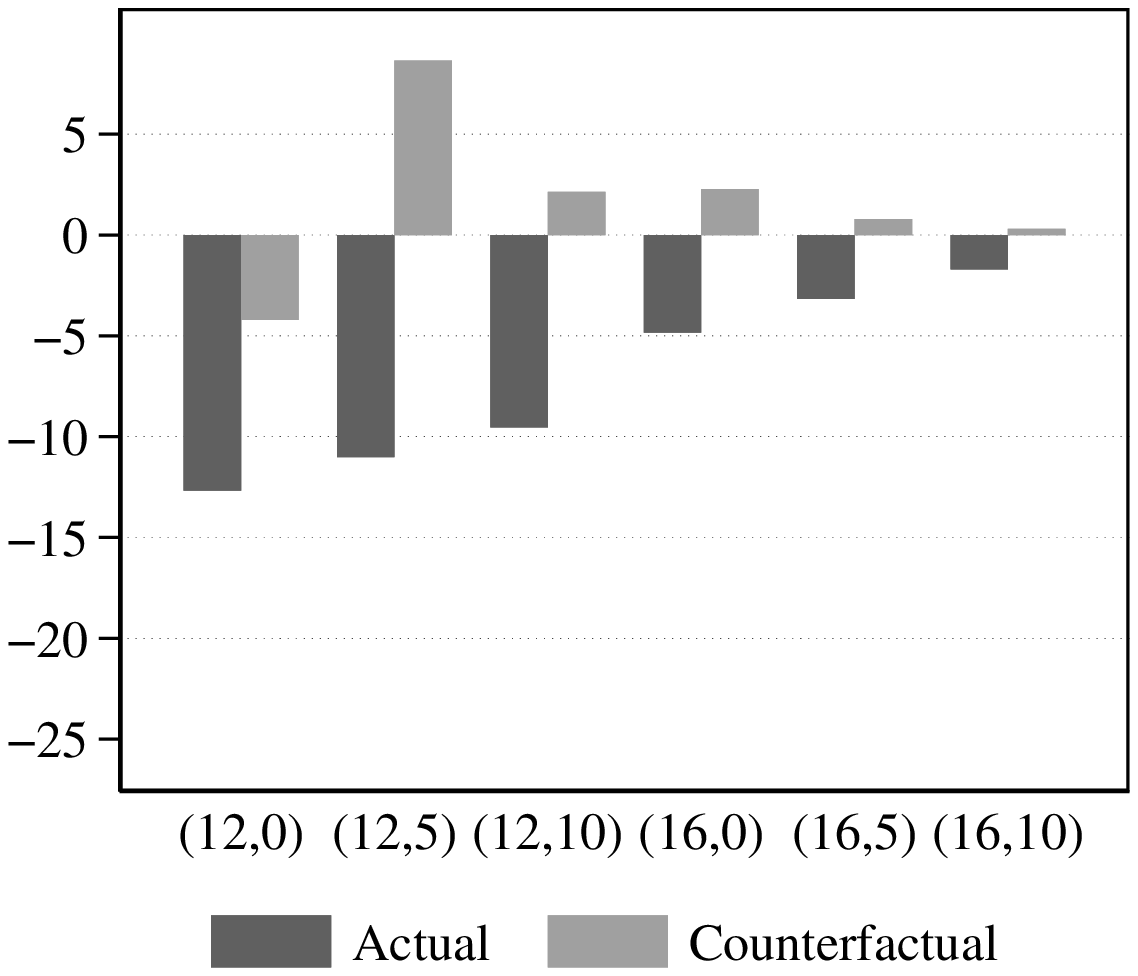}}
\par\end{centering}
\begin{centering}
\subfloat[50/20, males]{
\centering{}\includegraphics[scale=0.6]{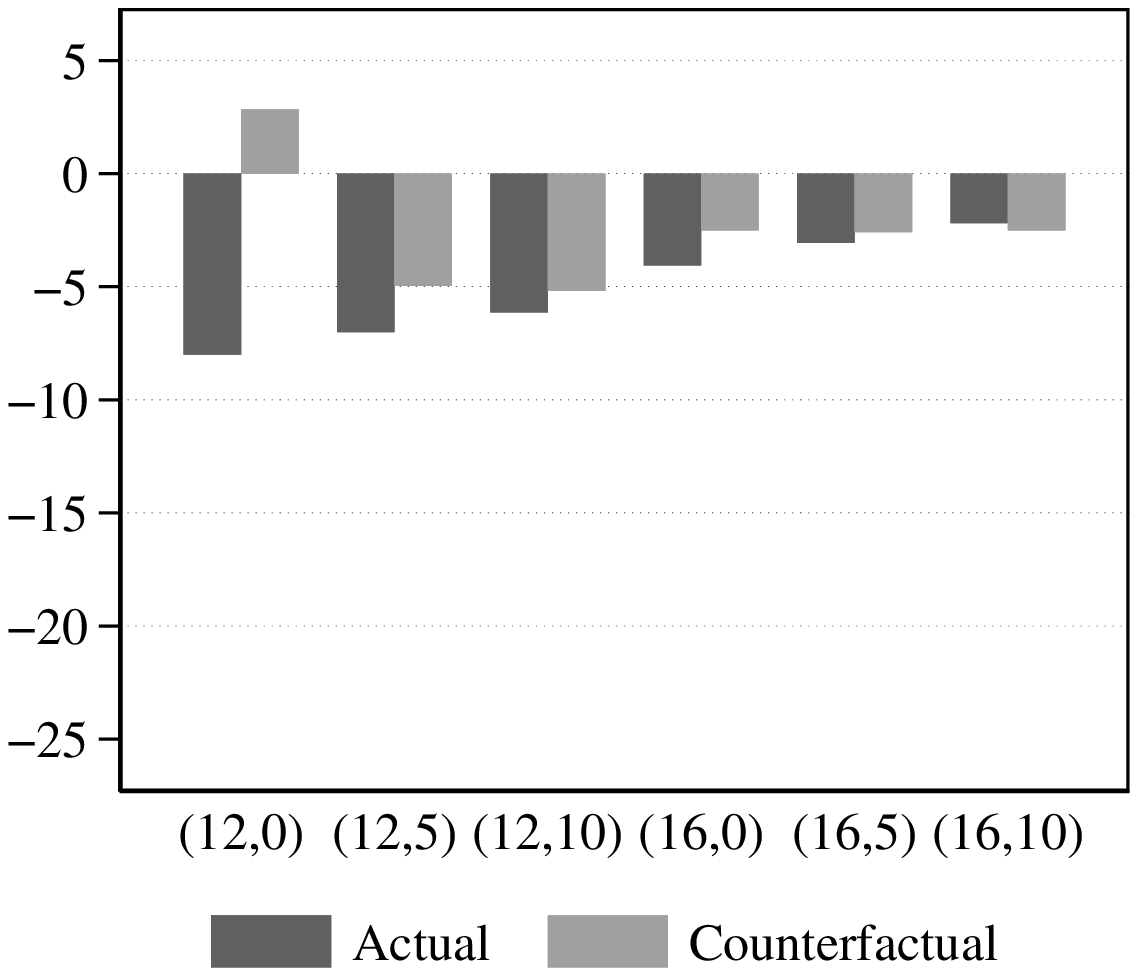}}\subfloat[50/20, females]{
\centering{}\includegraphics[scale=0.6]{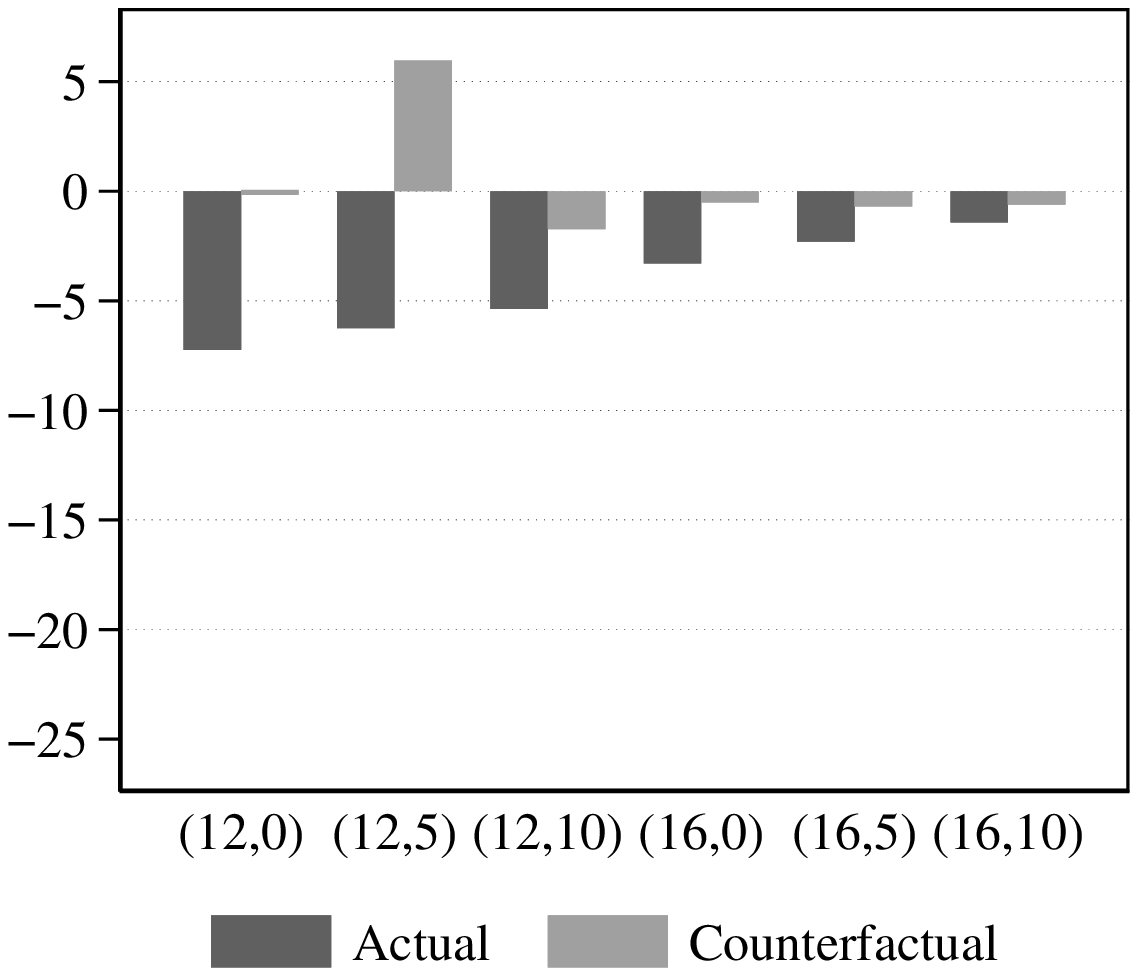}}
\par\end{centering}
\textit{\footnotesize{}Notes}{\footnotesize{}: Actual and counterfactual
log-point changes in the within-group wage differentials are obtained
from equations \eqref{eq: within1} and \eqref{eq: within2}.}{\footnotesize\par}
\end{figure}

The minimum wage contributes to a reduction in the 90/10 and 50/10
within-group wage differentials among workers with lower levels of
education and experience. The contribution of the minimum wage is
the same for changes in the 90/10 and 50/10 within-group wage differentials
except for female workers with 12 years of education and no experience.
The results reflect the fact that changes in the real value of the
minimum wage have no effect at the median or higher quantiles for
almost all groups. The minimum wage also contributes to a reduction
in the 50/20 within-group wage differential, but only moderately for
fewer groups. The contribution of the minimum wage to changes in  within-group
wage differentials is greater for less-educated, less-experienced,
female workers than more-educated, more-experienced, male workers.
For workers with 16 years of education and five or more years of experience,
the contribution of the minimum wage is close to zero.

The 90/10, 50/10, and 50/20 within-group wage differentials declined
during the period (Figure \ref{fig: diff_within_actual}). The 50/10
wage differential declined more than the 50/20 wage differential.
The magnitude of the decline in within-group wage differentials is
similar for male and female workers, but it is greater for less-educated,
less-experienced workers than more-educated, more-experienced workers.
If there were no increase in the minimum wage, however, the 50/10
and 50/20 wage differentials would change almost equally. Furthermore,
within-group wage differentials would decline similarly for less-educated,
less-experienced workers and more-educated, more-experienced workers,
while they would decline less for male workers and would not decline
but could increase for female workers. Our results indicate that the
minimum wage accounts mostly for the patterns of changes in within-group
wage differentials.

\section{Conclusion\label{sec: conclusion}}

We have examined the impact of the minimum wage on the wage structure
and evaluated the contribution of the minimum wage to changes in between-
and within-group inequality in the United States. In doing so, we
have addressed the issues of heterogeneity, censoring, and missing
wages by combining three quantile regression approaches.

We have shown that changes in the real value of the minimum wage over
recent decades have affected the relationship of hourly wages with
education, experience, and gender. In the literature, changes in between-group
wage differentials are typically attributed to skill-biased technological
change, compositional changes in the workforce, and changes related
to gender discrimination. Our results indicate that changes in the
real value of the minimum wage account in part for the patterns of
changes in the education, experience, and gender wage differentials.
If there were no increase in the real value of the minimum wage in
the 1990s and 2000s, the education and experience wage differentials
would increase more uniformly across quantiles, while the gender wage
differential would decline less uniformly across quantiles. Therefore,
when we interpret the patterns of changes in between-group wage differentials
through the lens of economic models, there is a need to adjust the
data taking into account the influence of the minimum wage.

We have further shown that the impact of the minimum wage is heterogeneous
across quantiles of workers' productivity not attributable to their
observed characteristics. In the literature, the sources of changes
in within-group wage differentials are less conclusive than those
of changes in between-group wage differentials. Our results indicate
that changes in the real value of the minimum wage account mostly
for the patterns of changes in within-group wage differentials for
workers with 10 or less years of experience. In particular, the decline
in the 50/10 and 50/20 within-group wage differential among female
workers for the years 1989 to 2012 is attributed almost entirely to
a rise in the real value of the minimum wage.

\clearpage{}

\bibliographystyle{ecta}
\bibliography{references}

\clearpage{}

\appendix

\section{Appendix}

\subsection{Conceptual framework\label{subsec: concept}}

We provide a simple conceptual framework to understand the role of
the minimum wage for the determination of the wage structure. Our
model is related to and builds on the model in \citet*{Bound_Johnson_AER92}
and \citet*{Autor_Katz_HLE99}. The key idea of the model is that
the actual wage can be decomposed into the competitive market wage
and the wedge. The wedge, which can be referred to as the rent, is
a deviation of the actual wage from the competitive market wage.

The actual wage, $W_{ist}$, for an individual $i$ in state $s$
and year $t$ can be expressed as the product of the competitive market
wage, $W_{t}^{c}$, in year $t$ and the rent, $R_{ist}$, for an
individual $i$ in state $s$ and year $t$.
\[
W_{ist}=W_{t}^{c}R_{ist}
\]
The log of the actual wage, $w_{ist}$, can be decomposed additively
into the log of the competitive market wage, $w_{t}^{c}$, and the
log of the rent, $r_{ist}$.
\[
w_{ist}=w_{t}^{c}+r_{ist}
\]

In general, the rent is determined by state-specific institutional
and non-competitive factors, $m_{st}$ and $\boldsymbol{x}_{st}$,
and individual-specific productivity factors, $\boldsymbol{z}_{ist}$.
Here, we consider the minimum wage to be a key institutional factor
and allow for its interactive effect with individual characteristics.
\[
R_{ist}=f\left(m_{st},x_{st},z_{ist}\right)=\exp\left[\left(z_{ist}^{\prime}\beta\right)m_{st}+\left(z_{ist}^{\prime}\otimes x_{st}^{\prime}\right)\gamma\right]
\]
Given this functional form, the log wage equation can be derived as:
\[
w_{ist}=z_{ist}^{\prime}\left[m_{st}\beta+\left(I_{J+1}\otimes x_{st}^{\prime}\right)\gamma\right],
\]
where $I$ is an identity matrix, and $w_{t}^{c}$ is subsumed into
$x_{st}$. The equation can be extended to allow for random coefficients.
\[
w_{ist}=z_{ist}^{\prime}\left[m_{st}\beta\left(u\right)+\left(I_{J+1}\otimes x_{st}^{\prime}\right)\gamma\left(u\right)\right],
\]
where $u$ represents unobserved individual characteristics, distributed
uniformly from zero to one. This random coefficients model is an alternative
representation of equations \eqref{eq: quantile1} and \eqref{eq: mean1}.

\subsection{Estimation and imputation procedures\label{subsec: procedure}}

We describe the procedures for the censored quantile regression estimation
and the quantile imputation. We implement the procedures for each
state $s=1,\ldots,50$, each year $t=1979,1980,\ldots,2012$, and
each quantile $\tau=0.04,0.05,\ldots,0.97$. In this section, we suppress
the subscripts $s$ and $t$ for notational simplicity.

\subsubsection{Censored quantile regression\label{subsec: censored_quantile}}

The estimation proceeds in three steps \citep*{Chernozhukov_Hong_JASA02}.
In the first and second steps, we select the sample to be used for
estimation. In the third step, we estimate the quantile regression
model using the selected sample.

\paragraph{Step 1.}

We estimate the probabilities of not being left- and right-censored
for each individual. When we partition the support of $\boldsymbol{z}_{i}$
into $\boldsymbol{\mathcal{Z}}_{1},\ldots,\mathcal{\boldsymbol{\mathcal{Z}}}_{H}$,
we can nonparametrically estimate the probabilities of not being left-
and right-censored from the empirical probabilities: $\widehat{p}^{L}\left(\boldsymbol{z}_{i}\right)\coloneqq\sum_{h=1}^{H}\widehat{p}_{h}^{L}\left\{ \boldsymbol{z}_{i}\in\boldsymbol{\mathcal{Z}}_{h}\right\} $
and $\widehat{p}^{R}\left(\boldsymbol{z}_{i}\right)\coloneqq\sum_{h=1}^{H}\widehat{p}_{h}^{R}\left\{ \boldsymbol{z}_{i}\in\boldsymbol{\mathcal{Z}}_{h}\right\} $,
respectively, where for each $h$
\[
\widehat{p}_{h}^{L}\left(\boldsymbol{z}_{i}\right)\coloneqq\frac{\sum_{i=1}^{N}\mathbbm{1}\left\{ w_{i}>m,\boldsymbol{z}_{i}\in\boldsymbol{\mathcal{Z}}_{h}\right\} }{\sum_{i=1}^{N}\mathbbm{1}\left\{ \boldsymbol{z}_{i}\in\boldsymbol{\mathcal{Z}}_{h}\right\} },\qquad\text{and}\qquad\widehat{p}_{h}^{R}\left(\boldsymbol{z}_{i}\right)\coloneqq\frac{\sum_{i=1}^{N}\mathbbm{1}\left\{ w_{i}>c,\boldsymbol{z}_{i}\in\boldsymbol{\mathcal{Z}}_{h}\right\} }{\sum_{i=1}^{N}\mathbbm{1}\left\{ \boldsymbol{z}_{i}\in\boldsymbol{\mathcal{Z}}_{h}\right\} }.
\]
We partition the support of $\boldsymbol{z}_{i}$ by years of education
(0\textendash 12, 12+), years of experience (0\textendash 9, 10\textendash 19,
20\textendash 29, 30+), and gender. Using the empirical probabilities,
we select the sample:
\[
\mathcal{I}_{1}\coloneqq\left\{ i\in\left\{ 1,\ldots,N\right\} :1-\widehat{p}^{L}\left(\boldsymbol{z}_{i}\right)+\eta^{L}<\tau<\widehat{p}^{R}\left(\boldsymbol{z}_{i}\right)-\eta^{R}\right\} ,
\]
where $\eta^{L}$ and $\eta^{R}$ are small positive constants to
accommodate possible specification and estimation errors. Following
\citet*{Chernozhukov_Hong_JASA02}, we set $\eta^{L}$ and $\eta^{R}$
at the 0.1th quantiles of the empirical probabilities of not being
censored given $1-\widehat{p}^{L}\left(\boldsymbol{z}_{i}\right)<\tau$
and $\tau<\widehat{p}^{R}\left(\boldsymbol{z}_{i}\right)$, respectively.

\paragraph{Step 2.}

We estimate the quantile regression model using the selected sample
$\mathcal{I}_{1}$. Using a set of estimated coefficients $\widetilde{\boldsymbol{\alpha}}\left(\tau\right)$,
we select the sample:
\[
\mathcal{I}_{2}\coloneqq\left\{ i\in\left\{ 1,\ldots,N\right\} :m+\zeta^{L}<\boldsymbol{z}_{i}^{\prime}\widetilde{\boldsymbol{\alpha}}\left(\tau\right)<c-\zeta^{R}\right\} ,
\]
where $\zeta^{L}$ and $\zeta^{R}$ are small positive constants.
Following \citet{Chernozhukov_FernandezVal_Kowalski_JoE15}, we set
$\eta^{L}$ and $\eta^{R}$ at the 0.03th quantiles of the positive
fitted values of $\boldsymbol{z}_{i}^{\prime}\widetilde{\boldsymbol{\alpha}}\left(\tau\right)-m$
and $c_{i}-\boldsymbol{z}_{i}^{\prime}\widetilde{\boldsymbol{\alpha}}\left(\tau\right)$,
respectively.

\paragraph{Step 3.}

We estimate the quantile regression model using the selected sample
$\mathcal{I}_{2}$.

\subsubsection{Quantile imputation\label{subsec: imputation}}

The imputation proceeds in two steps \citep*{Wei_HQR17}.

\paragraph{Step 1.}

We estimate the censored quantile regression model \eqref{eq: quantile2}
using a sample of individuals for whom we can observe wages. We obtain
a set of estimated coefficients $\left\{ \widehat{\boldsymbol{\alpha}}\left(\tau\right):\tau\in\mathcal{T}^{\ast}\right\} $,
where $\mathcal{T}^{\ast}\coloneqq\left\{ 0.04,0.05,\ldots,0.49\right\} $.

\paragraph{Step 2.}

We draw a random variable, $u_{i}^{\ell}$, from a uniform distribution
over $\mathcal{T}^{\ast}$ independently 10 times for individuals
for whom we cannot their wages. For each realization of $u_{i}^{\ell}$,
we predict their wages using the quantile regression model:
\[
\widehat{w}_{i}^{\ell}\coloneqq\boldsymbol{z}_{i}^{\prime}\widehat{\boldsymbol{\alpha}}\bigl(u_{i}^{\ell}\bigr).
\]
If the predicted value is smaller than the minimum wage or greater
than the top-coded value, it is replaced with the minimum wage or
the top-coded value. We impute their wages by taking the mean of predicted
values. We calculate their weights using hours worked imputed by fitting
a fifth-order polynomial regression on wages.

\subsection{Additional Results}

\begin{figure}[H]
\caption{Impact of the minimum wage on the wage structure without imputation\label{fig: estimates0}}

\begin{centering}
\subfloat[Intercept\label{fig: intercept0}]{
\centering{}\includegraphics[scale=0.6]{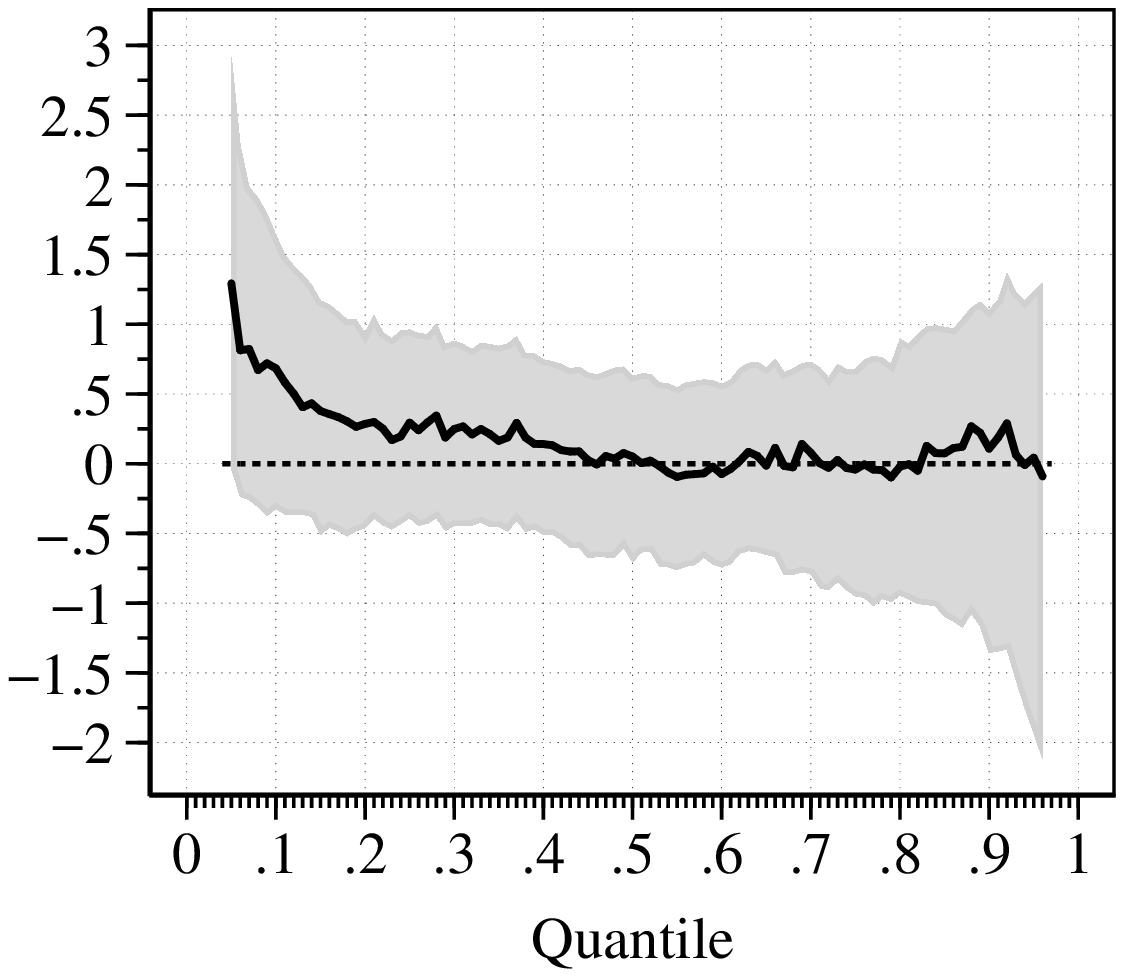}}\subfloat[Education\label{fig: education0}]{
\centering{}\includegraphics[scale=0.6]{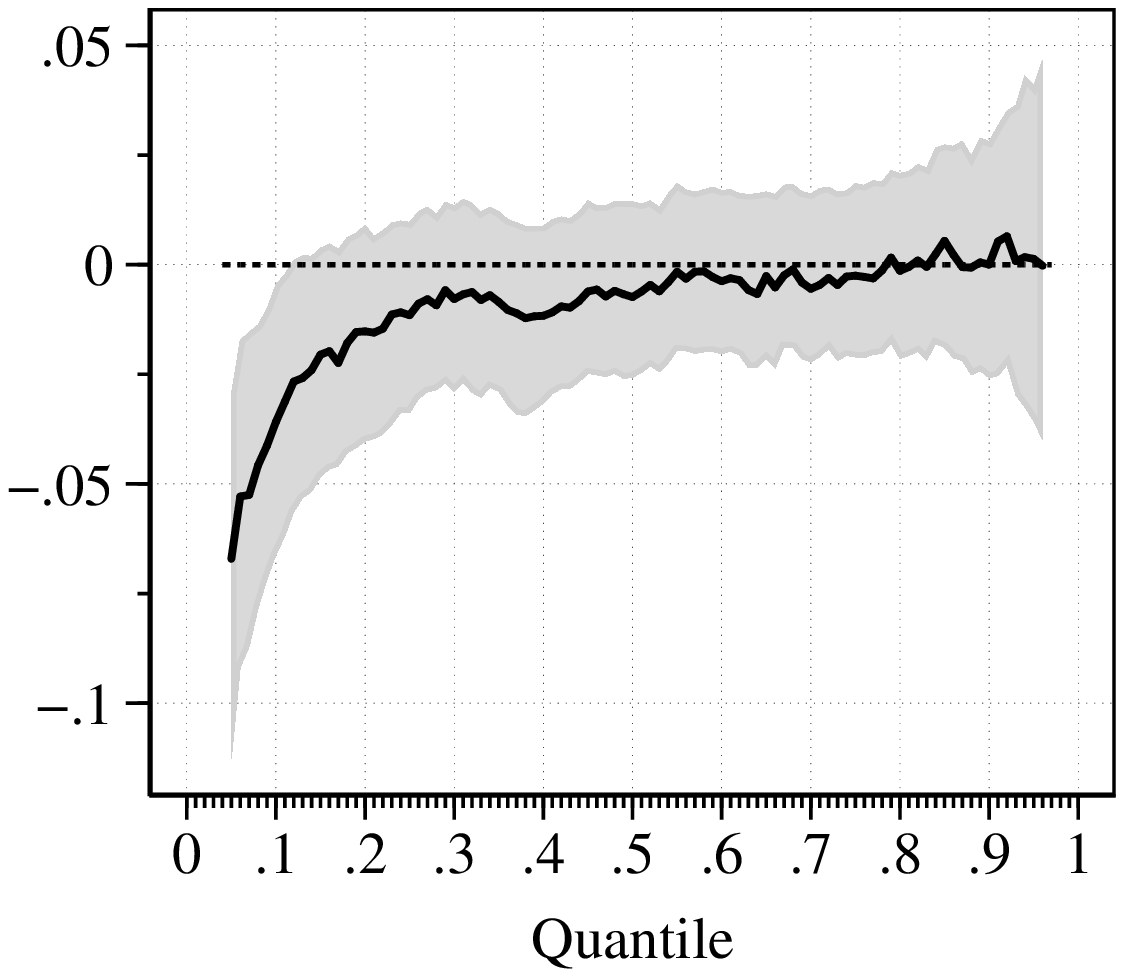}}
\par\end{centering}
\begin{centering}
\subfloat[Experience\label{fig: experience0}]{
\centering{}\includegraphics[scale=0.6]{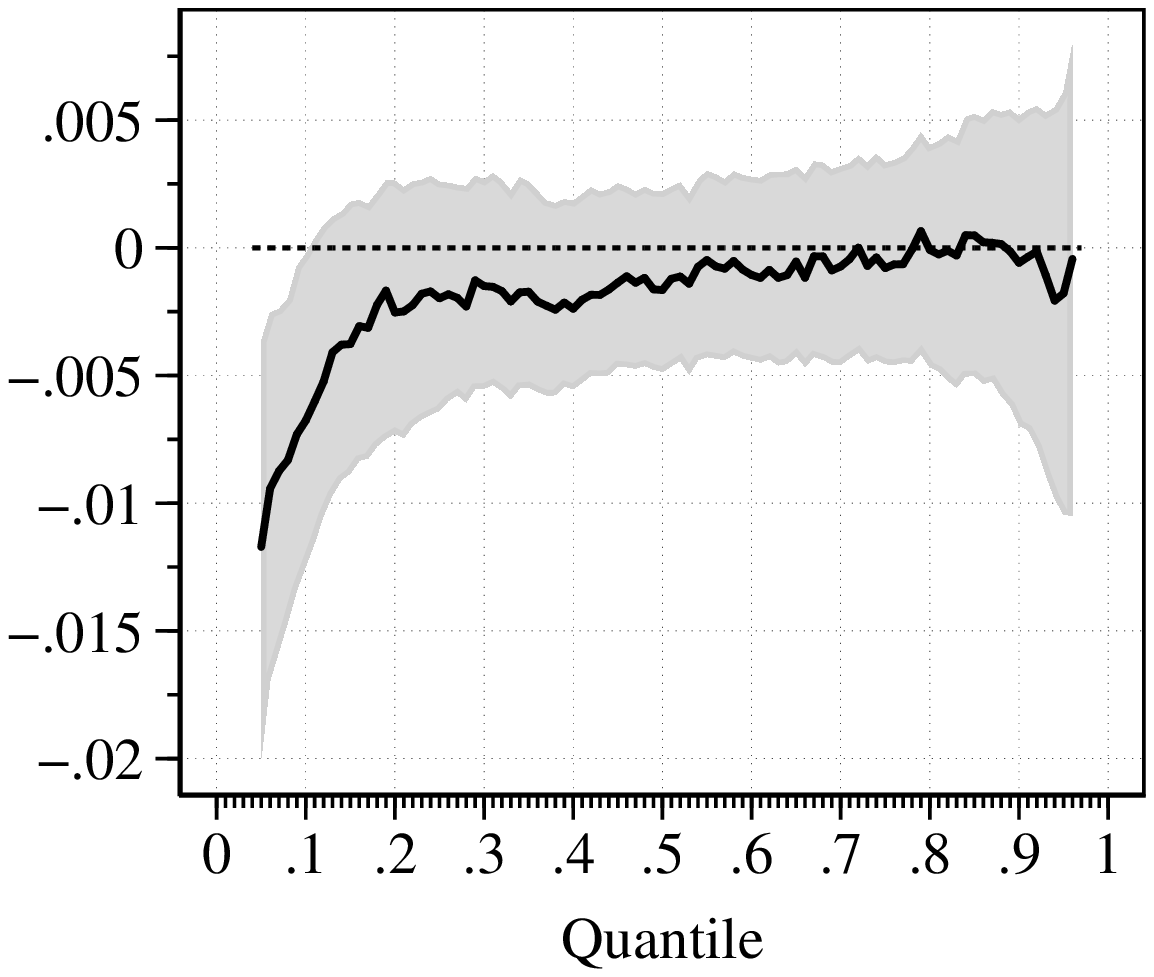}}\subfloat[Gender (male)\label{fig: gender0}]{
\centering{}\includegraphics[scale=0.6]{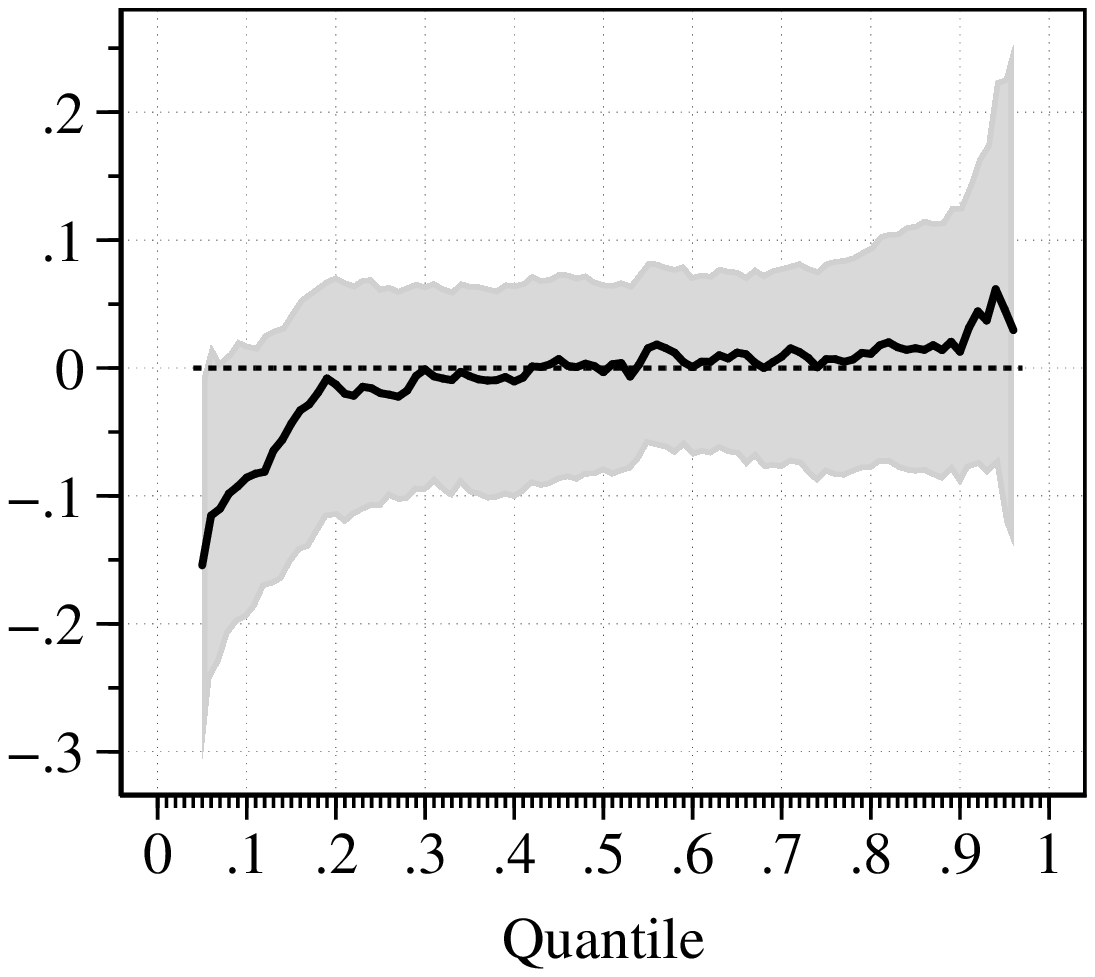}}
\par\end{centering}
\textit{\footnotesize{}Notes}{\footnotesize{}: Estimates of partial
effects in equation \eqref{eq: mean1} are reported. The shaded area
represents the 95 percent confidence interval.}{\footnotesize\par}
\end{figure}

\subsubsection{Impact on the wage structure\label{subsec: impact}}

Figure \ref{fig: estimates0} shows the impact of the minimum wage
on the intercept and slope coefficients in the wage equation across
quantiles, when we do not impute the wages of individuals for whom
we cannot observe wages. Both the intercept and slope coefficients
in the wage equation are affected by the real value of the minimum
wage in the same way as we see in Figure \ref{fig: estimates}.

\begin{figure}[H]
\caption{Impact of the minimum wage on the wage structure: confidence band\label{fig: estimates_band}}

\begin{centering}
\subfloat[Intercept]{
\centering{}\includegraphics[scale=0.6]{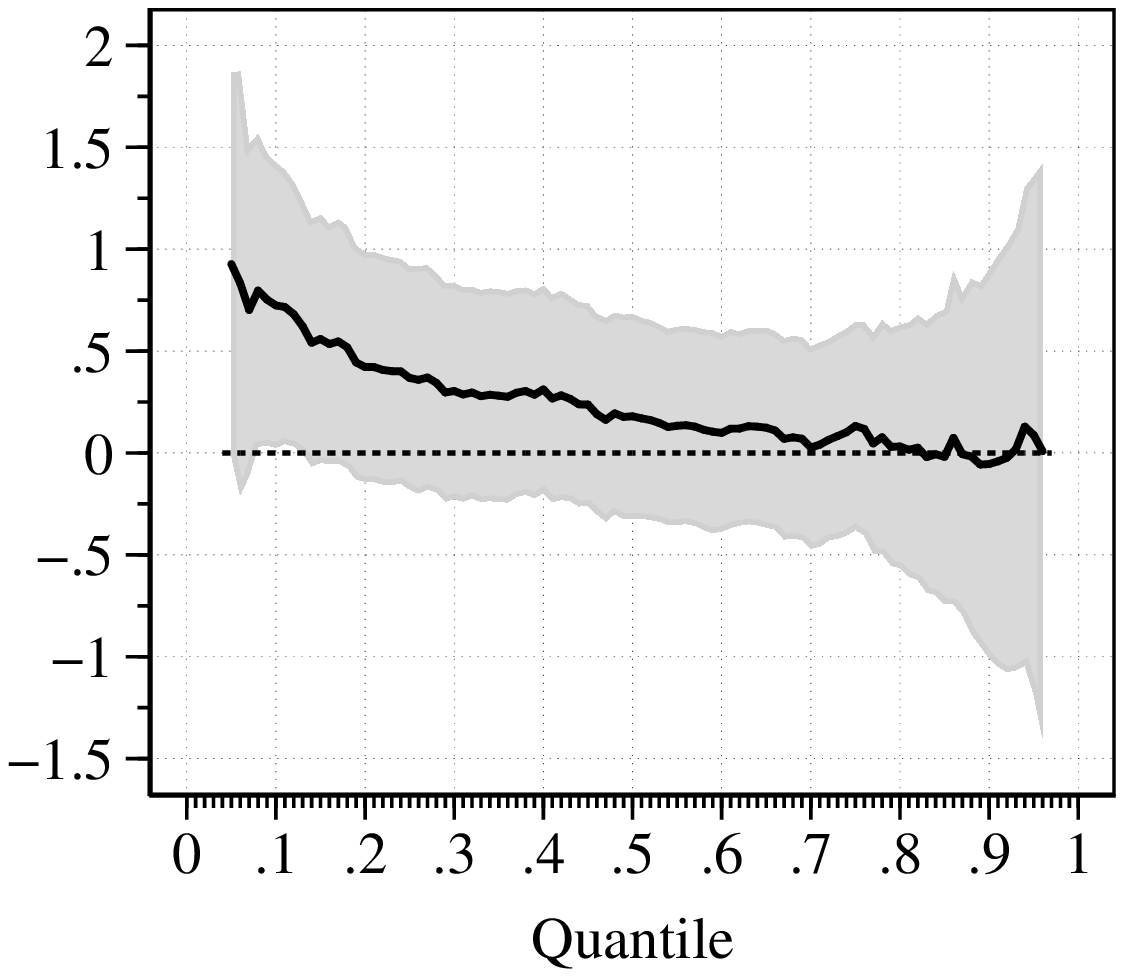}}\subfloat[Education]{
\centering{}\includegraphics[scale=0.6]{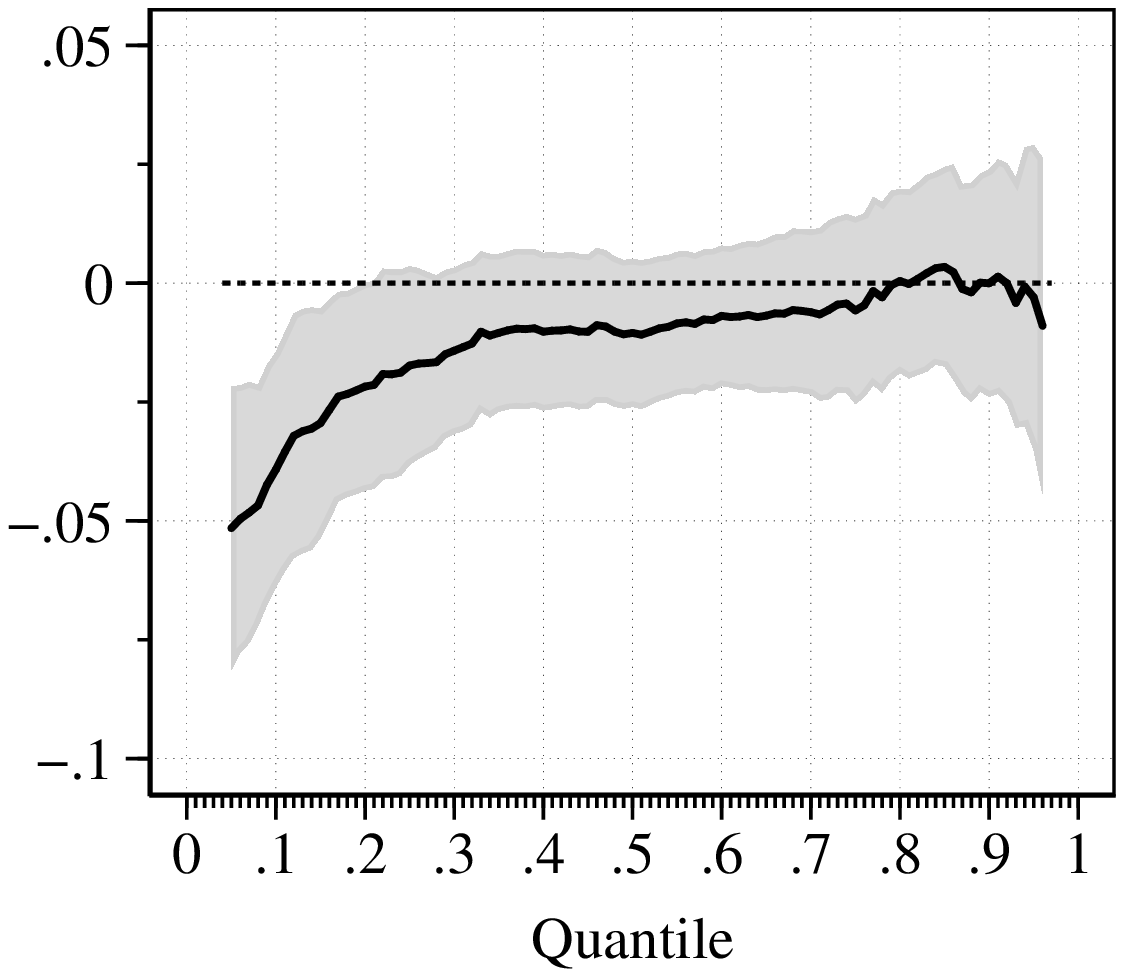}}
\par\end{centering}
\begin{centering}
\subfloat[Experience]{
\centering{}\includegraphics[scale=0.6]{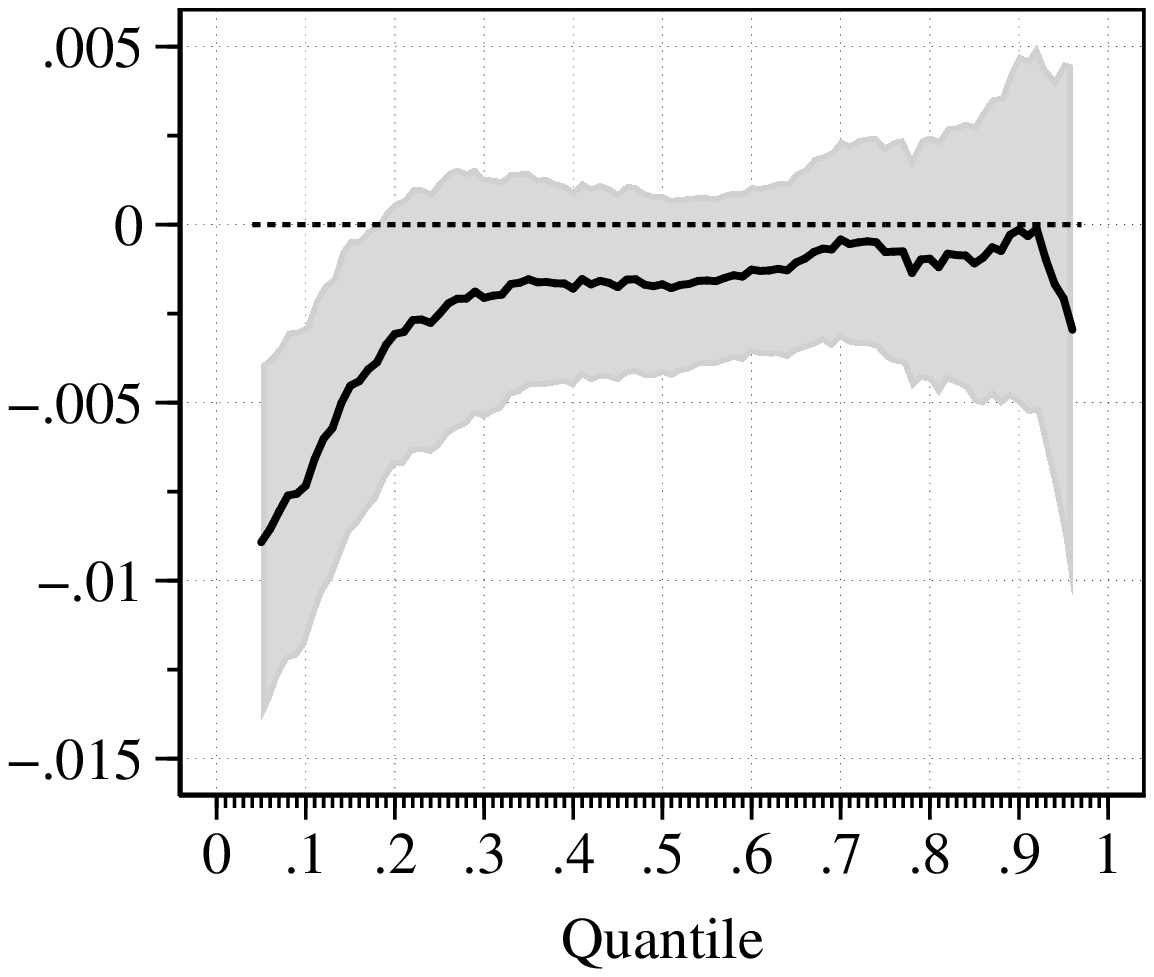}}\subfloat[Gender (male)]{
\centering{}\includegraphics[scale=0.6]{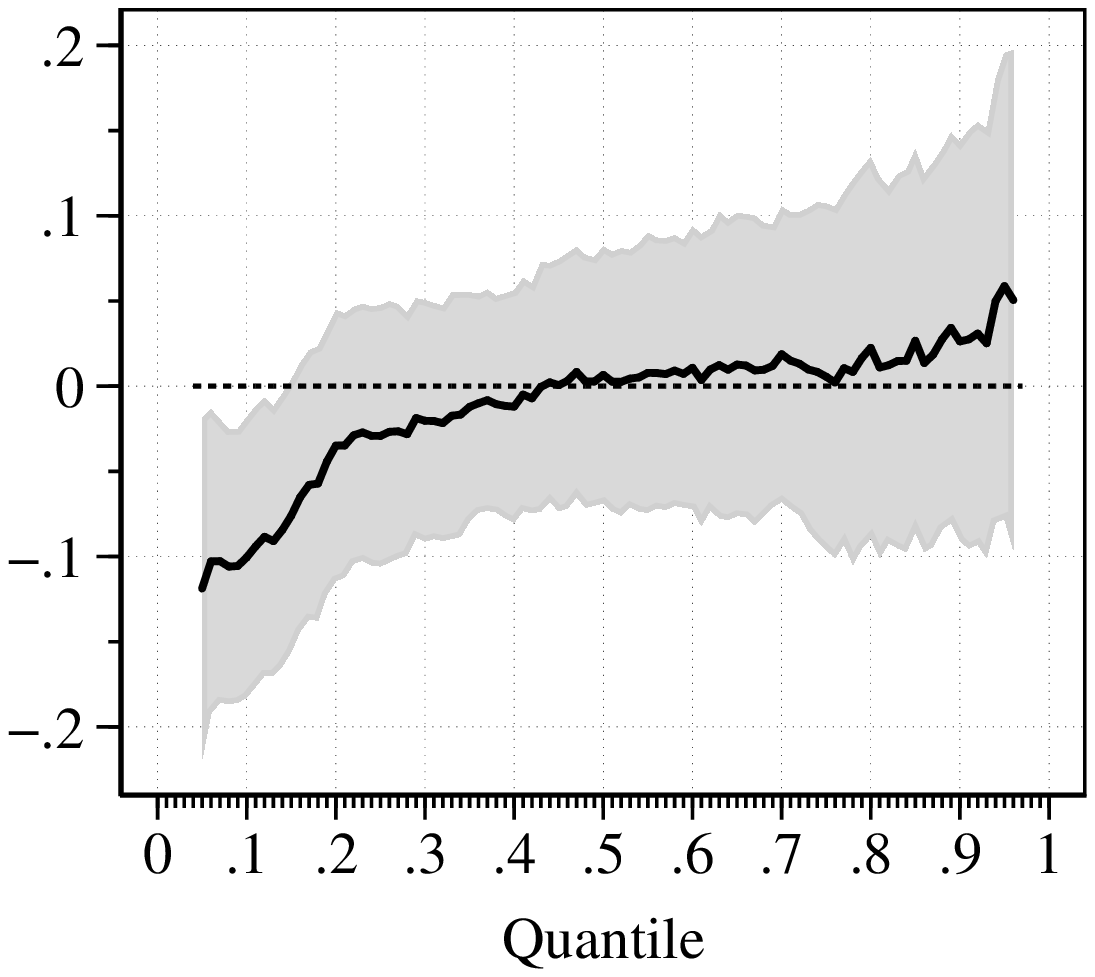}}
\par\end{centering}
\textit{\footnotesize{}Notes}{\footnotesize{}: Estimates of partial
effects in equation \eqref{eq: mean1} are reported. The shaded area
represents the 90 percent uniform confidence band.}{\footnotesize\par}
\end{figure}

\begin{figure}[H]
\caption{Long-term effect of the minimum wage on the wage structure: confidence
band\label{fig: estimates_lag_band}}

\begin{centering}
\subfloat[Intercept]{
\centering{}\includegraphics[scale=0.6]{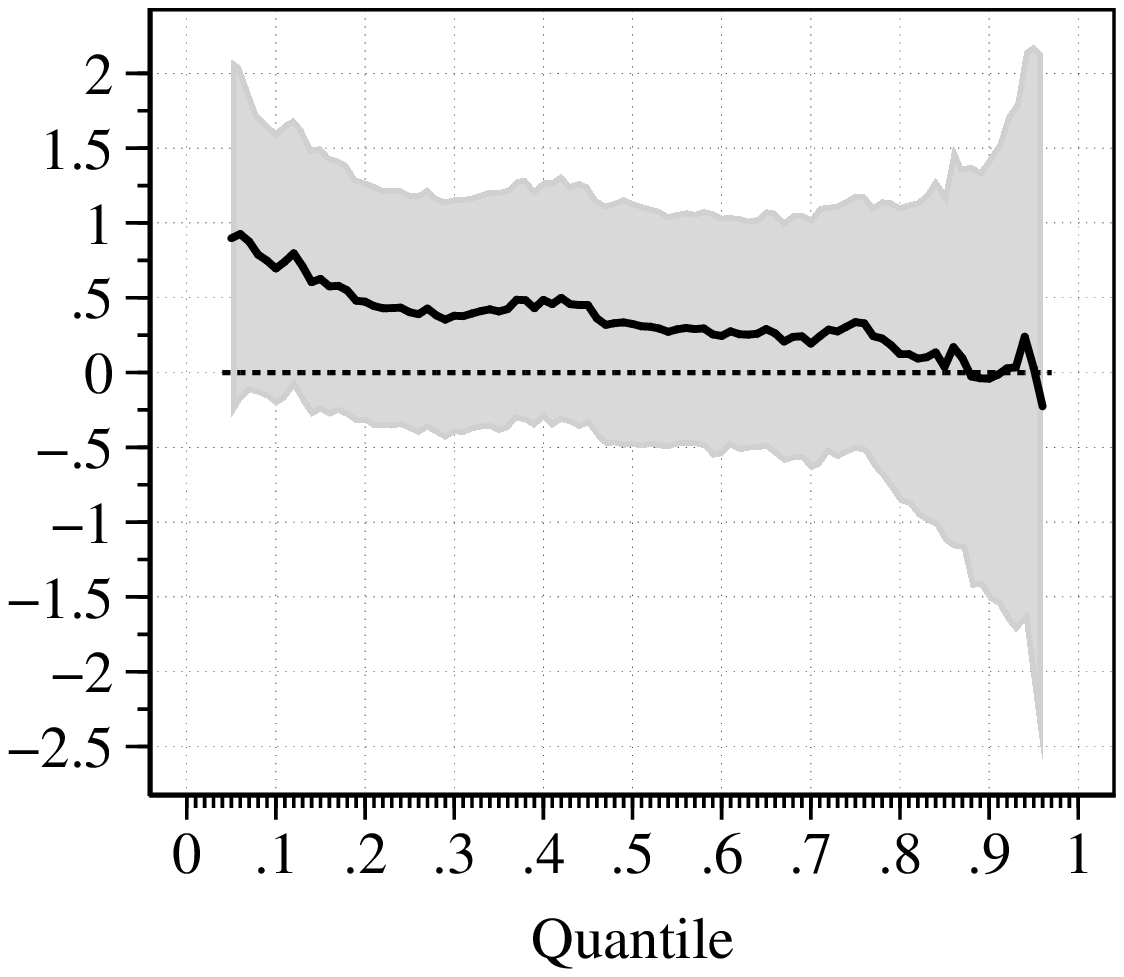}}\subfloat[Education]{
\centering{}\includegraphics[scale=0.6]{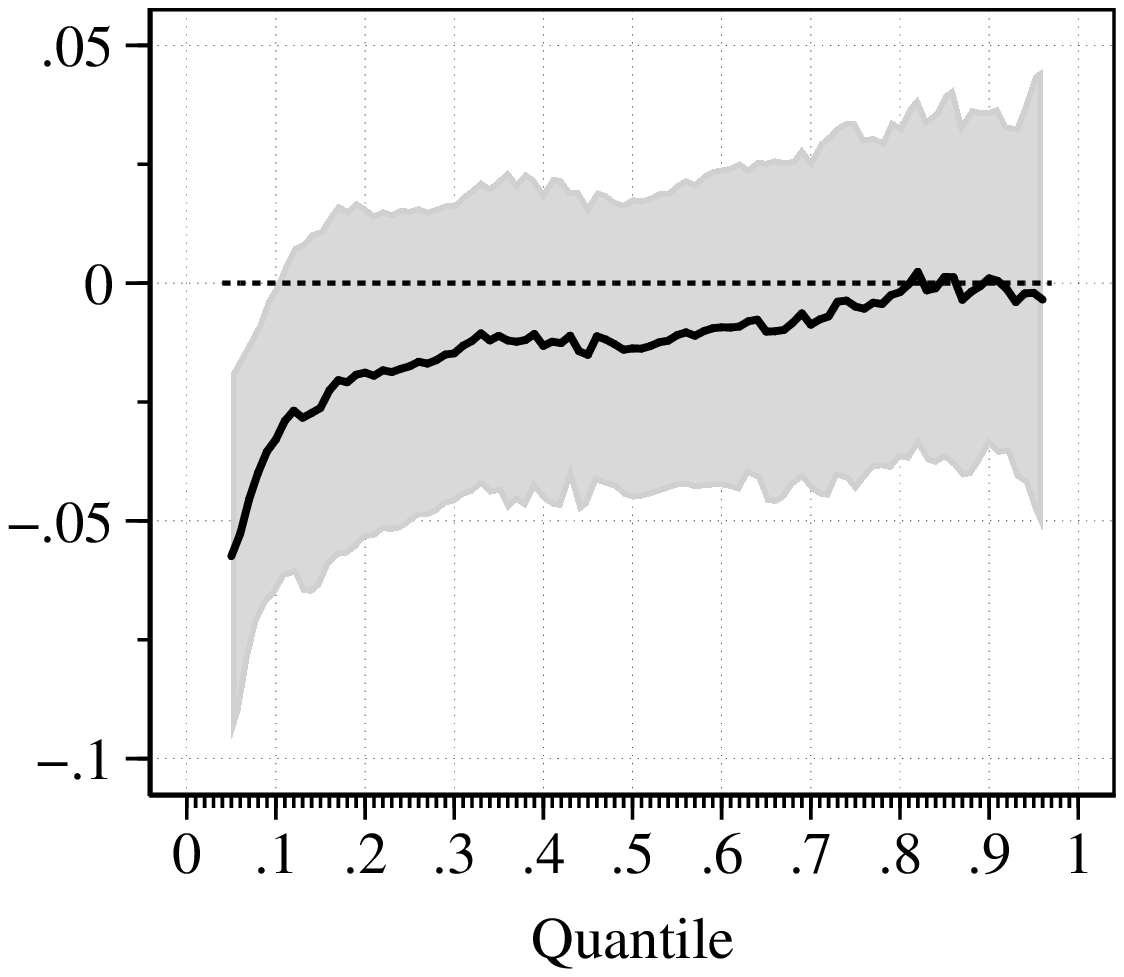}}
\par\end{centering}
\begin{centering}
\subfloat[Experience]{
\centering{}\includegraphics[scale=0.6]{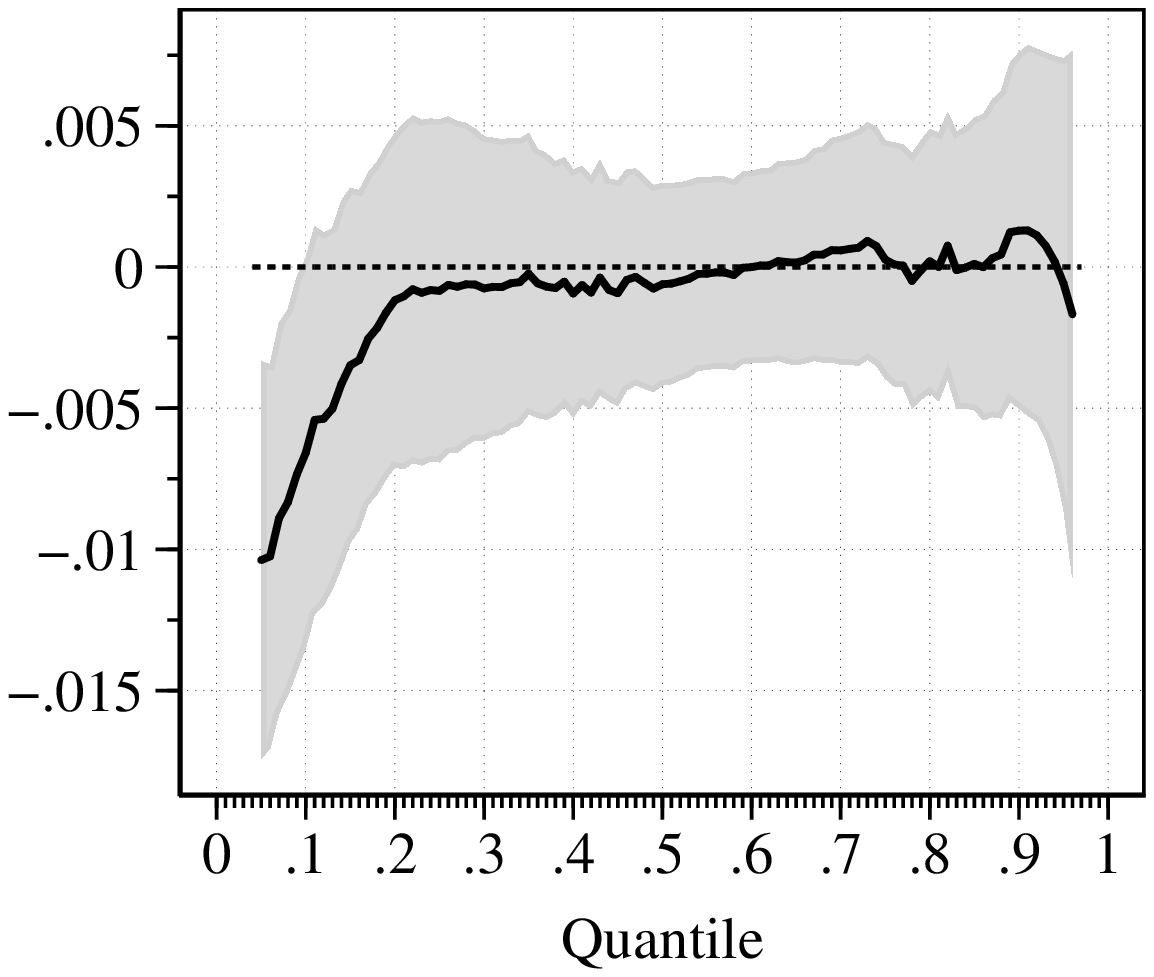}}\subfloat[Gender (male)]{
\centering{}\includegraphics[scale=0.6]{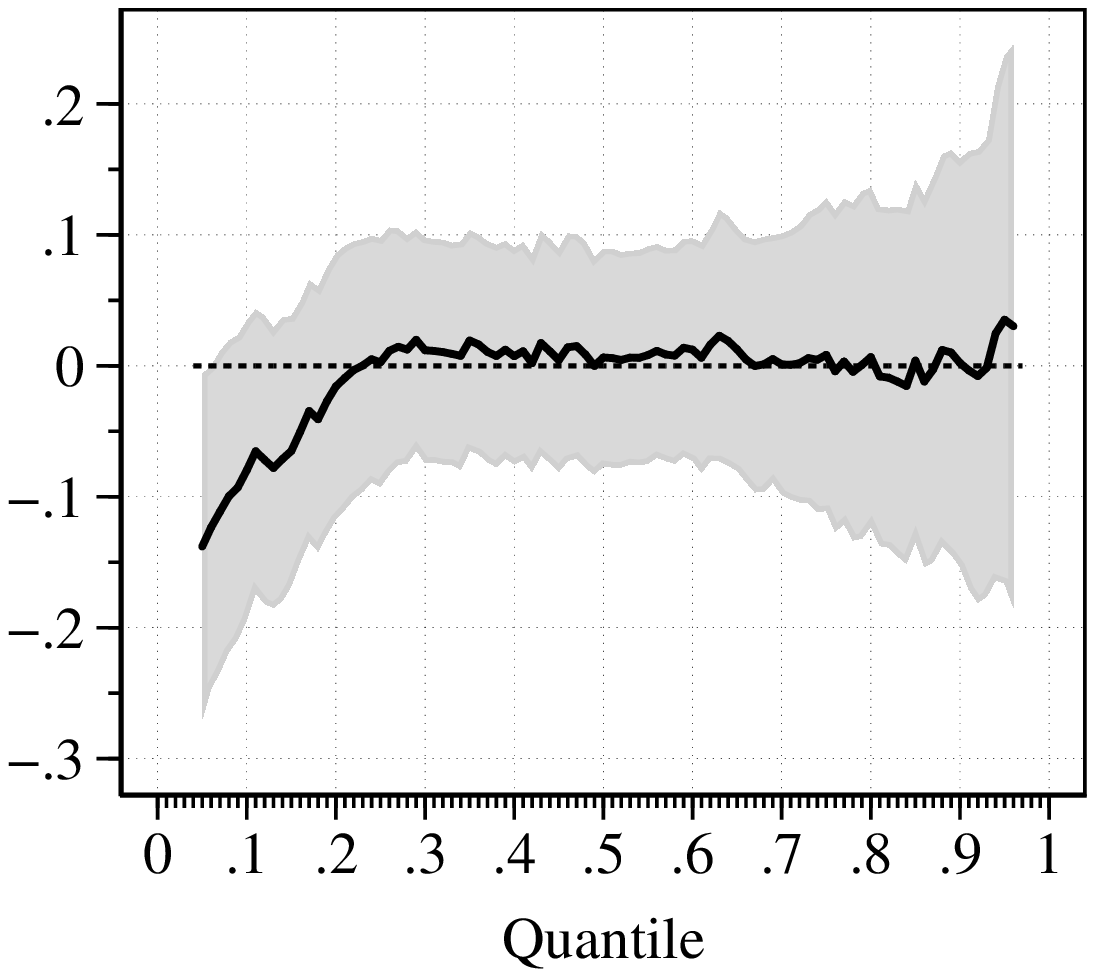}}
\par\end{centering}
\textit{\footnotesize{}Notes}{\footnotesize{}: Estimates of the long-term
effects in equation \eqref{eq: mean2} are reported. The shaded area
represents the 90 percent uniform confidence band.}{\footnotesize\par}
\end{figure}

\begin{figure}[H]
\caption{Placebo effect on the wage structure: confidence band\label{fig: estimates_lead_band}}

\begin{centering}
\subfloat[Intercept]{
\centering{}\includegraphics[scale=0.6]{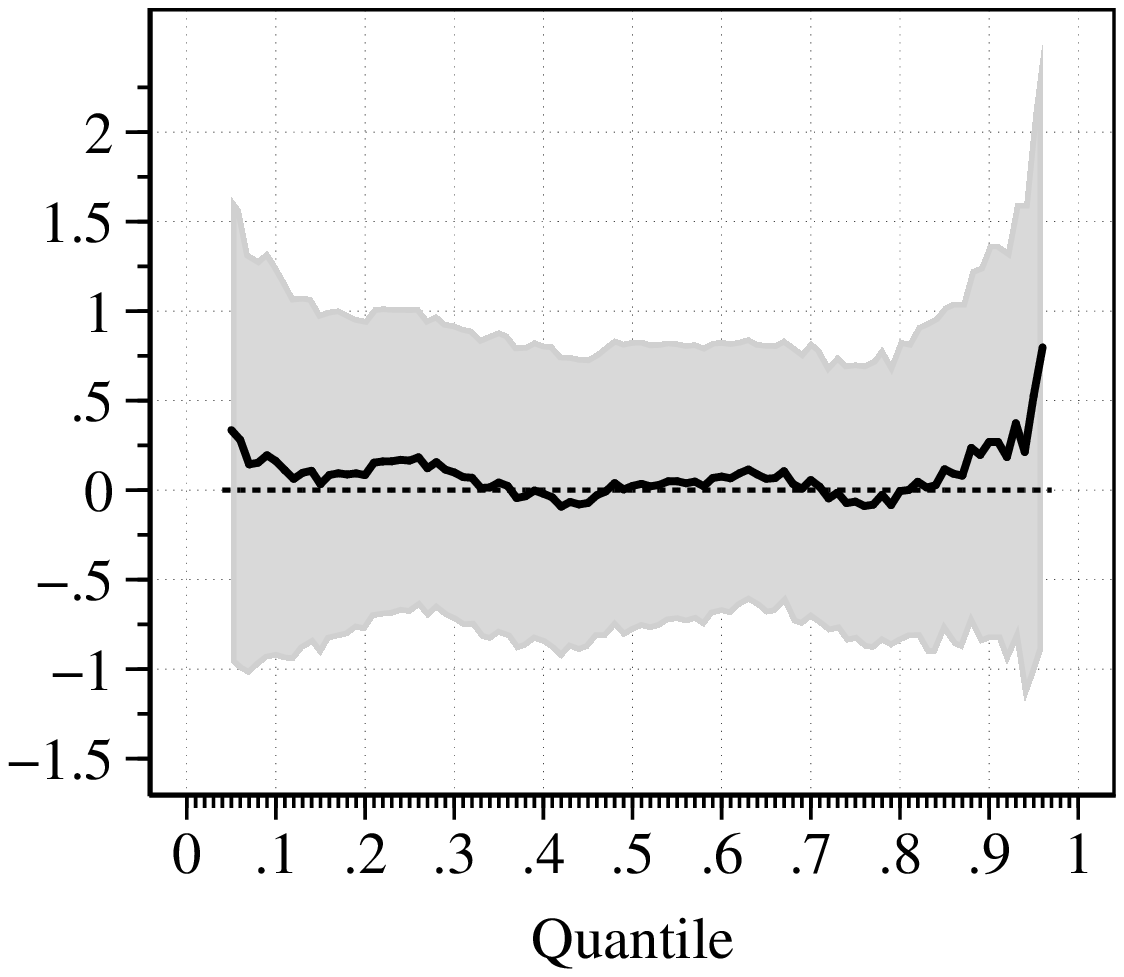}}\subfloat[Education]{
\centering{}\includegraphics[scale=0.6]{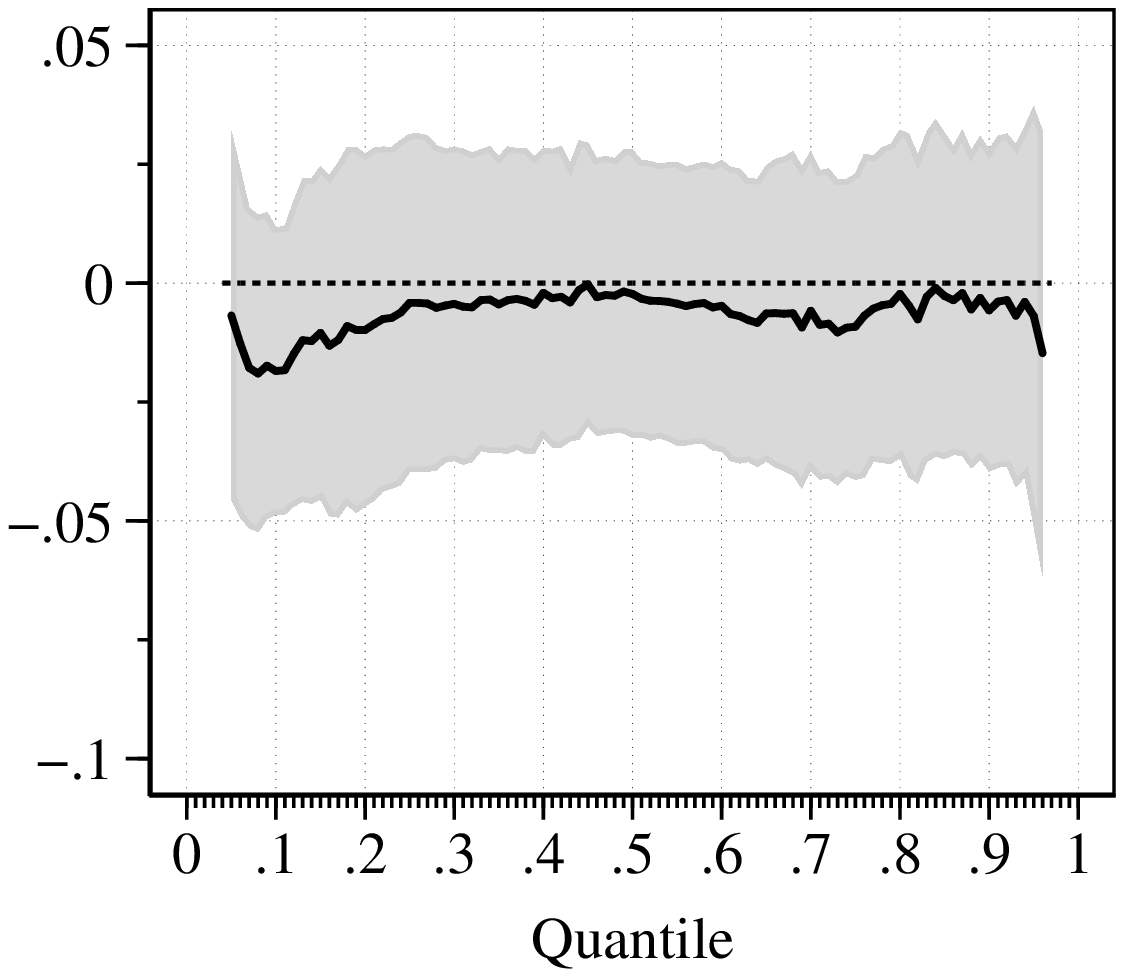}}
\par\end{centering}
\begin{centering}
\subfloat[Experience]{
\centering{}\includegraphics[scale=0.6]{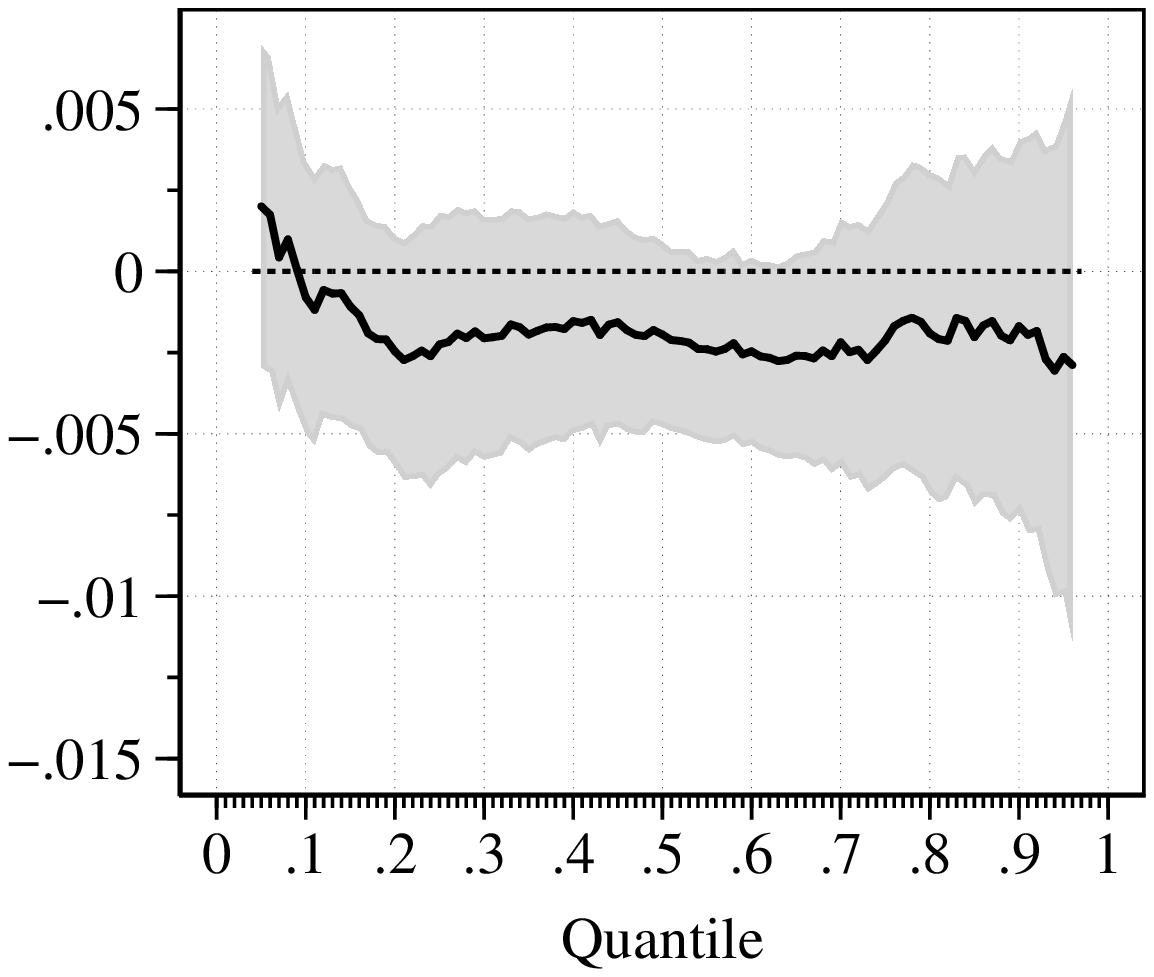}}\subfloat[Gender (male)]{
\centering{}\includegraphics[scale=0.6]{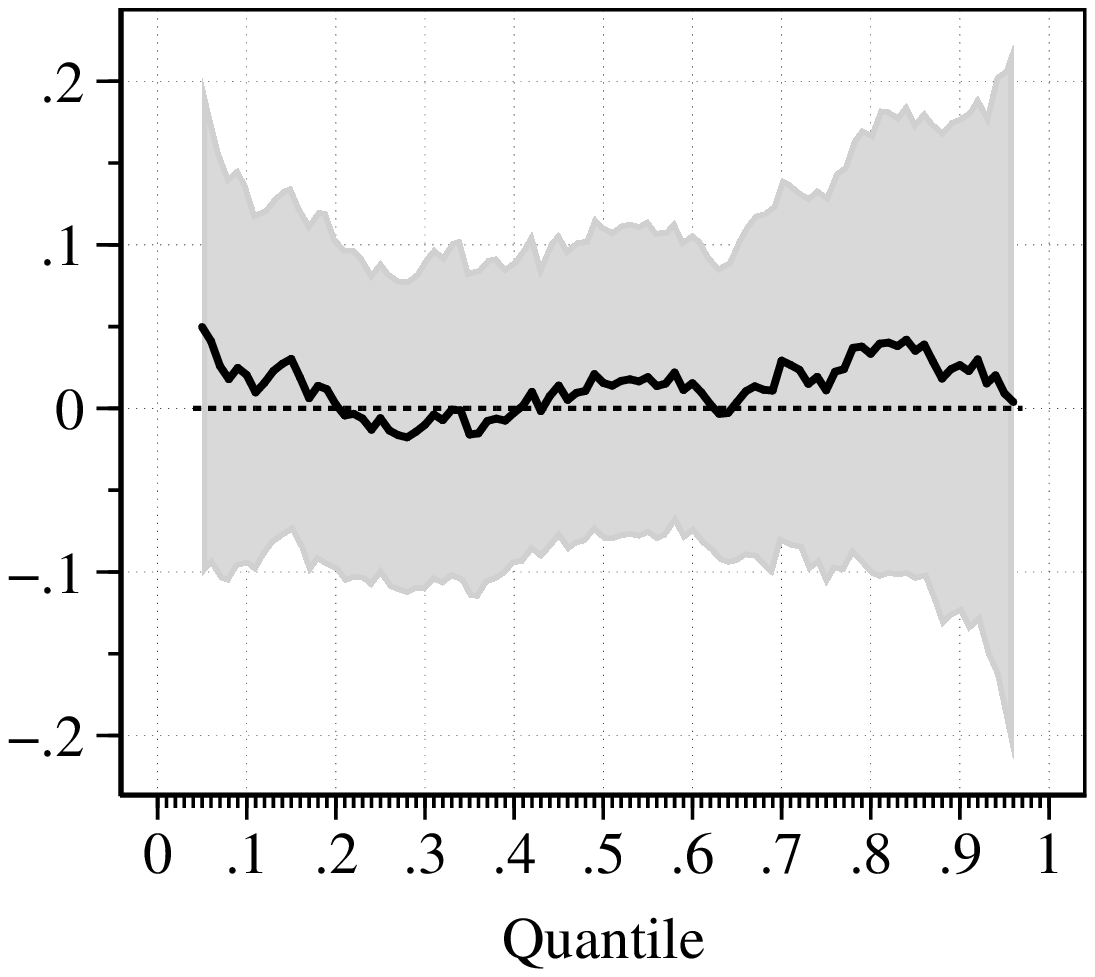}}
\par\end{centering}
\textit{\footnotesize{}Notes}{\footnotesize{}: Estimates of the leading
effects in equation \eqref{eq: mean2} are reported. The shaded area
represents the 90 percent uniform confidence band.}{\footnotesize\par}
\end{figure}

Figures \ref{fig: estimates_band}, \ref{fig: estimates_lag_band},
and \ref{fig: estimates_lead_band} show uniform confidence bands
of the estimates in Figures \ref{fig: estimates}, \ref{fig: estimates_lag},
and \ref{fig: estimates_lead}, respectively. We follow \citet{Chernozhukov_FernandezVal_Melly_EM13}
in obtaining uniform confidence bands. Naturally, uniform confidence
bands are wider than pointwise confidence intervals. However, we cannot
reject the hypothesis of no effect of the minimum wage.

\subsubsection{Changes in  between- and within-group wage differentials, 1979\textendash 1989\label{subsec: changes_1989}}

\begin{figure}[H]
\caption{Changes in the educational wage differential (16 versus 12 years of
education), 1979\textendash 1989\label{fig: diff_educ_actual_1989}}

\begin{centering}
\subfloat[5 years of experience, males]{
\centering{}\includegraphics[scale=0.6]{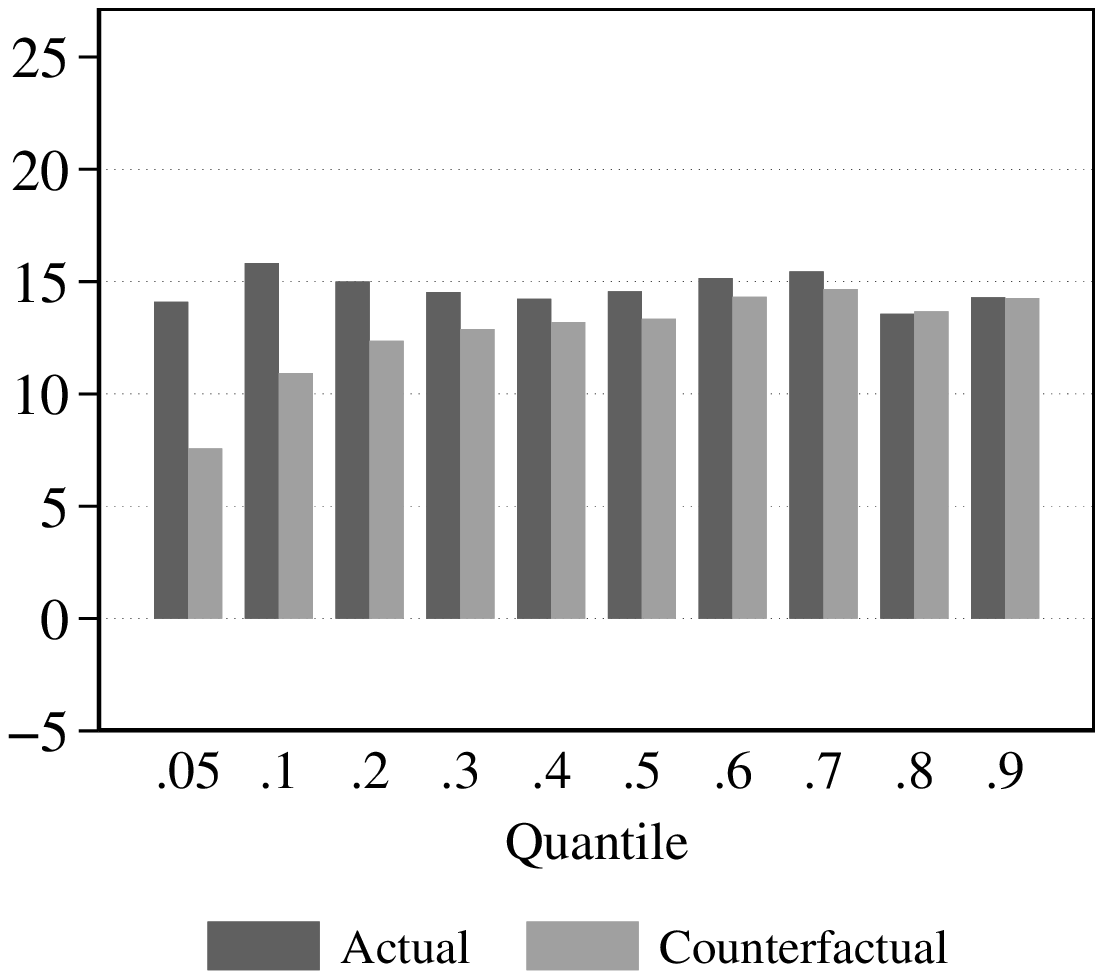}}\subfloat[10 years of experience, males]{
\centering{}\includegraphics[scale=0.6]{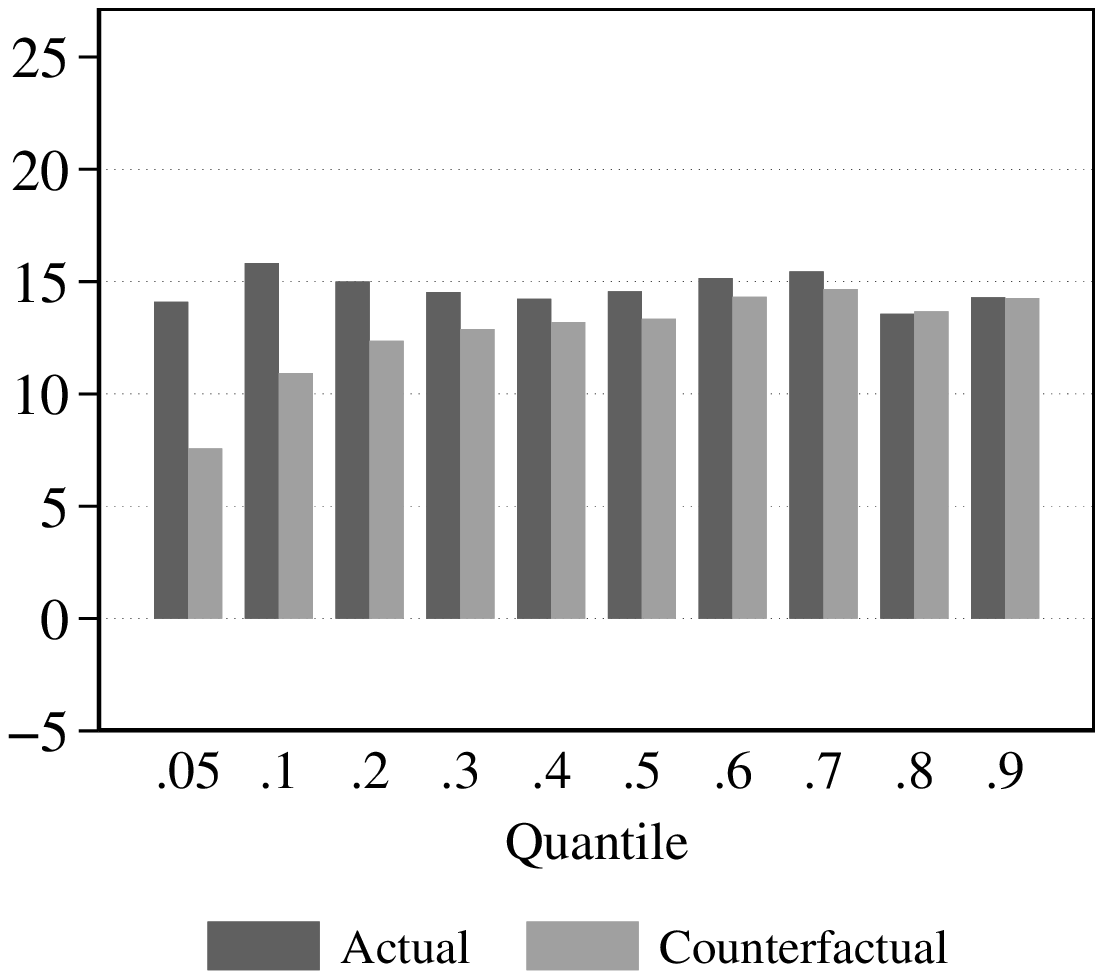}}
\par\end{centering}
\centering{}\subfloat[5 years of experience, females]{
\centering{}\includegraphics[scale=0.6]{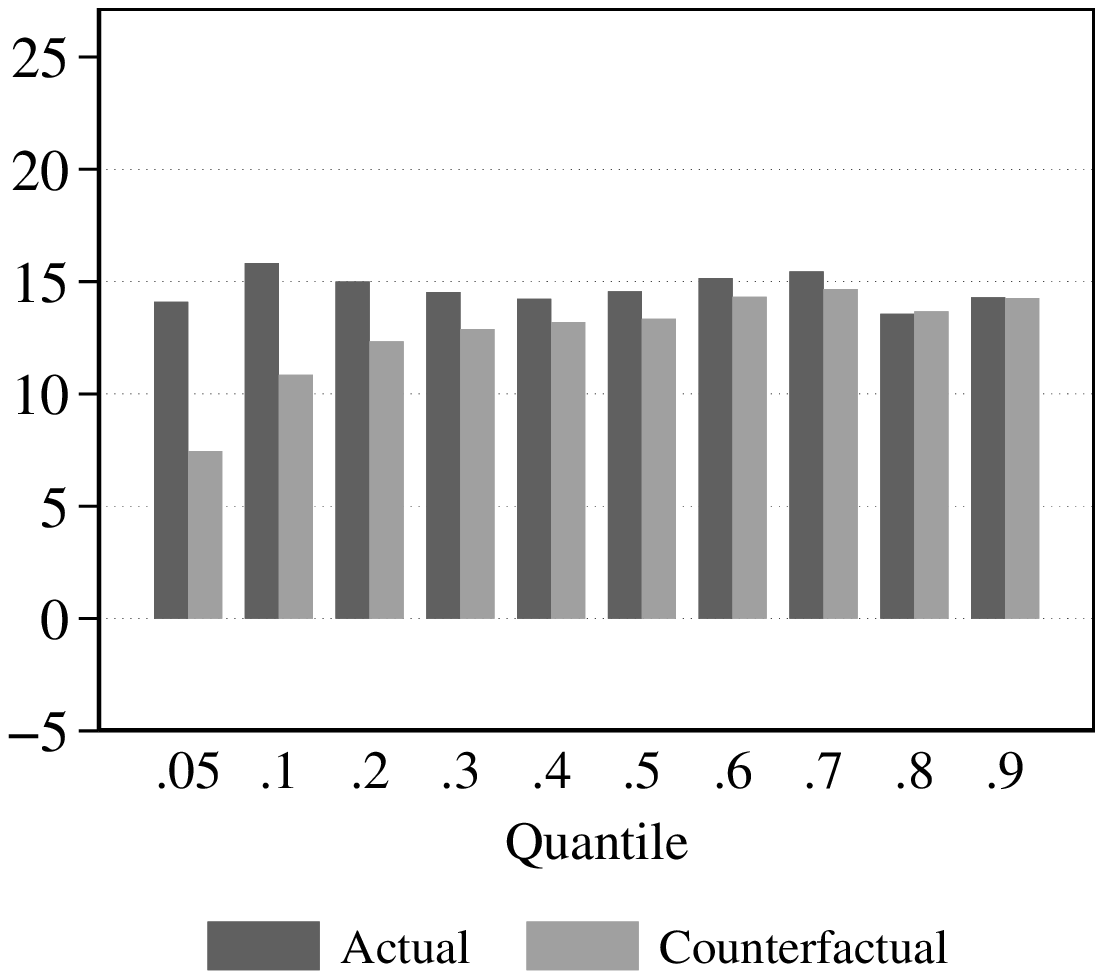}}\subfloat[10 years of experience, females]{
\centering{}\includegraphics[scale=0.6]{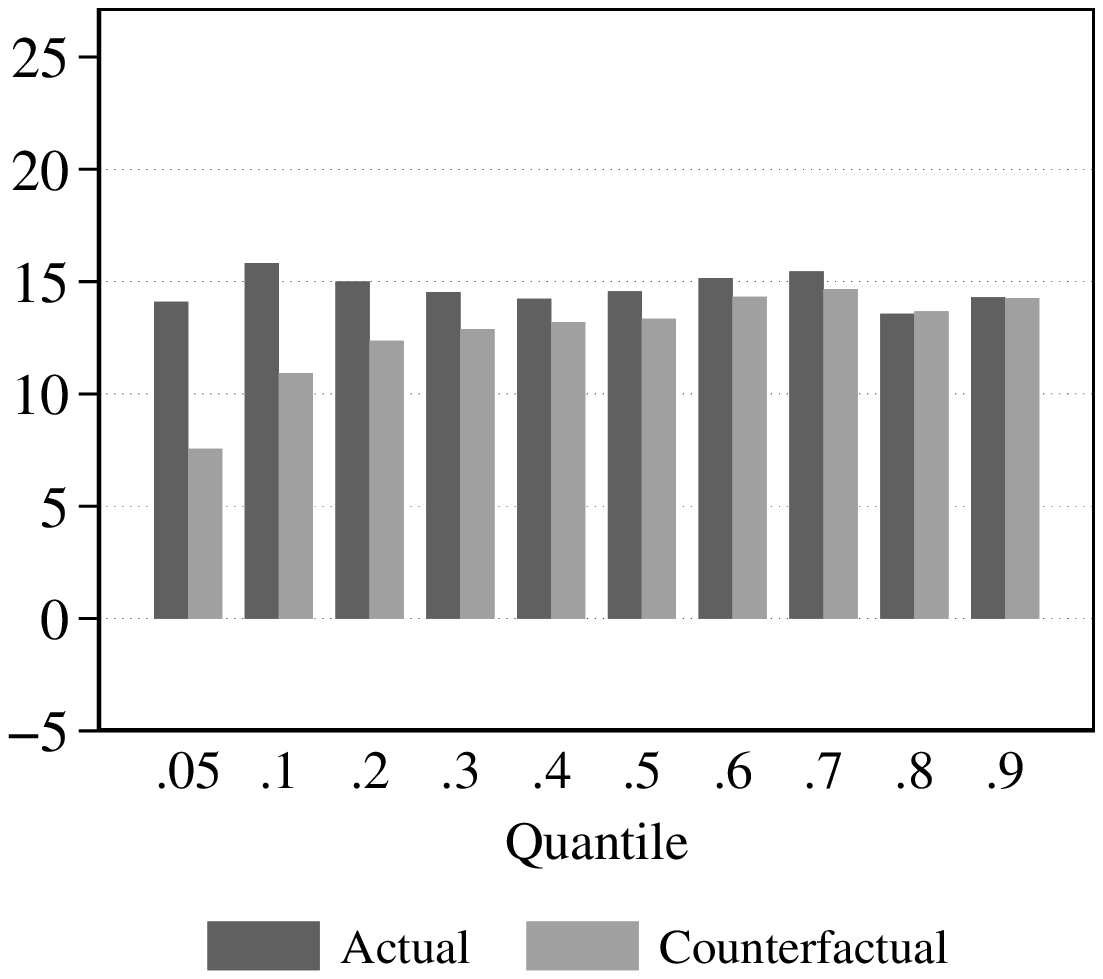}}
\end{figure}

\begin{figure}[H]
\caption{Changes in the experience wage differential (25 versus 5 years of
experience), 1979\textendash 1989\label{fig: diff_exper_actual_1989}}

\begin{centering}
\subfloat[12 years of education, males]{
\centering{}\includegraphics[scale=0.6]{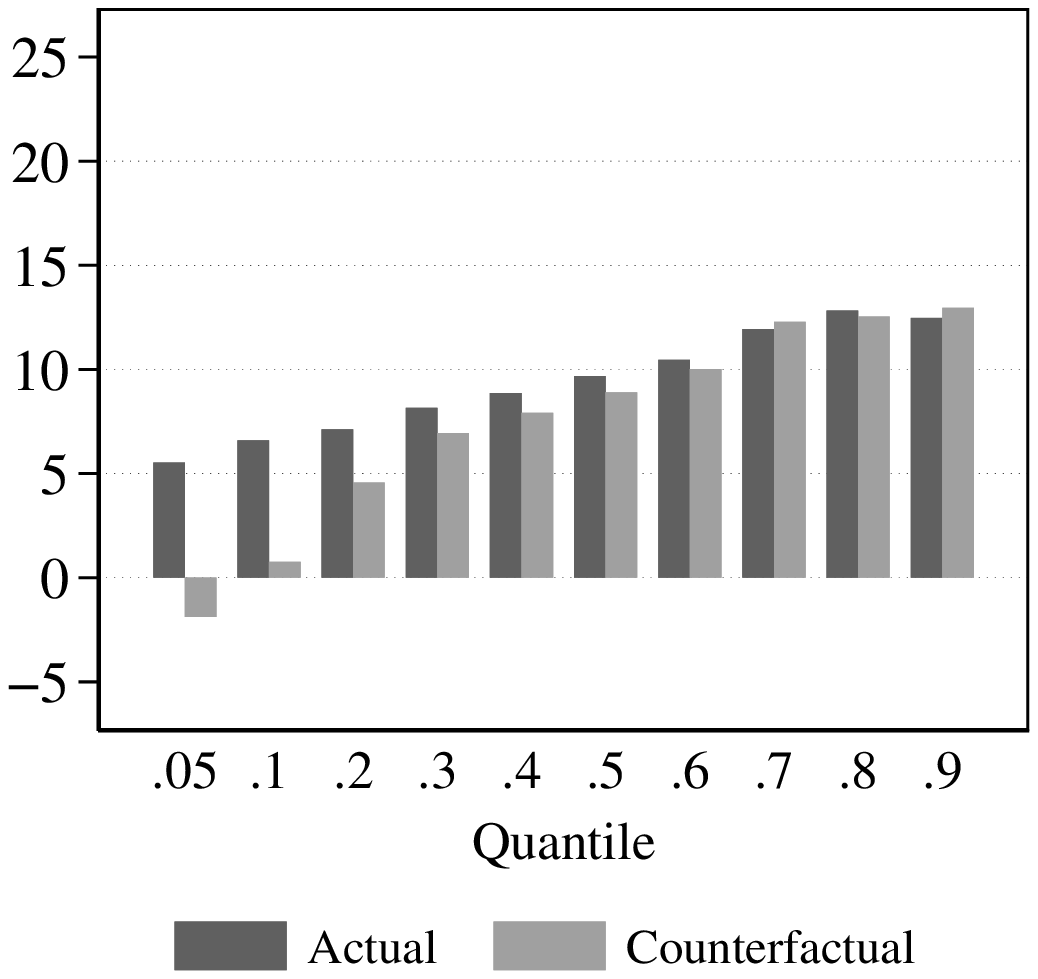}}\subfloat[16 years of education, males]{
\centering{}\includegraphics[scale=0.6]{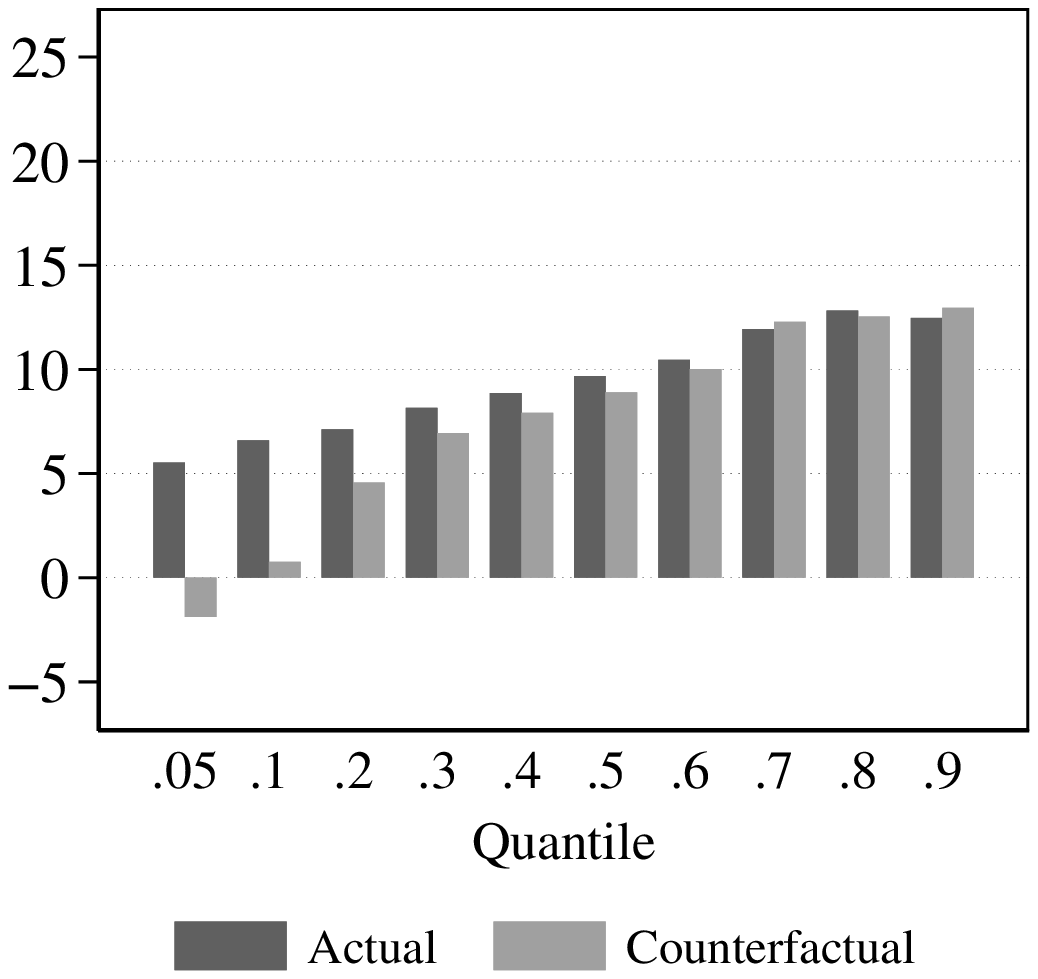}}
\par\end{centering}
\centering{}\subfloat[12 years of education, females]{
\centering{}\includegraphics[scale=0.6]{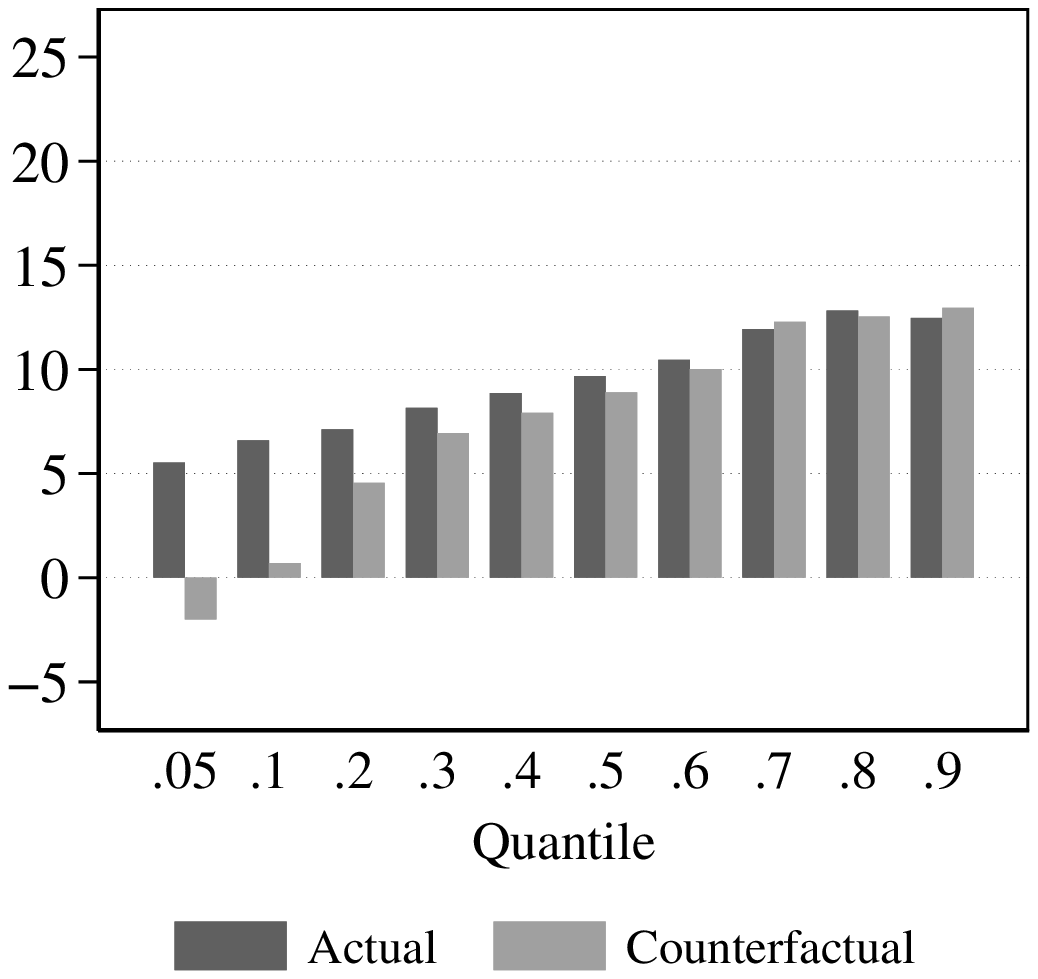}}\subfloat[16 years of education, females]{
\centering{}\includegraphics[scale=0.6]{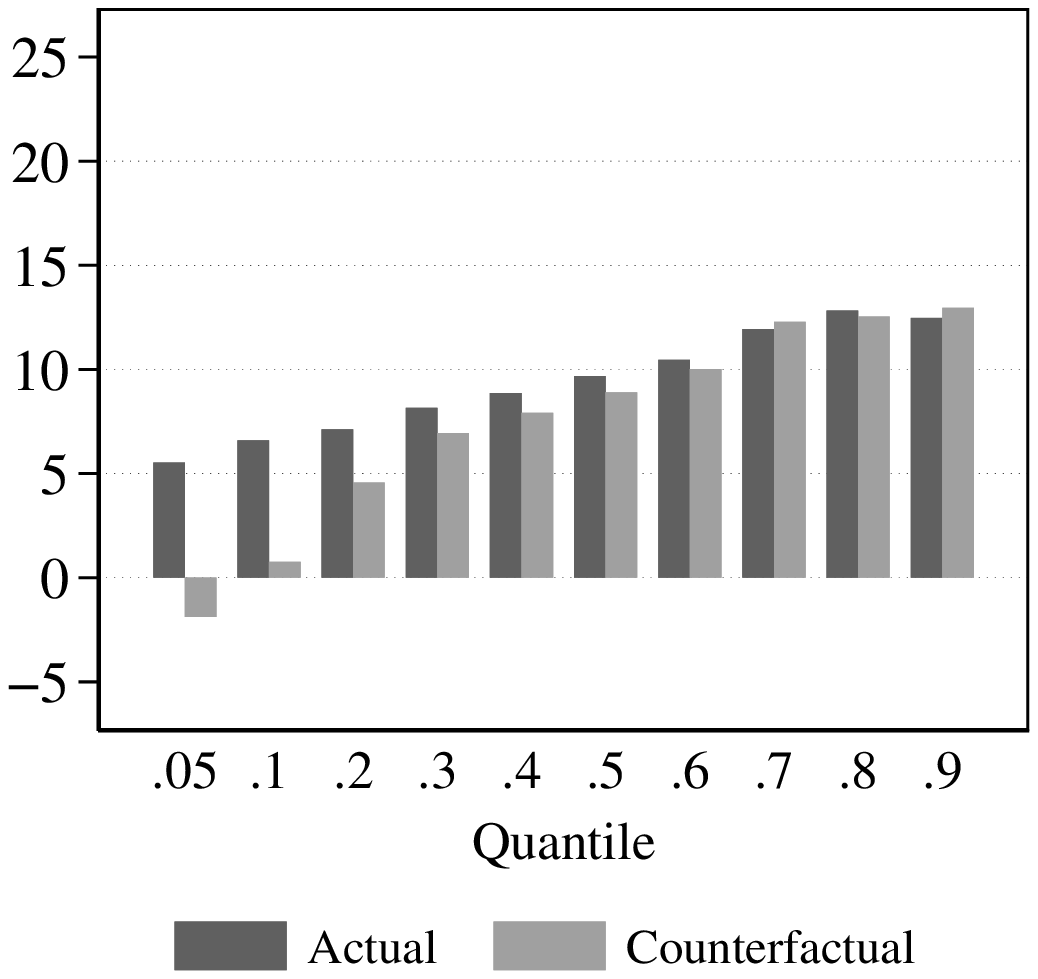}}
\end{figure}

\begin{figure}[H]
\caption{Changes in the gender wage differential (males versus females), 1979\textendash 1989\label{fig: diff_gender_actual_1989}}

\begin{centering}
\subfloat[12 years of education, 5 years of experience]{
\centering{}\includegraphics[scale=0.6]{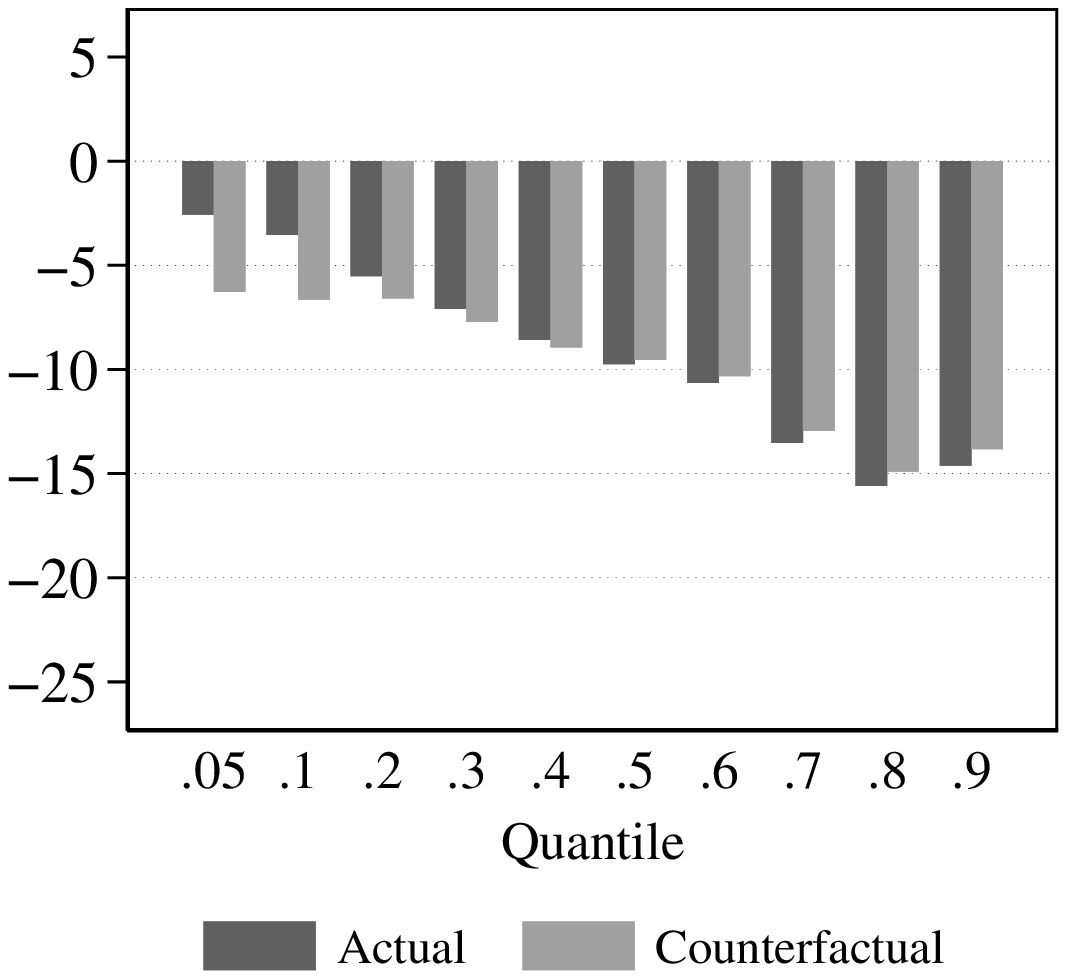}}\subfloat[12 years of education, 10 years of experience]{
\centering{}\includegraphics[scale=0.6]{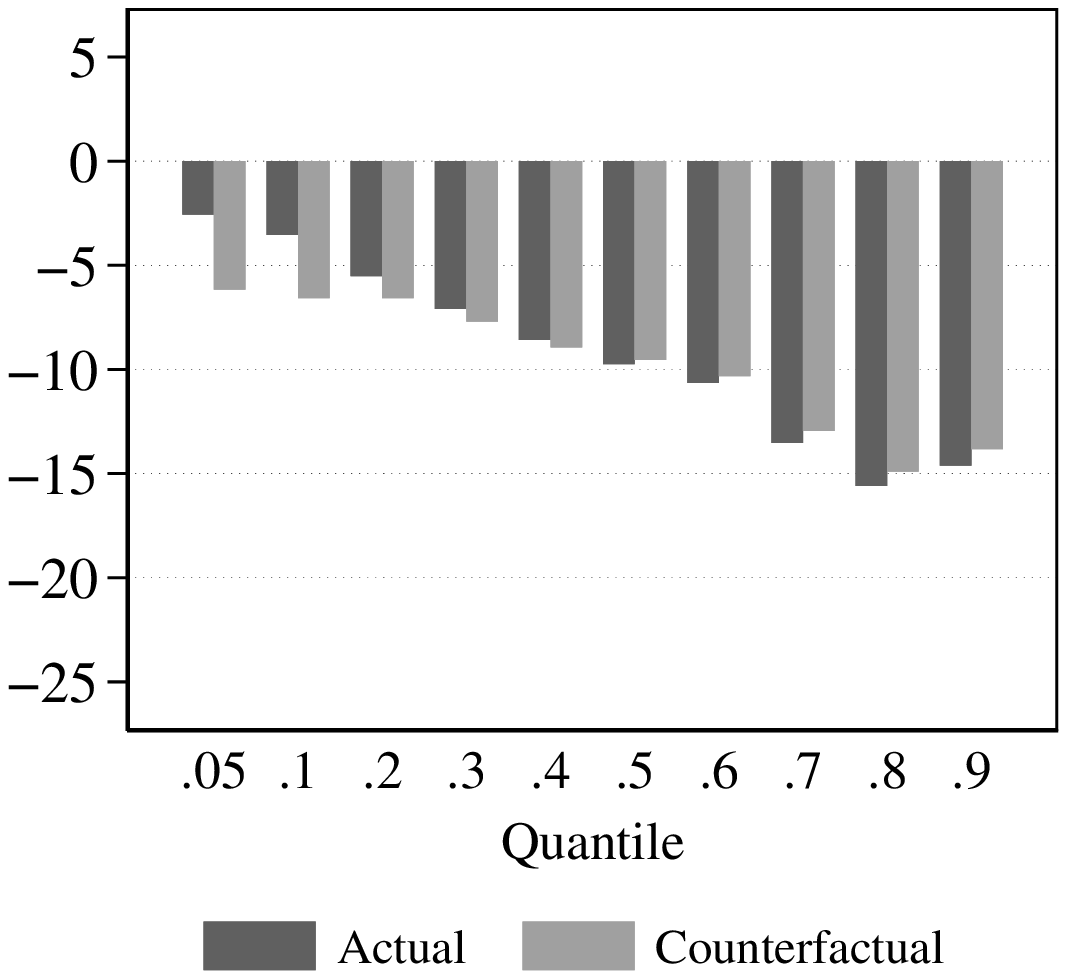}}
\par\end{centering}
\centering{}\subfloat[16 years of education, 5 years of experience]{
\centering{}\includegraphics[scale=0.6]{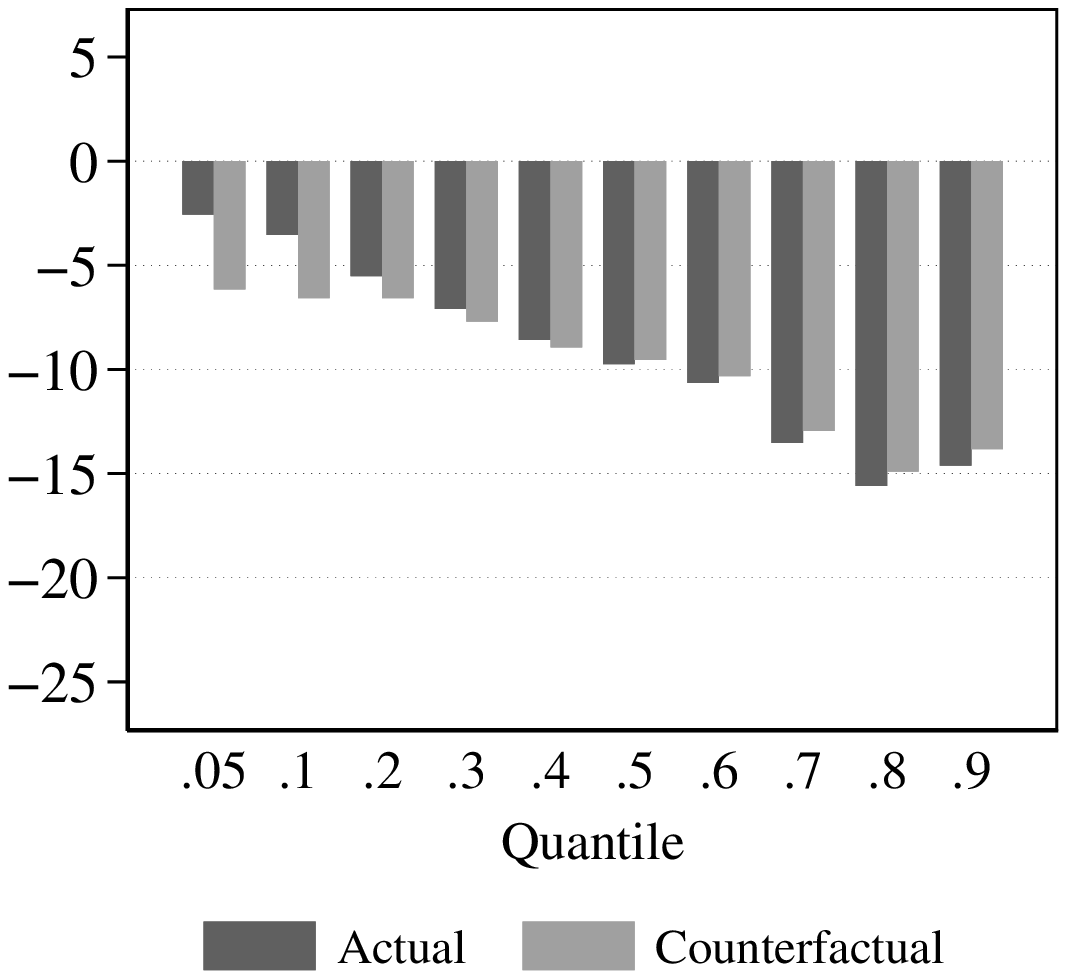}}\subfloat[16 years of education, 10 years of experience]{
\centering{}\includegraphics[scale=0.6]{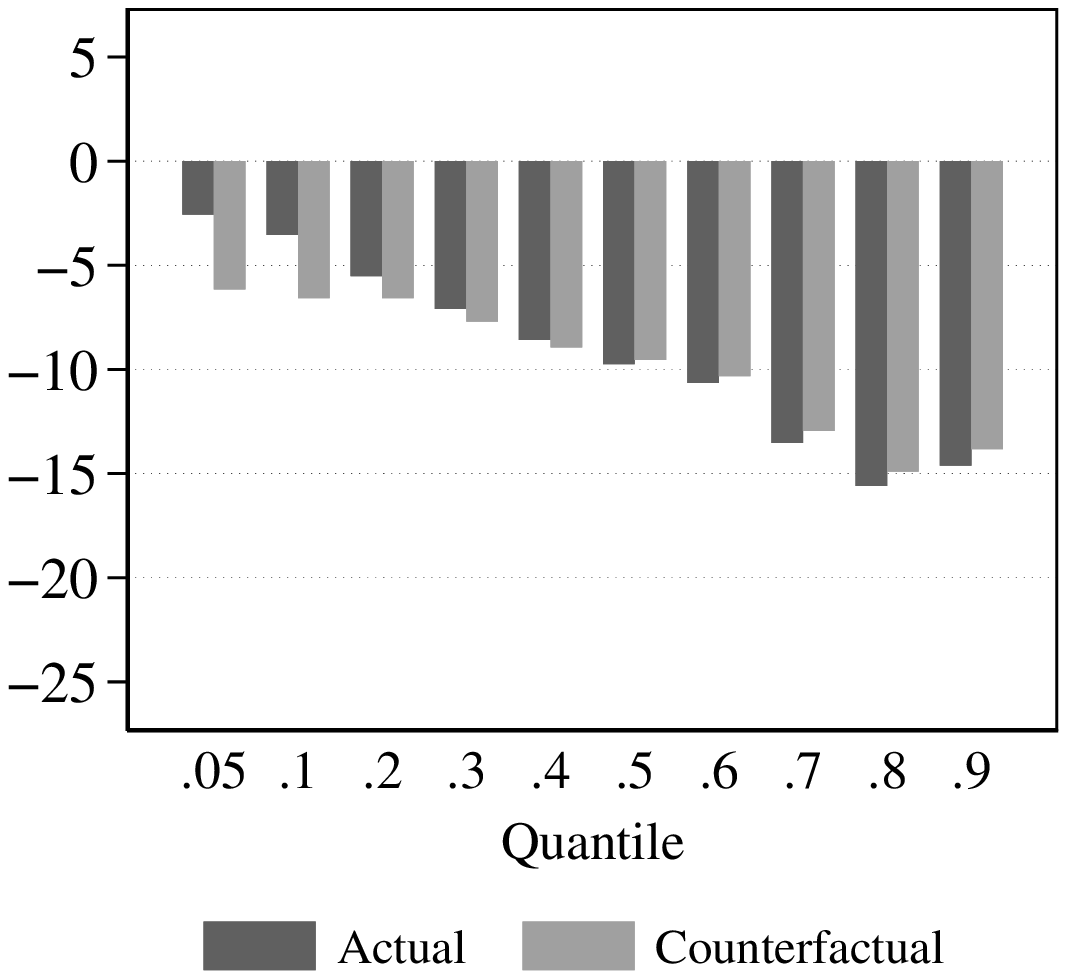}}
\end{figure}

\begin{figure}[H]
\caption{Changes in the 90/10, 50/10, and 50/20 within-group differentials,
1979\textendash 1989\label{fig: diff_within_actual_1989}}

\begin{centering}
\subfloat[90/10, males]{
\centering{}\includegraphics[scale=0.6]{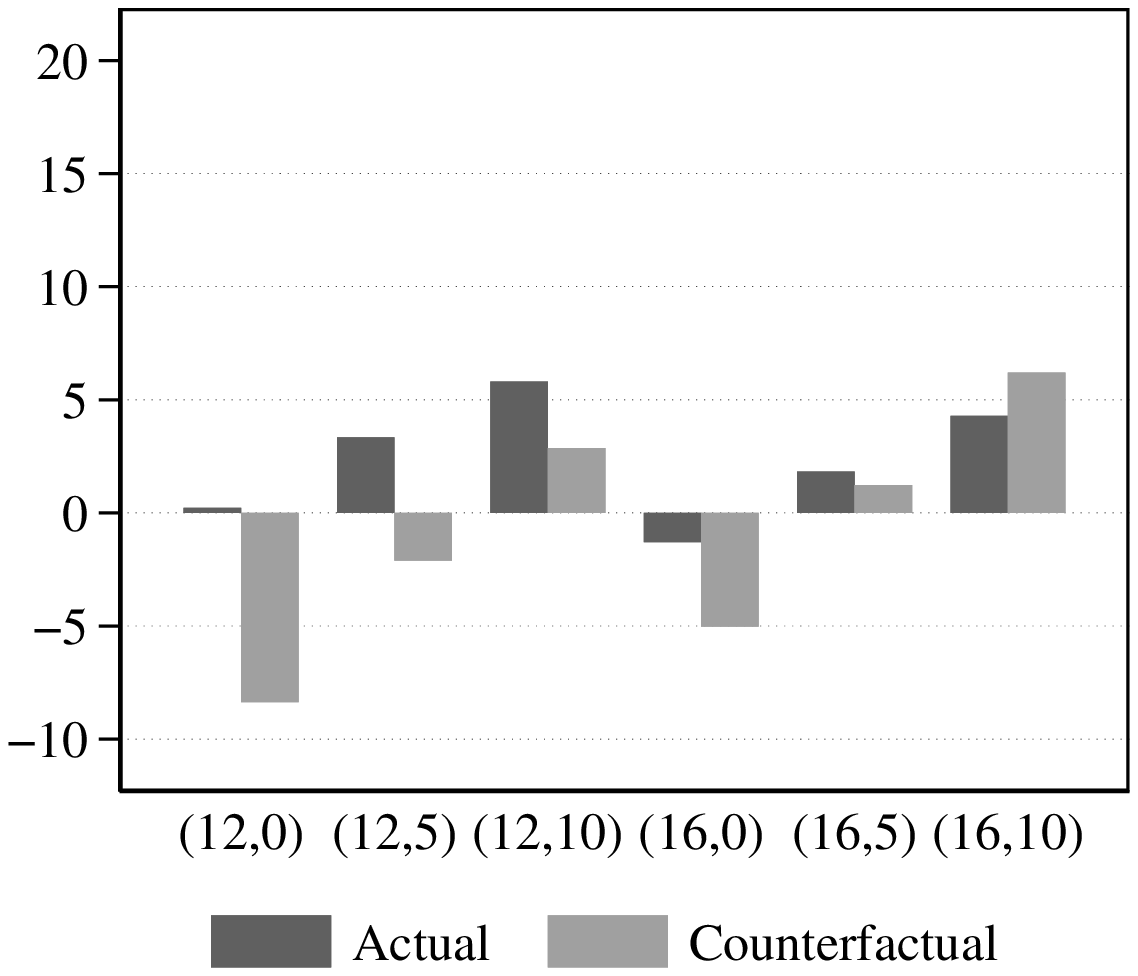}}\subfloat[90/10, females]{
\centering{}\includegraphics[scale=0.6]{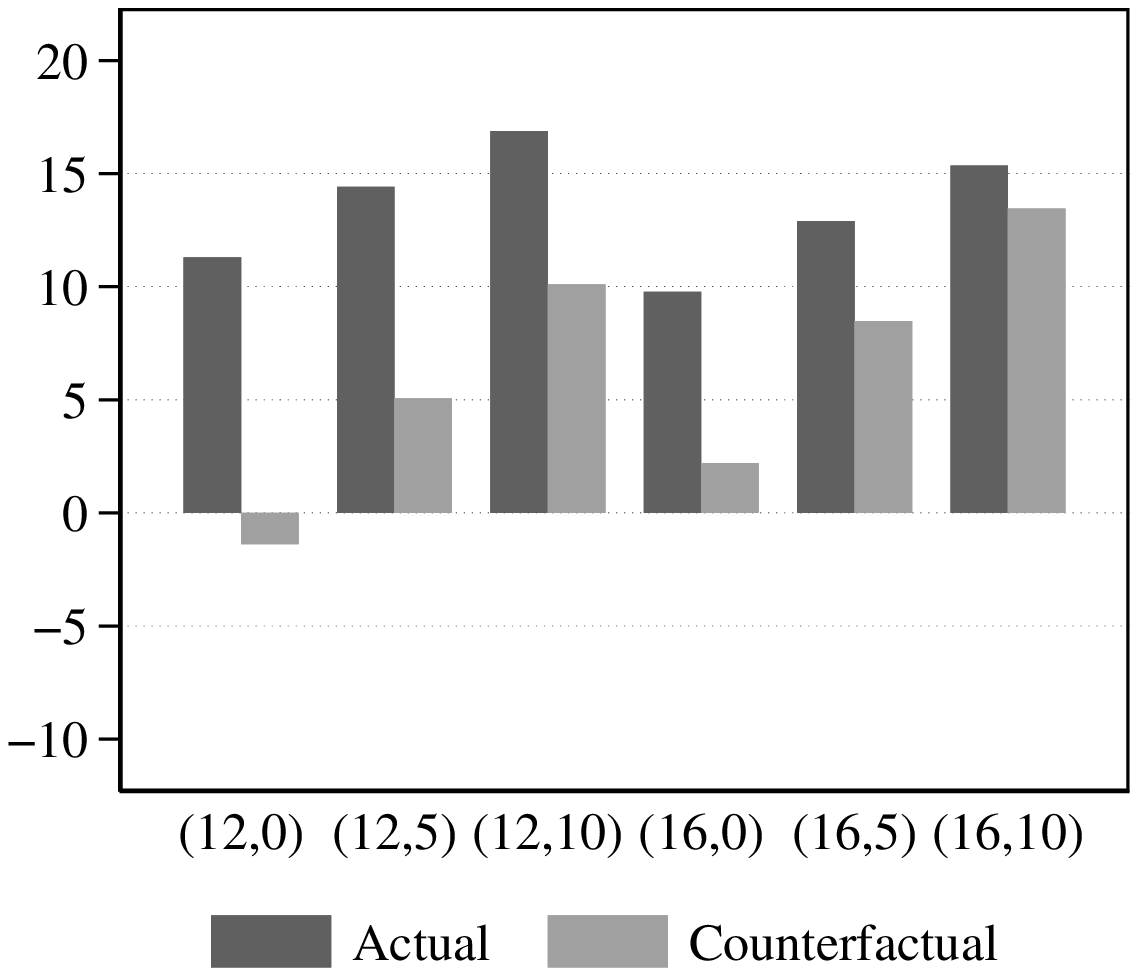}}
\par\end{centering}
\begin{centering}
\subfloat[50/10, males]{
\centering{}\includegraphics[scale=0.6]{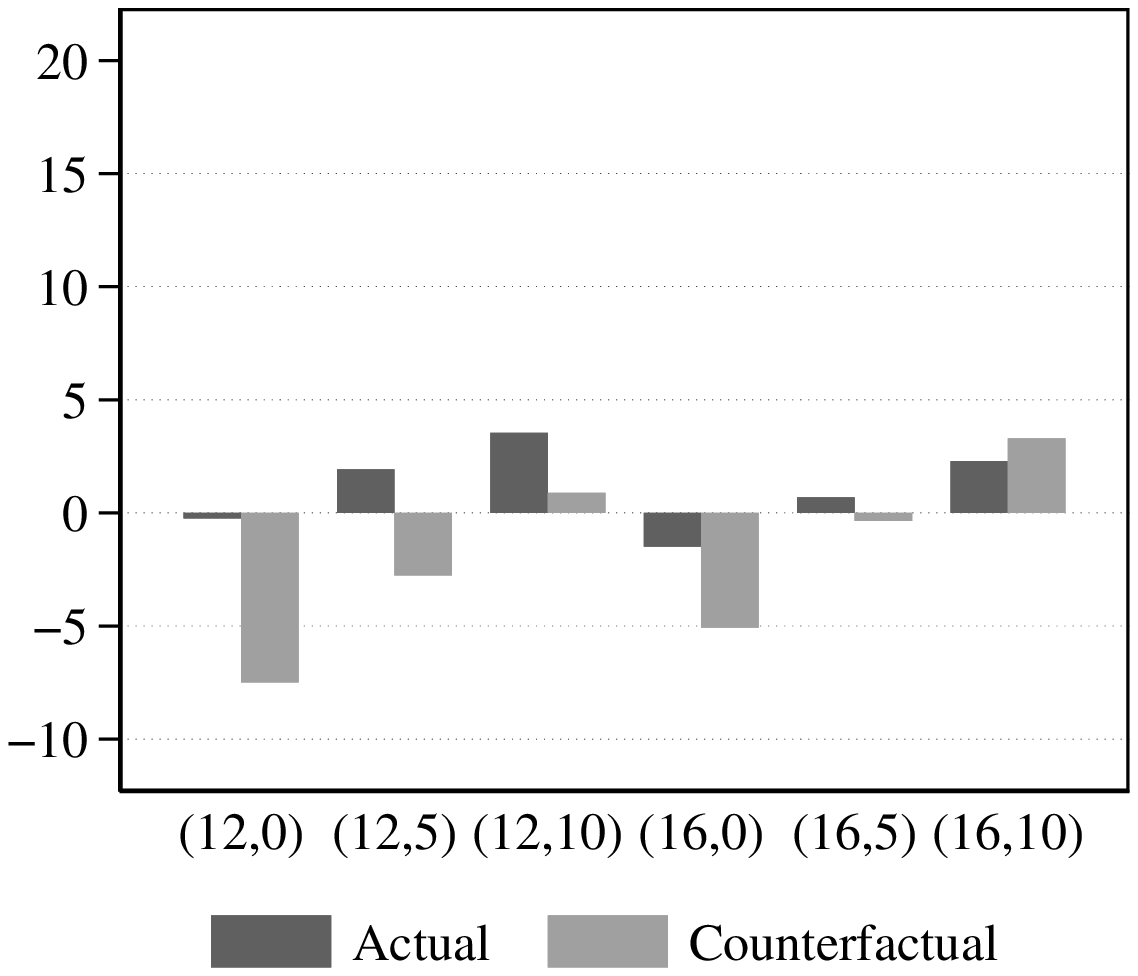}}\subfloat[50/10, females]{
\centering{}\includegraphics[scale=0.6]{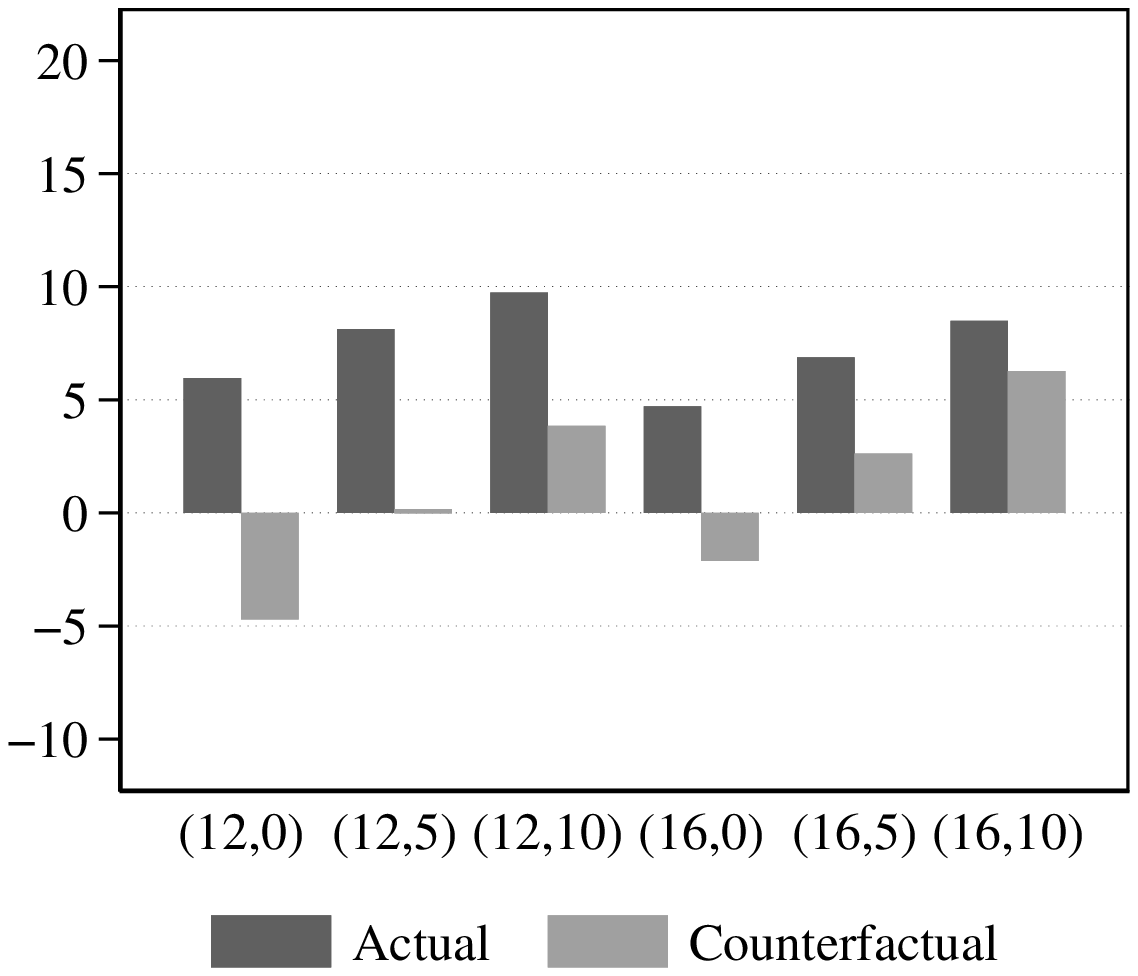}}
\par\end{centering}
\centering{}\subfloat[50/20, males]{
\centering{}\includegraphics[scale=0.6]{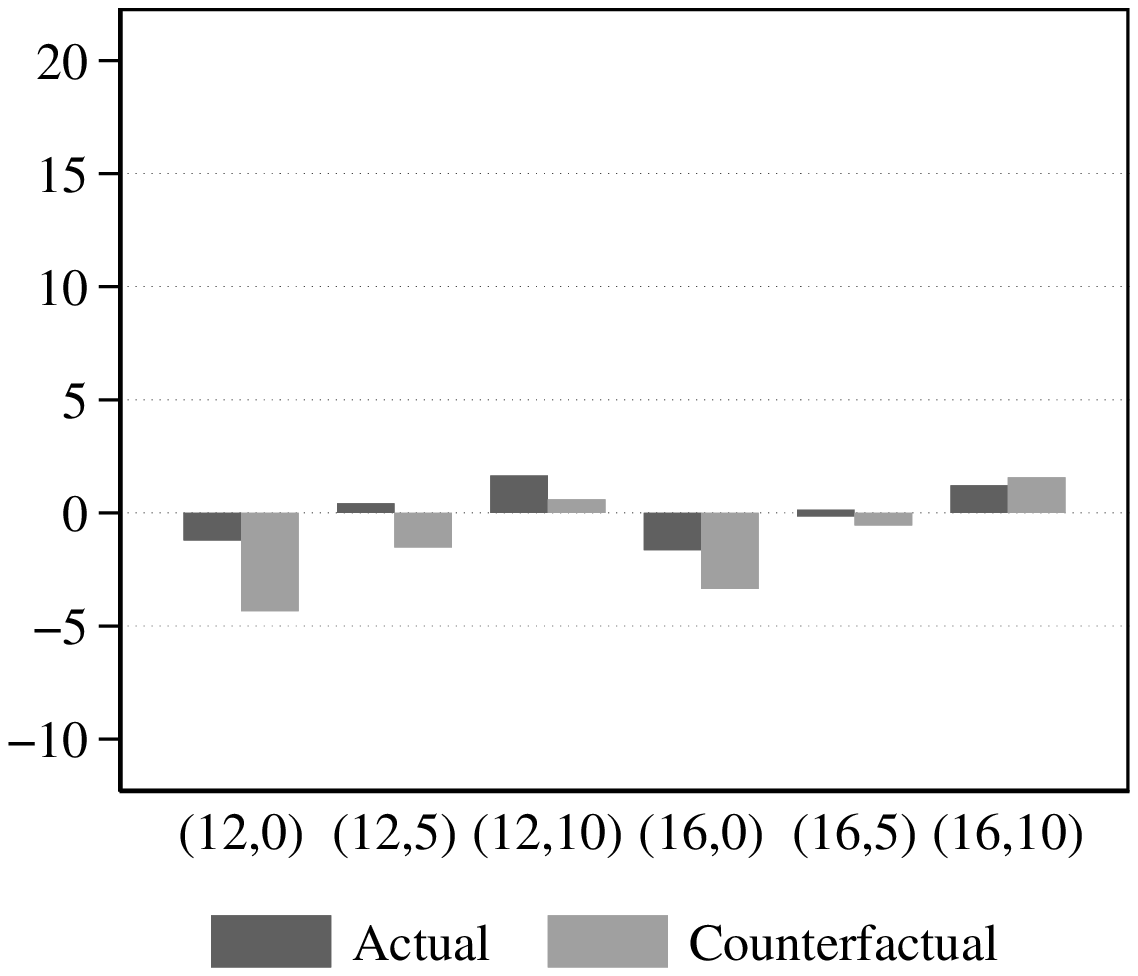}}\subfloat[50/20, females]{
\centering{}\includegraphics[scale=0.6]{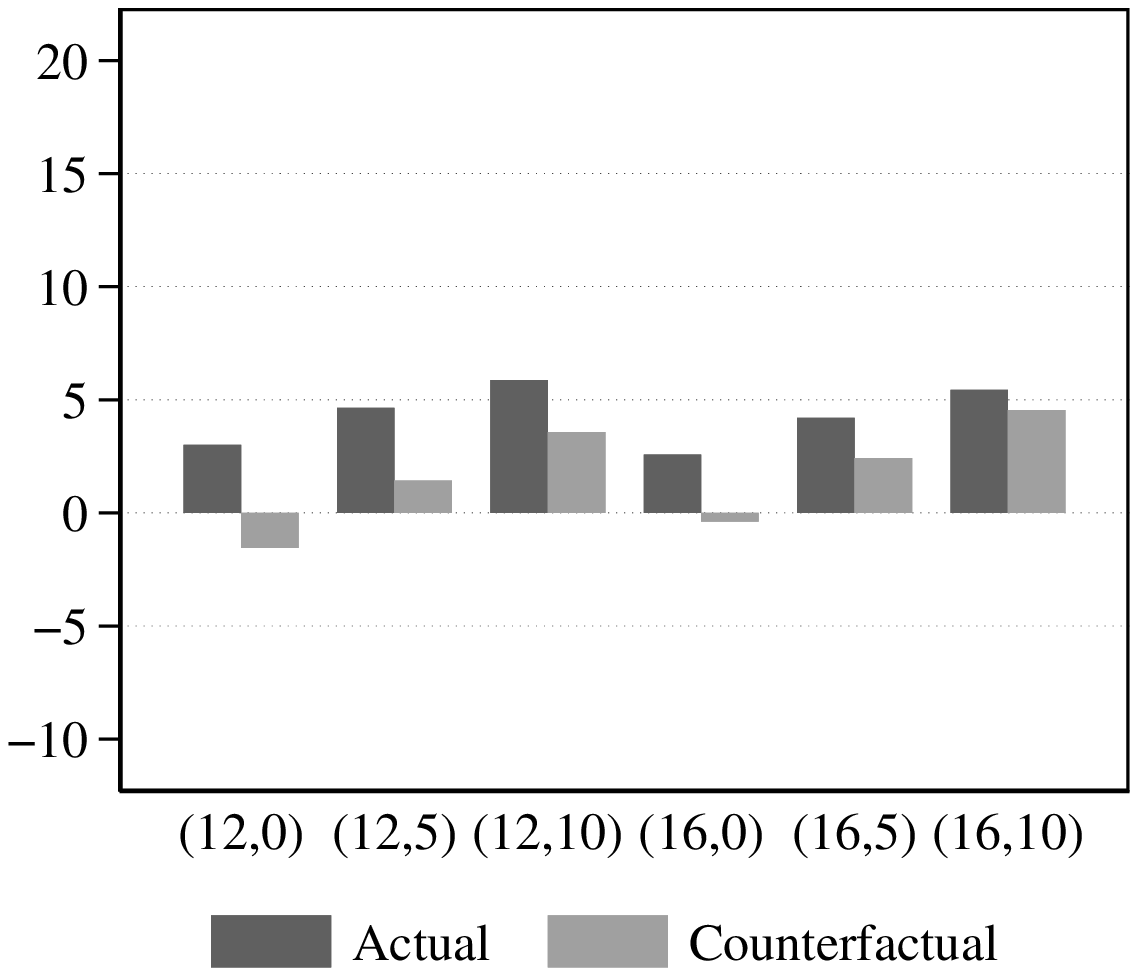}}
\end{figure}

Figures \ref{fig: diff_educ_actual_1989} to \ref{fig: diff_within_actual_1989}
show actual and counterfactual changes in between- and within-group
wage differentials for the years 1979 to 1989.

During the 1979\textendash 1989 period, the educational wage differentials
increased almost uniformly across quantiles (Figures \ref{fig: diff_educ_actual_1989}),
as also shown by \citet*{Buchinsky_EM94} and \citet{Angrist_Chernozhukov_FernandezVal_EM06}.
If there were no decrease in the real value of the minimum wage, however,
the educational wage differentials would increase less uniformly across
quantiles.

The experience wage differentials also increased roughly uniformly,
although they increased slightly more in the higher quantiles than
the lower quantiles (Figures \ref{fig: diff_exper_actual_1989}).
If there were no decrease in the real value of the minimum wage, however,
the experience wage differentials would increase more differently
across quantiles.

The gender wage differential declined more in the higher quantiles
than the lower quantiles. If there were no decrease in the real value
of the minimum wage, however, the gender wage differential would decline
more uniformly across quantiles.

The 90/10, 50/10, and 50/20 within-group wage differentials changed
little for male workers and increased for female workers. For female
workers, the magnitude of the increase in within-group wage differentials
is similar for less-educated and more-educated workers but greater
for more-experienced than less-experienced workers. If there were
no decrease in the real value of the minimum wage, however, within-group
wage differentials would increase much less especially for workers
with 5 or less years of experience.

\end{document}